\pdfoutput=1
\documentclass[aps,prb,twocolumn,superscriptaddress,showpacs]{revtex4-1}

\usepackage{amsmath,amssymb,latexsym,multirow}
\usepackage[pdftex]{graphicx}
\usepackage{txfonts}
\DeclareSymbolFont{symbols}{OMS}{cmsy}{m}{n}
\DeclareSymbolFont{largesymbols}{OMX}{cmex}{m}{n}
\renewcommand{\bm}[1]{\boldsymbol #1}
\newcommand{\lh}{\mbox{$\neg$}}
\newcommand{\rh}{\reflectbox{$\neg$}}
\renewcommand{\bar}{\overline}

\begin{document}

\title{
Nonequilibrium dynamical mean-field theory based on weak-coupling perturbation expansions:
Application to dynamical symmetry breaking in the Hubbard model
}

\author{Naoto Tsuji}
\affiliation{Department of Physics, University of Tokyo, 113-0033 Tokyo, Japan}
\author{Philipp Werner}
\affiliation{Department of Physics, University of Fribourg, 1700 Fribourg, Switzerland}

\begin{abstract}
We discuss the general formalism and validity of weak-coupling perturbation theory as an impurity solver for nonequilibrium dynamical mean-field theory.
The method is implemented and tested in the Hubbard model, using expansions up to fourth order for the paramagnetic phase at half filling
and third order for the antiferromagnetic and paramagnetic phase away from half filling.
We explore various types of weak-coupling expansions and examine the accuracy and applicability of the methods for equilibrium and nonequilibrium problems.
We find that in most cases an expansion of local self-energy diagrams 
including all the tadpole diagrams with respect to the Weiss Green's function (bare-diagram expansion) gives
more accurate results than other schemes such as self-consistent perturbation theory using the fully interacting Green's function
(bold-diagram expansion). In the paramagnetic phase at half filling, the fourth-order bare expansion improves the result of the second-order expansion
in the weak-coupling regime, 
while both expansions suddenly fail at some intermediate interaction strength. 
The higher-order bare perturbation is especially advantageous in the antiferromagnetic phase near half filling.
We use the third-order bare perturbation expansion within the nonequilibrium dynamical mean-field theory
to study dynamical symmetry breaking from the paramagnetic to the antiferromagnetic phase
induced by an interaction ramp in the Hubbard model. The results show that the order parameter, after an initial exponential growth,
exhibits an amplitude oscillation around a nonthermal value followed by a slow drift toward the thermal value.
The transient dynamics seems to be governed by a nonthermal critical point, associated with a nonthermal universality class, which is distinct from the conventional Ginzburg-Landau theory.
\end{abstract}

\date{\today}

\pacs{71.10.Fd, 64.60.Ht}

\maketitle

\section{Introduction}

The study of nonequilibrium phenomena in correlated quantum systems is an active and rapidly expanding field,
which is driven by the progress of time-resolved spectroscopy experiments
in solids \cite{Ogasawara2000, Iwai2003, Perfetti2006, Okamoto2007, Wall2011}
and experiments on ultracold atoms trapped in an optical lattice.\cite{BlochDalibardZwerger2008, Joerdens2008, Schneider2008}
Recent studies are revealing ultrafast dynamics of phase transitions and order parameters, 
which include the melting of charge density waves (CDW),\cite{Schmitt2008,Hellmann2010,Petersen2011}
nonequilibrium dynamics of superconductivity,\cite{Matsunaga2012}
photoinduced transient transitions to superconductivity,\cite{Fausti2011,Kaiser2012} 
and the observation of the amplitude mode in CDW materials\cite{Demsar1999,Yusupov2010,Torchinsky2013} 
and the Higgs mode in an $s$-wave superconductor.\cite{Matsunaga2013}
Such experiments offer a testing ground for the study of dynamical phase transitions and dynamical symmetry breaking\cite{Kibble1976,Zurek1985} in real materials.
They also raise important theoretical issues related to the description of nonequilibrium phenomena in correlated systems.
One is the possible appearance of nonthermal quasistationary states that are inaccessible in equilibrium,
such as prethermalized states,\cite{BergesBorsanyiWetterich2004, MoeckelKehrein2008, EcksteinKollarWerner2009}
which can be interpreted as states controlled by nonthermal fixed points.\cite{BergesRothkoptSchmidt2008, WernerTsujiEckstein2012, TsujiEcksteinWerner2012} 
For example, it has been suggested that a symmetry-broken ordered state can survive for a long time in a nonthermal situation
in which the excitation energy corresponds to a temperature higher than the thermal critical temperature.\cite{WernerTsujiEckstein2012, TsujiEcksteinWerner2012} 
Such a state does not exist in equilibrium, so that the concept of nonthermal fixed points drastically extends the possibility for the presence of long-range order.
Another aspect is the long-standing theoretical issue of how to characterize a nonequilibrium phase transition 
and its critical behavior.\cite{HohenbergHalperin1977,Polkovnikov2011}

Since the dynamical phase transition that we are interested in occurs very far from equilibrium, where the temporal variation
of the order parameter is not particularly slow, we need a theoretical description of nonequilibrium many-body systems based on a ``microscopic theory,''
without employing a macroscopic coarsening or a phenomenological description (e.g., the time-dependent Ginzburg-Landau equation).
The nonequilibrium dynamical mean-field theory (DMFT) \cite{SchmidtMonien2002, FreericksTurkowskiZlatic2006, 
AokiTsujiEcksteinKollarOkaWerner2013} is one such approach, which has been recently developed. 
It is a nonequilibrium generalization of the equilibrium DMFT \cite{GeorgesKotliarKrauthRozenberg1996}
that maps a lattice model onto an effective local impurity problem embedded in a dynamical mean-field bath.
It takes account of dynamical correlation effects, while spatial correlations are ignored. 
The formalism becomes exact in the large dimensional limit.\cite{MetznerVollhardt1989} 
Furthermore, it can describe the dynamics of symmetry-broken states with a long-range (commensurate) order.\cite{WernerTsujiEckstein2012,TsujiEcksteinWerner2012}
Since DMFT is based on a mean-field description, it allows to treat directly the thermodynamic limit (i.e., the calculations are free from finite-size effects).

To implement the nonequilibrium DMFT, one requires an impurity solver. Previously, several approaches have been employed, including
the continuous-time quantum Monte Carlo (QMC) method,\cite{MuehlbacherRabani2008,WernerOkaMillis2009,SchiroFabrizio2009,GullMillisLichtensteinRubtsovTroyerPhilipp2011}
the noncrossing approximation and its generalizations (strong-coupling perturbation theory),\cite{EcksteinWerner2010}
and the exact diagonalization.\cite{Arrigoni2013,Gramsch2013}
In this paper, we explore the weak-coupling perturbation theory as an impurity solver for the nonequilibrium DMFT.
Our aim is to establish a method that is applicable to relatively long-time simulations of nonequilibrium impurity problems in the weak-coupling regime,
where a lot of interesting nonequilibrium physics remains unexplored. In particular, our interest lies in simulating dynamical symmetry breaking toward ordered states
such as the antiferromagnetic (AFM) phase. QMC is numerically exact, but suffers from a dynamical sign problem,\cite{WernerOkaMillis2009}
which prohibits sufficiently long simulation times. An approximate diagrammatic approach, such as the weak-coupling perturbation theory, allows one
to let the system evolve up to times which are long enough to capture order-parameter dynamics.

Perturbation theory is a standard and well-known diagrammatic technique \cite{AbrikosovGorkovDzyaloshinskiBook,RammerBook, KamenevBook}
to solve quantum many-body problems
in the weak-coupling regime. It has been successfully applied to the study of the equilibrium Anderson impurity model.
\cite{YosidaYamada,HorvaticZlatic}
Although it is an expansion with respect to the ratio ($U/\pi\Delta$) between the interaction strength $U$ and the hybridization to a conduction bath $\Delta$, 
it has turned out to be a very good approximation up to moderate $U/\pi\Delta$.
Later the weak-coupling perturbation theory was employed as an impurity solver for the equilibrium DMFT.
\cite{GeorgesKotliar1992,ZhangRozenbergKotliar1993,
Freericks1994,FreericksJarrell1994,GebhardJeckelmannMahlertNishimotoNoack2003}
Especially the bare second-order perturbation [which is usually referred to as the iterated perturbation theory (IPT)]
was found to accidentally reproduce the strong-coupling limit and the Mott insulator-metal transition.
\cite{ZhangRozenbergKotliar1993,GeorgesKotliarKrauthRozenberg1996}
It was also applied to nonequilibrium quantum impurity problems.\cite{HershfieldDaviesWilkins,FujiiUeda} 
The nonequilibrium DMFT has been solved by the second-order perturbation theory
in the paramagnetic (PM) phase of the Hubbard model at half filling
\cite{EcksteinKollarWerner2010,AronKotliarWeber2012,Amaricci2012,TsujiOkaAokiWerner2012}
and by the third-order perturbation theory in the AFM phase.\cite{TsujiEcksteinWerner2012}
However, a thorough investigation of weak-coupling perturbation theory, including higher orders, as a nonequilibrium DMFT solver has been lacking so far.

The paper is organized as follows. In Sec.~\ref{nonequilibrium dmft afm}, we give an overview of the nonequilibrium DMFT formalism,
putting an emphasis on the treatment of the AFM phase. In Sec.~\ref{weak-coupling perturbation theory}, 
we present a general formulation of the nonequilibrium weak-coupling perturbation theory following the Kadanoff-Baym \cite{KadanoffBaymBook} 
and Keldysh \cite{Keldysh1964} formalism. We discuss various issues of the perturbation theory related to bare- and bold-diagram expansions,
symmetrization of the interaction term, and the treatment of the Hartree diagram. 
After testing various implementations of the perturbation theory for the equilibrium phases of the Hubbard model in Sec.~\ref{application to equilibrium}, 
we examine the applicability of the method to nonequilibrium problems without long-range order in Sec.~\ref{quench paramagnetic}.
Finally, in Sec.~\ref{ramp symmetry breaking}, we apply the third-order perturbation theory to the nonequilibrium DMFT
to study dynamical symmetry breaking to the AFM phase of the Hubbard model 
induced by an interaction ramp. By comparing the results with those of the phenomenological Ginzburg-Landau theory
and time-dependent Hartree approximation, we find that the order parameter does not directly thermalize but is ``trapped'' to a nonthermal value
around which an amplitude oscillation occurs. We show that the transient dynamics of the order parameter is governed by a 
nonthermal critical point,\cite{TsujiEcksteinWerner2012} 
and we characterize the associated nonthermal universality class. 
In the Appendix \ref{appendix:Dyson}, we provide details of the numerical implementation of the nonequilibrium Dyson equation
that must be solved in the nonequilibrium DMFT calculations.

\section{Nonequilibrium dynamical mean-field theory for the antiferromagnetic phase}
\label{nonequilibrium dmft afm}

We first review the formulation of the nonequilibrium DMFT including the antiferromagnetically
ordered state.\cite{WernerTsujiEckstein2012,TsujiEcksteinWerner2012} 
It is derived in a straightforward way by extending the ordinary nonequilibrium DMFT for the PM phase 
to one having an $AB$ sublattice dependence. The general structure of the formalism is analogous to other symmetry-broken phases with a commensurate long-range order.
For demonstration, we take the single-band Hubbard model, 
\begin{align}
H(t)
&=
\sum_{\bm k\sigma} \epsilon_{\bm k}(t) c_{\bm k\sigma}^\dagger c_{\bm k\sigma}
-\mu\sum_{i\sigma}\hat{n}_{i\sigma}
+U(t)\sum_i \hat{n}_{i\uparrow}\hat{n}_{i\downarrow},
\label{Hubbard}
\end{align}
where $\epsilon_{\bm k}$ is the band dispersion, $c_{\bm k\sigma}^\dagger$ ($c_{\bm k\sigma}$) is the creation (annihilation) operator,
$\hat{n}_{i\sigma}=c_{i\sigma}^\dagger c_{i\sigma}$ is the density operator,
$\mu$ is the chemical potential, and $U$ is the on-site interaction strength. $\epsilon_{\bm k}$ and $U$ may have a time dependence.
Let us assume that the lattice structure that we are interested in is a bipartite lattice, which has an $AB$ sublattice distinction.

\begin{figure}[tbp]
\begin{center}
\includegraphics[width=8cm]{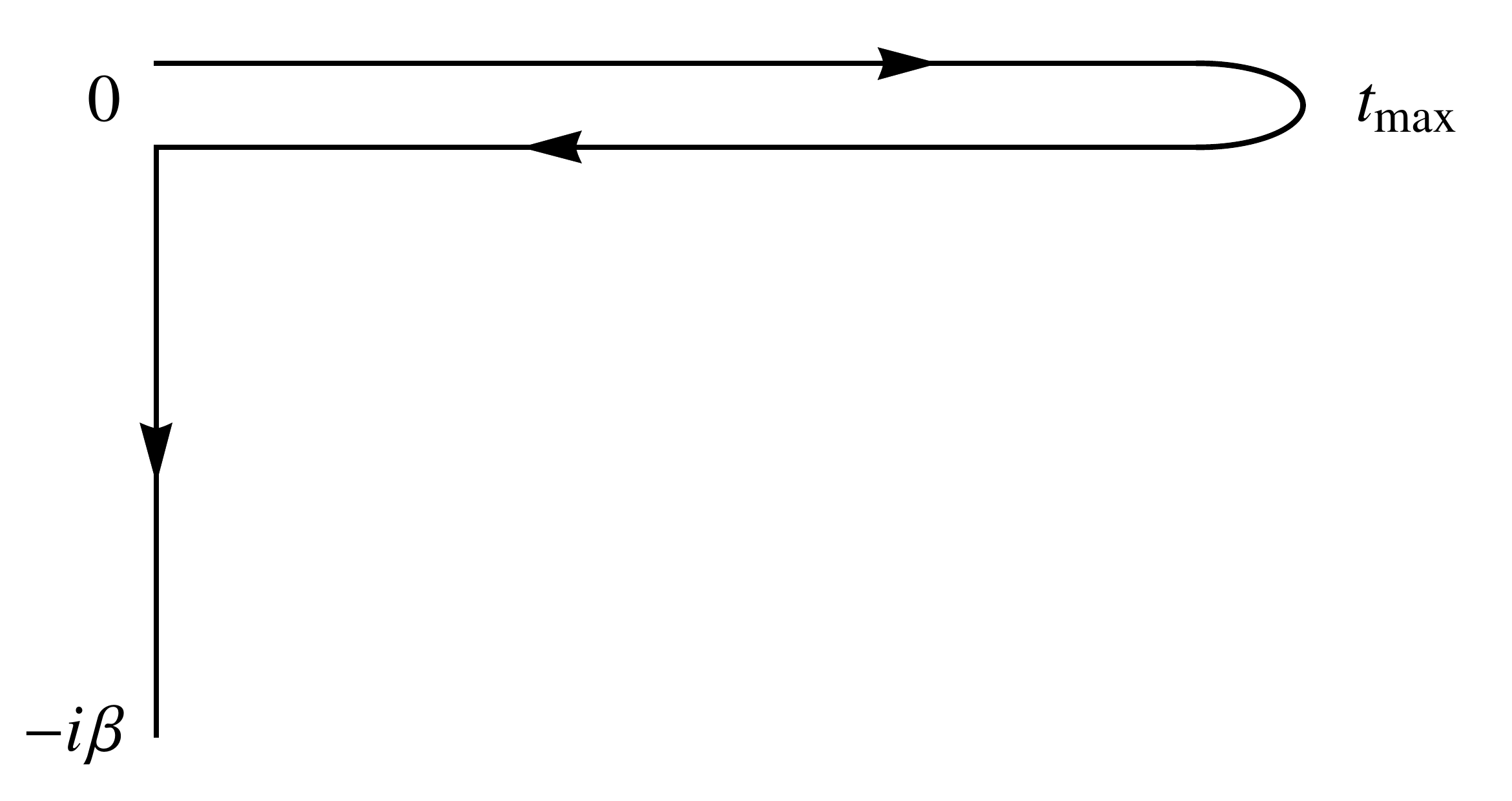}
\caption{The Kadanoff-Baym contour $\mathcal C$.}
\label{contour}
\end{center}
\end{figure}
In the DMFT construction, one maps the lattice model (\ref{Hubbard}) onto an effective single-site impurity model. 
In principle, one has to consider 
two independent impurity problems depending on whether the impurity site
corresponds to a lattice site on the $A$ or the $B$ sublattice.
The impurity action for the $a=A, B$ sublattice is defined by
\begin{align}
\mathcal S_a^{\rm imp}[\Delta]
&=
\int_{\mathcal C} dt \int_{\mathcal C}dt' \sum_\sigma d_{\sigma}^\dagger(t)\Delta_{a\sigma}(t,t') d_{\sigma}(t')
\nonumber
\\
&\quad
-\int_{\mathcal C} dt \sum_\sigma \mu\hat{n}_{\sigma}(t)
+\int_{\mathcal C} dt\, U(t)\hat{n}_{\uparrow}(t)\hat{n}_{\downarrow}(t).
\label{S^imp}
\end{align}
Here $d_\sigma^\dagger$ ($d_\sigma$) is the creation (annihilation) operator for the impurity energy levels, 
$\Delta_{a\sigma}(t,t')$ is the hybridization function on the $a$ sublattice, which is self-consistently determined in DMFT, 
$\hat{n}_{\sigma}=d_\sigma^\dagger d_\sigma$, and $\mathcal{C}$ is the Kadanoff-Baym contour depicted in Fig.~\ref{contour}.
The contour runs in the time domain from $t=0$ to $t_{\rm max}$, up to which the system time evolves,
comes back to $t=0$, and proceeds to $-i\beta$, which corresponds to the initial thermal equilibrium state with temperature $\beta^{-1}$.
Using the impurity action (\ref{S^imp}), one can define the nonequilibrium Green's function as
\begin{align}
G_{a\sigma}^{\rm imp}(t,t')
  &=
    -i\,{\rm Tr}\left(\mathcal{T}_{\mathcal{C}}\, e^{-i\mathcal S_a^{\rm imp}[\Delta]} d_{\sigma}(t) d_{\sigma}^\dagger(t')\right)/Z_a^{\rm imp},
\label{impurity Green}
\end{align}
where $\mathcal{T}_{\mathcal{C}}$ 
time orders the operators 
along the contour $\mathcal{C}$ represented by the arrows in Fig.~\ref{contour}, 
and $Z_a^{\rm imp}={\rm Tr}\, \mathcal{T}_{\mathcal{C}}\, e^{-i\mathcal S_a^{\rm imp}[\Delta]}$.

On the other hand, one has the lattice Green's function,
\begin{align}
G_{ij,\sigma}^{\rm lat}(t,t')
&=
-i\,{\rm Tr}\left(\mathcal T_{\mathcal C}\, e^{-i\mathcal S^{\rm lat}} c_{i\sigma}(t)c_{j\sigma}^\dagger(t')\right)/Z^{\rm lat},
\label{lattice Green}
\end{align}
with $\mathcal S^{\rm lat}=\int_{\mathcal C} dt\, H(t)$ and $Z^{\rm lat}=\,{\rm Tr}\,\mathcal T_{\mathcal C}\, e^{-i\mathcal S^{\rm lat}}$.
The hybridization function $\Delta_{a\sigma}$ is implicitly determined such that the local part of the lattice Green's function (\ref{lattice Green})
coincides with the impurity Green's function (\ref{impurity Green}),
\begin{align}
G_{ii,\sigma}^{\rm lat}[\Delta](t,t')=G_{a\sigma}^{\rm imp}(t,t') \quad (i\in a),
\label{G^lat=G^imp}
\end{align}
where $i\in a$ means that the lattice site labeled by $i$ belongs to the $a=A, B$ sublattice.
The essential ingredient of DMFT is the approximation that the lattice self-energy is local in space,
based on which one requires the local lattice self-energy to be identical to the impurity self-energy,
\begin{align}
\Sigma_{ij,\sigma}^{\rm lat}(t,t')=\delta_{ij}\Sigma_{a\sigma}^{\rm imp}(t,t') \quad (i\in a).
\label{S^lat=S^imp}
\end{align}
With this condition, the self-consistency relation between the lattice and impurity models is closed, 
and the nonequilibrium DMFT for the AFM phase is formulated.
In the following, we omit the labels ``lat'' and ``imp'' thanks to the identifications (\ref{G^lat=G^imp}) and (\ref{S^lat=S^imp}).

To implement the self-consistency condition in practice, one uses the Dyson equation.
In solving the lattice Dyson equation, it is efficient to work in momentum space, where the lattice Green's function
is Fourier transformed to
\begin{align}
G_{\bm k\sigma}^{ab}(t,t')
&=
N^{-1}\sum_{i\in a, j\in b} e^{-i\bm k\cdot (\bm R_i-\bm R_j)} G_{ij,\sigma}^{\rm lat}(t,t'),
\end{align}
with $N$ the number of sublattice sites. Then the lattice Dyson equation reads
\begin{align}
\begin{pmatrix}
i\partial_t+\mu-\Sigma_{A\sigma} & -\epsilon_{\bm k} \\
-\epsilon_{\bm k} & i\partial_t+\mu-\Sigma_{B\sigma}
\end{pmatrix}
\ast
\begin{pmatrix}
G_{\bm k\sigma}^{AA} & G_{\bm k\sigma}^{AB} \\
G_{\bm k\sigma}^{BA} & G_{\bm k\sigma}^{BB}
\end{pmatrix}
=
\begin{pmatrix}
\delta_{\mathcal C} & 0 \\
0 & \delta_{\mathcal C}
\end{pmatrix}.
\label{lattice Dyson}
\end{align}
Here $\ast$ represents a convolution on the contour $\mathcal C$, 
$\epsilon_{\bm k}(t,t')=\epsilon_{\bm k}(t)\delta_{\mathcal C}(t,t')$,
and $\delta_{\mathcal C}(t,t')$ is the $\delta$ function defined on $\mathcal C$.
The local Green's function is obtained from a momentum summation. If the system has an inversion symmetry (we consider only this case here),
the off-diagonal components of the local Green's function vanish, and we have
\begin{align}
\sum_{\bm k} G_{\bm k\sigma}^{ab}
&\equiv
G_{a\sigma}\delta_{ab}.
\label{momentum sum}
\end{align}
The local Green's function satisfies the Dyson equation for the impurity problem,
\begin{align}
G_{a\sigma}
=
\mathcal G_{0,a\sigma}+
\mathcal G_{0,a\sigma} \ast \Sigma_{a\sigma} \ast G_{a\sigma},
\label{impurity Dyson}
\end{align}
where 
\begin{align}
\mathcal G_{0,a\sigma}=(i\partial_t+\mu-\Delta_{a\sigma})^{-1}
\end{align}
is the Weiss Green's function. Thus we obtained a closed set of nonequilibrium DMFT self-consistency relations:
(\ref{lattice Dyson}), (\ref{momentum sum}), and (\ref{impurity Dyson}), for 
$\Sigma_{a\sigma}$, $\mathcal G_{0,a\sigma}$ (or $\Delta_{a\sigma}$), and $G_{a\sigma}$.
The calculation of $G_{a\sigma}$ from $\mathcal G_{0,a\sigma}$ is the task of the impurity solver.

Before finishing this section, let us comment on how to solve the lattice Dyson equation (\ref{lattice Dyson}).
Due to the existence of the AFM long-range order, it has a $2\times 2$ matrix structure, i.e., 
consists of coupled integral-differential equations.
However, as we show below, it can be decoupled to a set of integral-differential equations of the form
\begin{align}
[i\partial_t-\epsilon(t)] G(t,t')-\int_{\mathcal C} d\bar{t}\, \Sigma(t,\bar{t})G(\bar{t},t')=\delta_{\mathcal C}(t,t'),
\label{intdiff eq}
\end{align}
and integral equations of the form
\begin{align}
G(t,t')-\int_{\mathcal C} d\bar{t}\, K(t,\bar{t})G(\bar{t},t')=G_0(t,t').
\label{integral eq}
\end{align}
To see this, let us denote the lattice Green's function $G_{\bm k\sigma}^{aa}$ ($a=A, B$) for $\epsilon_{\bm k}=0$ by $g_{a\sigma}$.
It satisfies
\begin{align}
(i\partial_t+\mu-\Sigma_{a\sigma})\ast g_{a\sigma}=\delta_{\mathcal C}(t,t'),
\end{align}
which is in the form of Eq.~(\ref{intdiff eq}).
Using $g_{a\sigma}$, we can explicitly write the solution for Eq.~(\ref{lattice Dyson}),
\begin{align}
G_{\bm k\sigma}^{AA}
&=
(1-g_{A\sigma}\ast\epsilon_{\bm k}\ast g_{B\sigma}\ast\epsilon_{\bm k})^{-1} \ast g_{A\sigma},
\\
G_{\bm k\sigma}^{BB}
&=
(1-g_{B\sigma}\ast\epsilon_{\bm k}\ast g_{A\sigma}\ast\epsilon_{\bm k})^{-1} \ast g_{B\sigma},
\\
G_{\bm k\sigma}^{AB}
&=
G_{\bm k\sigma}^{AA}\ast \epsilon_{\bm k}\ast g_{B\sigma},
\\
G_{\bm k\sigma}^{BA}
&=
G_{\bm k\sigma}^{BB}\ast \epsilon_{\bm k}\ast g_{A\sigma}.
\end{align}
By substituting $F=g_{a\sigma}\ast\epsilon_{\bm k}\ast g_{\bar{a}\sigma}\ast\epsilon_{\bm k}$ 
($\bar{a}$ denotes the sublattice opposite to $a$) and $Q=g_{a\sigma}$,
we have $(1-F)\ast G_{\bm k\sigma}^{aa}=Q$, which is exactly of the form of Eq.~(\ref{integral eq}).
Since Eqs.~(\ref{intdiff eq}) and (\ref{integral eq}) are implemented in the standard nonequilibrium DMFT without long-range orders,
one can recycle those subroutines to solve Eq.~(\ref{lattice Dyson}).

For the case of the semicircular density of states (DOS), $D(\epsilon)=\sqrt{4\nu_\ast^2-\epsilon^2}/(2\pi\nu_\ast^2)$,
and $\epsilon_{\bm k}(t)=\epsilon_{\bm k}$ (time independent), 
one can analytically take the momentum summation for the lattice Green's function, resulting in the relation
\begin{align}
\Delta_{a\sigma}(t,t')=v_\ast^2 G_{\bar{a}\sigma}(t,t').
\end{align}
Thus, instead of solving Eq.~(\ref{lattice Dyson}), one can make use of
\begin{align}
(i\partial_t+\mu-v_\ast^2 G_{\bar{a}\sigma}) \ast \mathcal{G}_{0,a\sigma}
&=
\delta_{\mathcal C}(t,t')
\label{semicircular DOS}
\end{align}
as the DMFT self-consistency condition. The DMFT calculations in the rest of the paper are done for the semicircular DOS,
and we use $v_\ast$ ($v_\ast^{-1}$) as a unit of energy (time).
With the symmetry $G_{\bar{a}\sigma}=G_{a\bar{\sigma}}$ in the AFM phase, it is sufficient to consider the impurity problem
for one of the two sublattices so that we can drop the sublattice label $a$.

In a practical implementation of the nonequilibrium DMFT self-consistency, what one has to numerically solve are basically
equations of the forms (\ref{intdiff eq}) and (\ref{integral eq}).
These are Volterra integral(-differential) equations of the second kind.
Various numerical algorithms for them can be found in the literature.\cite{NumericalRecipesC,LinzBook,BrunnervanderHouwenBook}
Here we use the fourth-order implicit Runge-Kutta method (or the collocation method). The details of the implementation are presented in the Appendix \ref{appendix:Dyson}.

\section{Weak-coupling perturbation theory}
\label{weak-coupling perturbation theory}

In this section, we explain the general formalism of the weak-coupling perturbation theory for nonequilibrium quantum impurity problems.
It is explicitly implemented up to third order for the AFM phase at arbitrary filling and fourth order for the PM phase at half filling.
We discuss various technical details of the perturbation theory, including the symmetrization of the interaction term, bare and bold diagrams,
and the treatment of the Hartree term.

\subsection{General formalism}

To define the perturbation expansion for the nonequilibrium impurity problem,
we split the impurity action (\ref{S^imp}) into a noninteracting part $\mathcal S_{0}^{\rm imp}$ and an interacting part $\mathcal S_{1}^{\rm imp}$ 
($\mathcal S^{\rm imp}=\mathcal S_{0}^{\rm imp}+\mathcal S_{1}^{\rm imp}$),
\begin{align}
\mathcal S_{0}^{\rm imp}
&=
\int_{\mathcal C} dt \int_{\mathcal C}dt' \sum_\sigma d_{\sigma}^\dagger(t)\Delta_{\sigma}(t,t') d_{\sigma}(t')
\nonumber
\\
&\quad
-\int_{\mathcal C} dt\, \sum_\sigma \left[\mu-U(t)\alpha_{\bar{\sigma}}\right]\hat{n}_{\sigma}(t),
\label{S_0^imp}
\\
\mathcal S_{1}^{\rm imp}
&=
\int_{\mathcal{C}} dt\, U(t)\left(\hat{n}_{\uparrow}(t)-\alpha_\uparrow\right)
\left(\hat{n}_{\downarrow}(t)-\alpha_\downarrow\right).
\label{S_1^imp}
\end{align}
Here we have introduced auxiliary constants $\alpha_\sigma$ to symmetrize the interaction term.
Accordingly, the chemical potential in $\mathcal S_{0}^{\rm imp}$ is shifted,
and the Weiss Green's function is modified into
\begin{align}
\mathcal G_{0,\sigma}=(i\partial_t+\mu-U\alpha_{\bar{\sigma}}-\Delta_\sigma)^{-1}.
\end{align}
As a result, the self-consistency condition for the case of the semicircular DOS is changed from Eq.~(\ref{semicircular DOS}) to
\begin{align}
(i\partial_t+\mu-U\alpha_{\bar{\sigma}}-v_\ast^2 G_{\bar{\sigma}})\ast \mathcal G_{0,\sigma}=\delta_{\mathcal C}.
\label{modified semicircular DOS}
\end{align}
Physical observables should not, in principle, depend on the choice of $\alpha_\sigma$, 
whereas the quality of the approximation made by the perturbation theory may depend on it.
Such $\alpha_\sigma$ parameters have been used to suppress the sign problem in the continuous-time QMC method.\cite{Rubtsov2005,EcksteinKollarWerner2010,WernerOkaEcksteinMillis2010}

The weak-coupling perturbation theory for nonequilibrium problems is formulated in a straightforward way as a generalization
of the equilibrium perturbation theory in the Matsubara formalism.\cite{AbrikosovGorkovDzyaloshinskiBook,MahanBook}
We expand the exponential in Eq.~(\ref{impurity Green}) into a Taylor series with respect to the interaction term,
\begin{align}
G_{\sigma}(t,t')
&=
(-i)\frac{1}{Z^{\rm imp}}\sum_{n=0}^\infty \frac{(-i)^n}{n!} \int_{\mathcal{C}} dt_1 \cdots dt_n 
\nonumber
\\
&\quad\times
{\rm Tr}\left( \mathcal{T}_{\mathcal{C}}\, e^{-i\mathcal S_{0}^{\rm imp}} H_1(t_1)\cdots H_1(t_n) 
d_\sigma(t) d_\sigma^\dagger(t') \right),
\label{expanded Green function}
\end{align}
where $H_1(t)=U(t)(\hat{n}_\uparrow-\alpha_\uparrow)(\hat{n}_\downarrow-\alpha_\downarrow)$.
The linked cluster theorem ensures that all the disconnected diagrams that contribute to Eq.~(\ref{expanded Green function}) 
can be factorized to give a proportionality constant $Z^{\rm imp}/Z_{0}^{\rm imp}$
with $Z_{0}={\rm Tr}\,\mathcal{T}_{\mathcal{C}}\, e^{-i\mathcal S_{0}^{\rm imp}}$.
As a result, the expansion can be expressed in the simplified form 
\begin{align}
G_{\sigma}(t,t')
&=
(-i)\sum_{n=0}^\infty (-i)^n \int_{\mathcal{C}, t_1\prec\cdots\prec t_n} dt_1 \cdots dt_n
\nonumber
\\
&\quad\times
\langle \mathcal{T}_{\mathcal{C}}\, H_1(t_1)\cdots H_1(t_n) 
d_\sigma(t) d_\sigma^\dagger(t') \rangle_{0}^{\text{conn.}},
\label{connected Green function}
\end{align}
where $\langle\cdots\rangle_{0}$ denotes 
${\rm Tr}\, (\mathcal{T}_{\mathcal{C}}\, e^{-i\mathcal S_{0}^{\rm imp}}\cdots)/Z_{0}^{\rm imp}$, 
and ``conn.'' means that one only takes account of connected diagrams. The factor $n!$ is canceled by specifying the contour ordering
as $t_1\prec\cdots\prec t_n$ ($t_1$ comes first and $t_n$ last). Owing to Wick's theorem, one can evaluate each term in Eq.~(\ref{connected Green function})
using the Weiss Green's function, 
\begin{align}
\mathcal{G}_{0,\sigma}(t,t')=-i\langle \mathcal{T}_{\mathcal{C}}\, d_\sigma(t)d_\sigma^\dagger(t')\rangle_{0}.
\label{Weiss Green}
\end{align}

\begin{figure}[tbp]
\includegraphics[width=8cm]{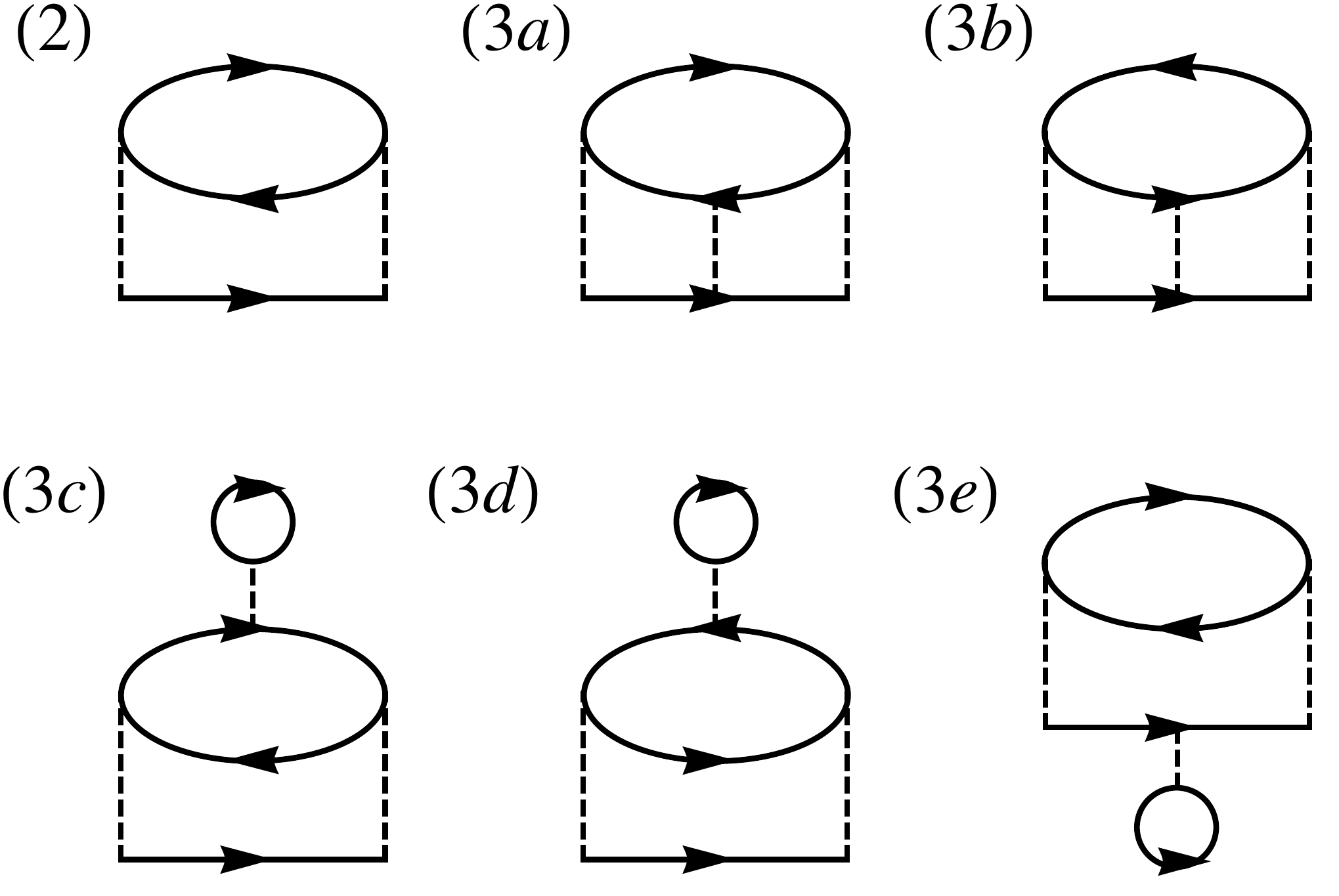}
\caption{The self-energy diagrams up to third order (except for the Hartree diagrams). 
The solid lines represent the fermion propagator, while the dashed lines are interaction vertices.}
\label{third-order diagrams}
\end{figure}

In the standard weak-coupling perturbation theory, one usually considers an expansion of the self-energy $\Sigma_\sigma(t,t')$ instead of the Green's function. 
This is because one can then take into account an infinite series of diagrams for the Green's function by solving the Dyson equation.
The self-energy consists of one-particle irreducible diagrams of the expansion (\ref{connected Green function}), i.e., the diagrams that cannot be disconnected 
by cutting a fermion propagator. Figure~\ref{third-order diagrams} shows examples of Feynman diagrams
for the self-energy. 
In addition, we have tadpole diagrams.
Since the quadratic terms in $H_1$ [$U(t)\alpha_{\bar{\sigma}}\hat{n}_\sigma$] play the role of counterterms to the tadpoles, each tadpole diagram 
amounts to $n_{0,\sigma}(t)-\alpha_\sigma$, where $n_{0,\sigma}(t)=-i\mathcal{G}_{0,\sigma}^<(t,t)$.
We summarize the Feynman rules to calculate the self-energy diagrams:
\begin{enumerate}
\item Draw topologically distinct one-particle irreducible diagrams.
\item Associate the Weiss Green's function $-i\mathcal{G}_{0,\sigma}(t,t')$ with each solid line.
\item Multiply $(-i)U(t)$ for each interaction vertex (dashed line).
\item Multiply $n_{0,\sigma}(t)-\alpha_\sigma$ for each tadpole diagram.
\item Multiply $(-1)$ for each Fermion loop.
\item Multiply an additional factor $(-i)$, coming from the definition of the Green's function (\ref{impurity Green}).
\item Carry out a contour integral along $\mathcal{C}$ for each internal vertex.
\end{enumerate}
One notices that, if the weak-coupling perturbation theory is employed as an impurity solver, $\Delta_\sigma$ does not explicitly appear in the DMFT calculation.
Instead, $\mathcal G_{0,\sigma}$ represents the dynamical mean field.

We show several examples of the application of the Feynman rules above in Sec.~\ref{third order} (third order) and \ref{fourth order} (fourth order).

\subsection{Self-consistent perturbation theory}

Instead of expanding the self-energy diagrams with respect to the Weiss Green's function $\mathcal G_{0,\sigma}$, one can also expand it with respect to the fully interacting Green's function $G_\sigma$.
In this expansion, each bare propagator $\mathcal G_{0,\sigma}(t,t')$ (depicted by a thin line) is replaced by the dressed propagator $G_\sigma(t,t')$ (bold line).
Since $G_\sigma$ itself already contains an infinite number of diagrams, which is recursively generated from the Dyson Eq.~(\ref{impurity Dyson}),
one can take account of many more diagrams than in the expansion with respect to $\mathcal G_{0,\sigma}$.
To avoid a double counting of diagrams in this expansion, we take the ``skeleton diagrams'' of the self-energy, i.e., two-particle irreducible diagrams
that cannot be disconnected by cutting two fermion propagators, which reduces the number of diagrams to be considered.

At first, $G_\sigma$ is not known, so that one starts with an initial guess of $G_\sigma$ (which is usually chosen to be $\mathcal G_{0,\sigma}$).
Using the perturbation theory, one evaluates the self-energy $\Sigma_\sigma$ from $G_\sigma$. Plugging $\Sigma_\sigma$ into the Dyson Eq.~(\ref{impurity Dyson}),
one obtains a new $G_\sigma$, which is again used to evaluate the self-energy. One iterates this procedure until $G_\sigma$ and $\Sigma_\sigma$ converge.
In this way, $G_\sigma$ and $\Sigma_\sigma$ are determined self-consistently within the perturbation theory (hence named the self-consistent perturbation theory).

In the self-consistent perturbation theory, there is a short cut in implementing the DMFT self-consistency. Since the self-energy is determined from 
the local Green's function $G_\sigma$, the Weiss Green's function $\mathcal G_{0,\sigma}$ does not explicitly appear in the calculation.
Thus, one can skip the evaluation of $\mathcal G_{0,\sigma}$ with the impurity Dyson Eq.~(\ref{impurity Dyson}).
For the case of the semicircular DOS, one can eliminate $\mathcal G_{0,\sigma}$ from Eqs.~(\ref{impurity Dyson}) and (\ref{semicircular DOS}) to obtain
\begin{align}
G_\sigma=(i\partial_t+\mu-U\alpha_{\bar{\sigma}}-v_\ast^2 G_{\bar{\sigma}}-\Sigma_\sigma)^{-1},
\end{align}
which defines the DMFT self-consistency condition.

Let us remark that the self-consistent perturbation theory is a ``conserving approximation'',\cite{BaymKadanoff1961} i.e., it automatically guarantees the conservation of global quantities
such as the total energy and the particle number. The perturbation theory defines the self-energy as a functional of $G$, $\Sigma=\Sigma[G]$,
which is a sufficient condition to preserve the conservation laws. It is important that the conservation law is satisfied in a simulation of the time evolution
to obtain physically meaningful results. 
However, this does not necessarily mean that the self-consistent perturbation theory is superior to a nonconserving approximation (such as the expansion with respect to $\mathcal G_{0,\sigma}$).
As we see in Sec.~\ref{application to equilibrium} and \ref{quench paramagnetic}, under some conditions the nonconserving approximation (despite small violations of conservation laws) reproduces the correct dynamics more accurately than the conserving approximation.

One can also consider a combination of bare- and bold-diagram expansions. An often used combination is to take the bold diagram for the Hartree term (Fig.~\ref{hartree}) and bare diagrams for the other parts of the self-energy.
This kind of expansion is necessarily a nonconserving approximation.
We examine this type of approximations in Sec.~\ref{application to equilibrium}.

\subsection{Treatment of the Hartree term}
\label{treatment of Hartree}

\begin{figure}[htbp]
\includegraphics[width=1.5cm]{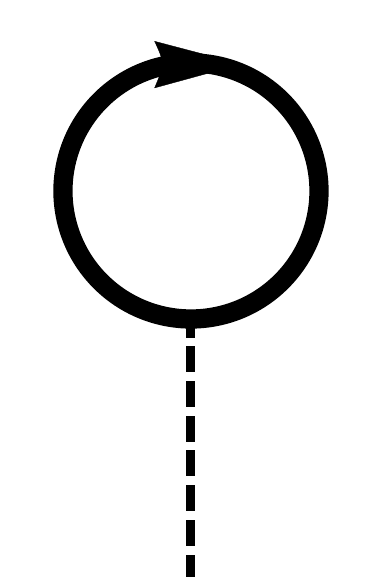}
\caption{The Hartree diagram. The bold line represents the interacting Green's function $G_\sigma$.}
\label{hartree}
\end{figure}

There is a subtle issue concerning the treatment of the Hartree term in the self-energy diagrams. The Hartree term is the portion of the self-energy $\Sigma(t,t')$
that is proportional to $\delta_{\mathcal C}(t,t')$. Let us denote it by 
\begin{align}
\Sigma_\sigma^{\rm Hartree}(t,t')=h_\sigma(t)\delta_{\mathcal C}(t,t').
\end{align}
The corresponding diagram, summed up to infinite order in $U$, is given by the bold tadpole shown in Fig.~\ref{hartree}. It reads
\begin{align}
h_\sigma(t)
&=
U(t)(n_{\bar{\sigma}}(t)-\alpha_{\bar{\sigma}}),
\label{hartree eq}
\end{align}
where $n_\sigma(t)=-iG_\sigma^<(t,t)$ is the physical density. Since the Hartree term (\ref{hartree eq}) is written with 
the interacting Green's function, it is determined self-consistently within the perturbation theory.
\begin{figure}[htbp]
\includegraphics[width=6.5cm]{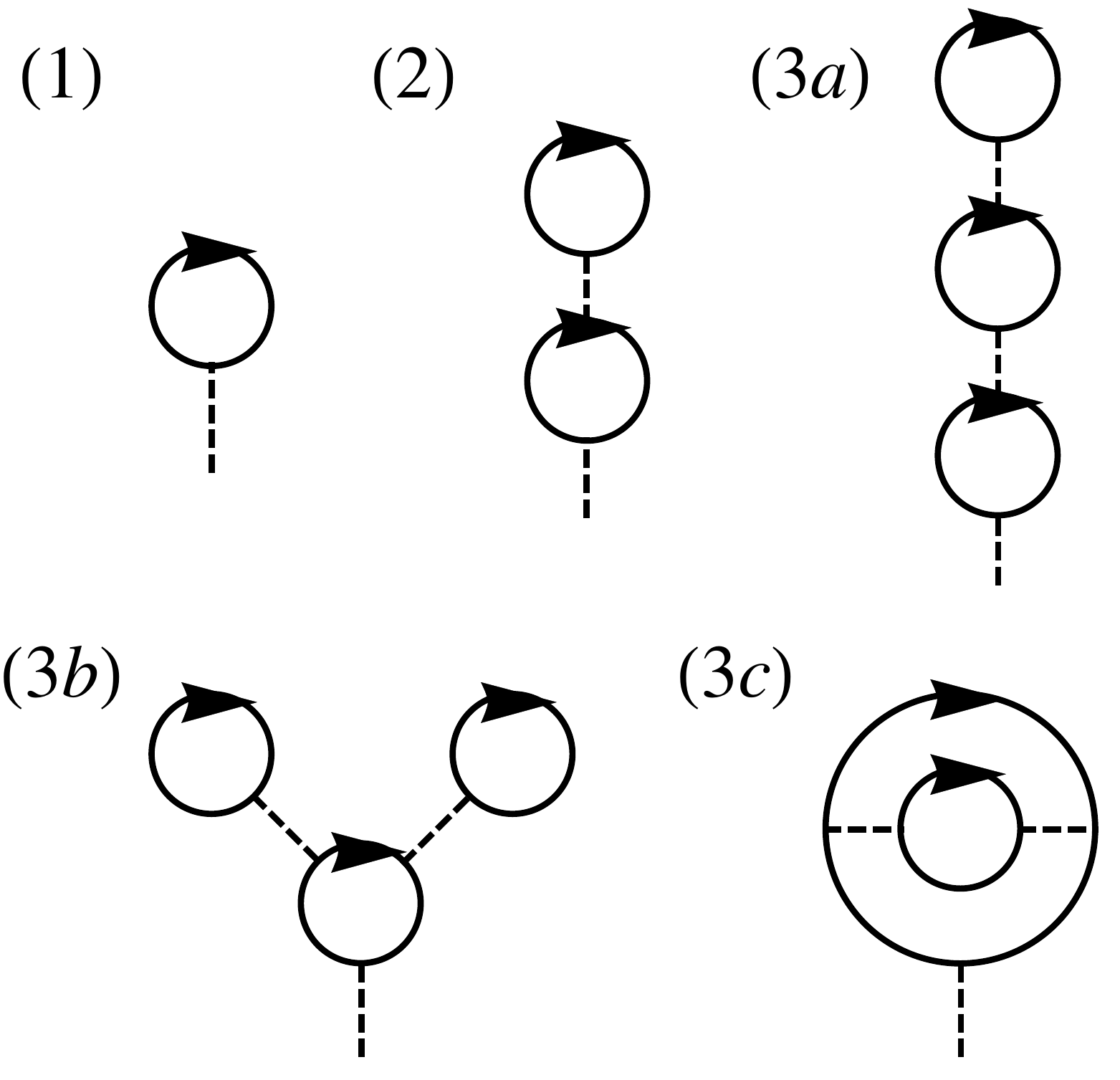}
\caption{An explicit expansion of the Hartree term in the bare Green's function $\mathcal G_{0,\sigma}$ up to third order.}
\label{hartree diagrams}
\end{figure}

Although the expression (\ref{hartree eq}) is exact up to infinite order in $U$ and it seems natural to use it, 
it is useful to consider an expansion of the Hartree diagram  
with respect to $\mathcal G_{0,\sigma}$. We show the resulting diagrams up to third order in Fig.~\ref{hartree diagrams}.
In this ``bare-diagram'' expansion, a lot of internal tadpoles are generated in the Hartree diagrams. 
Each tadpole gives a contribution of $n_{0,\sigma}(t)-\alpha_{\sigma}$ instead of $n_{\sigma}(t)-\alpha_{\sigma}$. 
The question is which is the better approximation.
[Note that Eq.~(\ref{hartree eq}) itself is exact, but if it is combined with other diagrams, it becomes an approximation.]
The bold Hartree term (Fig.~\ref{hartree}) includes many more diagrams than the bare expansion (Fig.~\ref{hartree diagrams});
however, the answer is not {\it a priori} obvious.

Moreover, we have the freedom to choose the constant $\alpha_\sigma$.
At half-filling and in the PM phase, it is natural to take $\alpha_\sigma=\frac{1}{2}$ because of the particle-hole symmetry.
It cancels all the tadpole diagrams since $n_\sigma-\alpha_\sigma=n_{0,\sigma}-\alpha_\sigma=0$. Due to the particle-hole symmetry 
[i.e., $\mathcal{G}_{0,\sigma}(t,t')=-\mathcal{G}_{0,\sigma}(t',t)$], all the odd-order diagrams vanish as well.
On the other hand, when the system is away from half filling or is spin-polarized (i.e., $n_\uparrow\neq n_\downarrow$), 
the particle-hole symmetry (for each spin) is lost, and we do not have a solid guideline to choose the value of $\alpha_\sigma$.
$\alpha_\sigma$ can be $n_\sigma$, $n_{0,\sigma}$, or some other fixed values (such as $\frac{1}{2}$). 
\begin{table}[b]
\begin{center}
\begin{tabular}{|c||c|c|c|c|c|}
\hline
& I & II & III & IV & V \\
\hline\hline
nontadpole & bare & bare & bare & bare & bold \\
\hline
tadpole & bare & bare & bold & bold & bold \\
\hline
$\alpha_\sigma$ & 1/2 & $n_{0,\sigma}$ & 1/2 & $n_\sigma$ & $\alpha_\sigma$ \\
\hline
\multirow{2}{1.9cm}{contribution of tadpole} & \multirow{2}{*}{$n_{0,\sigma}-1/2$} & \multirow{2}{*}{0} & \multirow{2}{*}{$n_\sigma-1/2$} & \multirow{2}{*}{0} & \multirow{2}{*}{$n_\sigma-\alpha_\sigma$} \\
& & & & & 
\\
\hline
\end{tabular}
\caption{Choices of the nontadpole and tadpole diagrams (bare or bold) 
and the constant $\alpha_\sigma$. The bottom row shows the contribution of each tadpole diagram. For the case (V), $\alpha_\sigma$ can be arbitrary.}
\label{n-alpha}
\end{center}
\end{table}

Later, in Sec.~\ref{application to equilibrium}, we examine these issues for the Hartree term by considering five representative cases summarized in Table~\ref{n-alpha}.
It might look better if one sets $\alpha_\sigma=n_\sigma$ for the bold diagrams or $\alpha_\sigma=n_{0,\sigma}$ for the bare diagrams,
since all the tadpole diagrams are then shifted into the propagator $G_\sigma$ or $\mathcal G_{0,\sigma}$.
However, it turns out that this choice is not a particularly good approximation. Due to cancellations among different diagrams, the naive expectation that more diagrams means better results is misleading.

\subsection{Third-order perturbation theory}
\label{third order}

In the case of the spin-polarized phase or away from half filling, when the particle-hole symmetry [$G_\sigma(t,t')=-G_\sigma(t',t)$] is lost, 
it becomes important to take into account the odd-order diagrams.
Here we consider the third-order weak-coupling perturbation theory. First, we look at the bare-diagram expansion.
We have shown topologically distinct Feynman diagrams of the self-energy up to third order in Fig.~\ref{third-order diagrams}
and the bare Hartree diagrams up to third order in Fig.~\ref{hartree diagrams}. 

Using the Feynman rules presented in Sec.~\ref{treatment of Hartree}, we can explicitly write the contribution of each diagram.
The self-energy at second order [Fig.~\ref{third-order diagrams}(2)] is given by
\begin{align}
\Sigma_\sigma^{(2)}(t,t')
  &=
    U(t)U(t')\mathcal{G}_{0,\sigma}(t,t')\mathcal{G}_{0,\bar{\sigma}}(t',t)\mathcal{G}_{0,\bar{\sigma}}(t,t'),
\end{align}
and the first two of the self-energy diagrams at third order [Figs.~\ref{third-order diagrams}(3a) and \ref{third-order diagrams}(3b)] are given by
\begin{align}
&\Sigma_\sigma^{(3a)}(t,t')
=
iU(t)U(t')\mathcal{G}_{0,\bar{\sigma}}(t,t')
\nonumber
\\
&\times
\int_{\mathcal{C}} d\bar{t}\, U(\bar{t})\mathcal{G}_{0,\sigma}(t,\bar{t}) \mathcal{G}_{0,\sigma}(\bar{t},t')
\mathcal{G}_{0,\bar{\sigma}}(t',\bar{t}) \mathcal{G}_{0,\bar{\sigma}}(\bar{t},t),
\\
&\Sigma_\sigma^{(3b)}(t,t')
=
iU(t)U(t') \mathcal{G}_{0,\bar{\sigma}}(t',t)
\nonumber
\\
&\times
\int_{\mathcal{C}} d\bar{t}\, U(\bar{t})\mathcal{G}_{0,\sigma}(t,\bar{t}) \mathcal{G}_{0,\sigma}(\bar{t},t')
\mathcal{G}_{0,\bar{\sigma}}(t,\bar{t}) \mathcal{G}_{0,\bar{\sigma}}(\bar{t},t').
\end{align}
To write the rest of the self-energy diagrams at third order [Figs.~\ref{third-order diagrams}(3c)-\ref{third-order diagrams}(3e)],
it is convenient to define a contour-ordered function,
\begin{align}
\chi_{1,\sigma}^{(3)}(t,t')
&\equiv
\int_{\mathcal C} d\bar{t}\, U(\bar{t})\left(n_{\bar{\sigma}}(\bar{t})-\alpha_{\bar{\sigma}}\right)
\mathcal G_{0,\sigma}(t,\bar{t})\mathcal G_{0,\sigma}(\bar{t},t'),
\end{align}
which takes care of the internal tadpoles. With $\chi_{1,\sigma}^{(3)}$,
we can write the self-energy diagrams [Figs.~\ref{third-order diagrams}(3c)-\ref{third-order diagrams}(3e)] as
\begin{align}
\Sigma_\sigma^{(3c)}(t,t')
&=
U(t)U(t')\mathcal G_{0,\sigma}(t,t')\mathcal G_{0,\bar{\sigma}}(t',t)\chi_{1,\bar{\sigma}}^{(3)}(t,t'),
\\
\Sigma_\sigma^{(3d)}(t,t')
&=
U(t)U(t')\mathcal G_{0,\sigma}(t,t')\mathcal G_{0,\bar{\sigma}}(t,t')\chi_{1,\bar{\sigma}}^{(3)}(t',t),
\\
\Sigma_\sigma^{(3e)}(t,t')
&=
U(t)U(t')\mathcal G_{0,\bar{\sigma}}(t,t')\mathcal G_{0,\bar{\sigma}}(t',t)\chi_{1,\sigma}^{(3)}(t,t').
\end{align}

In the bare-diagram expansion, we need to evaluate the Hartree diagrams (Fig.~\ref{hartree diagrams}).
To this end, we define another contour-ordered function,
\begin{align}
\chi_{2,\sigma}^{(3)}(t,t')
&=
\int_{\mathcal C}d\bar{t}\,U(\bar{t})\mathcal G_{0,\sigma}(t,\bar{t})\mathcal G_{0,\sigma}(\bar{t},t)
\mathcal G_{0,\bar{\sigma}}(t,\bar{t})\mathcal G_{0,\bar{\sigma}}(\bar{t},t').
\end{align}
With $\chi_{1,\sigma}^{(3)}$ and $\chi_{2,\sigma}^{(3)}$, each Hartree diagram in Fig.~\ref{hartree diagrams} reads
\begin{align}
h_\sigma^{(1)}(t)
&=
U(t)(n_{0,\bar{\sigma}}(t)-\alpha_{\bar{\sigma}}),
\\
h_\sigma^{(2)}(t)
&=
-iU(t)\chi_{1,\bar{\sigma}}(t,t),
\\
h_\sigma^{(3a)}(t)
&=
-U(t)\int_{\mathcal C}d\bar{t}\, U(\bar{t})\mathcal G_{0,\bar{\sigma}}(t,\bar{t})\mathcal G_{0,\bar{\sigma}}(\bar{t},t)\chi_{1,\sigma}^{(3)}(\bar{t},\bar{t}),
\\
h_\sigma^{(3b)}(t)
&=
-iU(t)\int_{\mathcal C}d\bar{t}\, U(\bar{t})(n_{0,\sigma}(\bar{t})-\alpha_\sigma)\mathcal G_{0,\bar{\sigma}}(t,\bar{t})\chi_{1,\bar{\sigma}}^{(3)}(\bar{t},t),
\\
h_\sigma^{(3c)}(t)
&=
-iU(t)\int_{\mathcal C}d\bar{t}\, U(\bar{t})\mathcal G_{0,\bar{\sigma}}(t,\bar{t})\chi_{2,\sigma}^{(3)}(\bar{t},t).
\end{align}

One can see that in the third-order perturbation the calculation of each self-energy diagram includes a single contour integral at most.
This means that the computational cost of the impurity solution is of $O(N^3)$ with $N$ the number of discretized time steps. It is thus of the same order as solving
the Dyson equation [in the form of Eq.~(\ref{intdiff eq}) or Eq.~(\ref{integral eq})] or calculating a convolution of two contour-ordered functions.
This means that the impurity problem can be solved with a cost comparable to the DMFT self-consistency part, which is crucial 
for simulating the long-time evolution.

In the self-consistent version of the third-order perturbation theory, we consider the two-particle irreducible diagrams.
The diagrams of Figs.~\ref{third-order diagrams}(2), \ref{third-order diagrams}(3a), and \ref{third-order diagrams}(3b) are two-particle irreducible, whereas the others in Fig.~\ref{third-order diagrams} are reducible. The Hartree term is given 
by the bold diagram (Fig.~\ref{hartree}). The equations to represent each diagram are the same as those for the bare diagrams,
except that all $\mathcal G_{0,\sigma}$ are replaced by $G_\sigma$.

\subsection{Fourth-order perturbation theory for the paramagnetic phase at half filling}
\label{fourth order}

\begin{figure}[htbp]
\includegraphics[width=8cm]{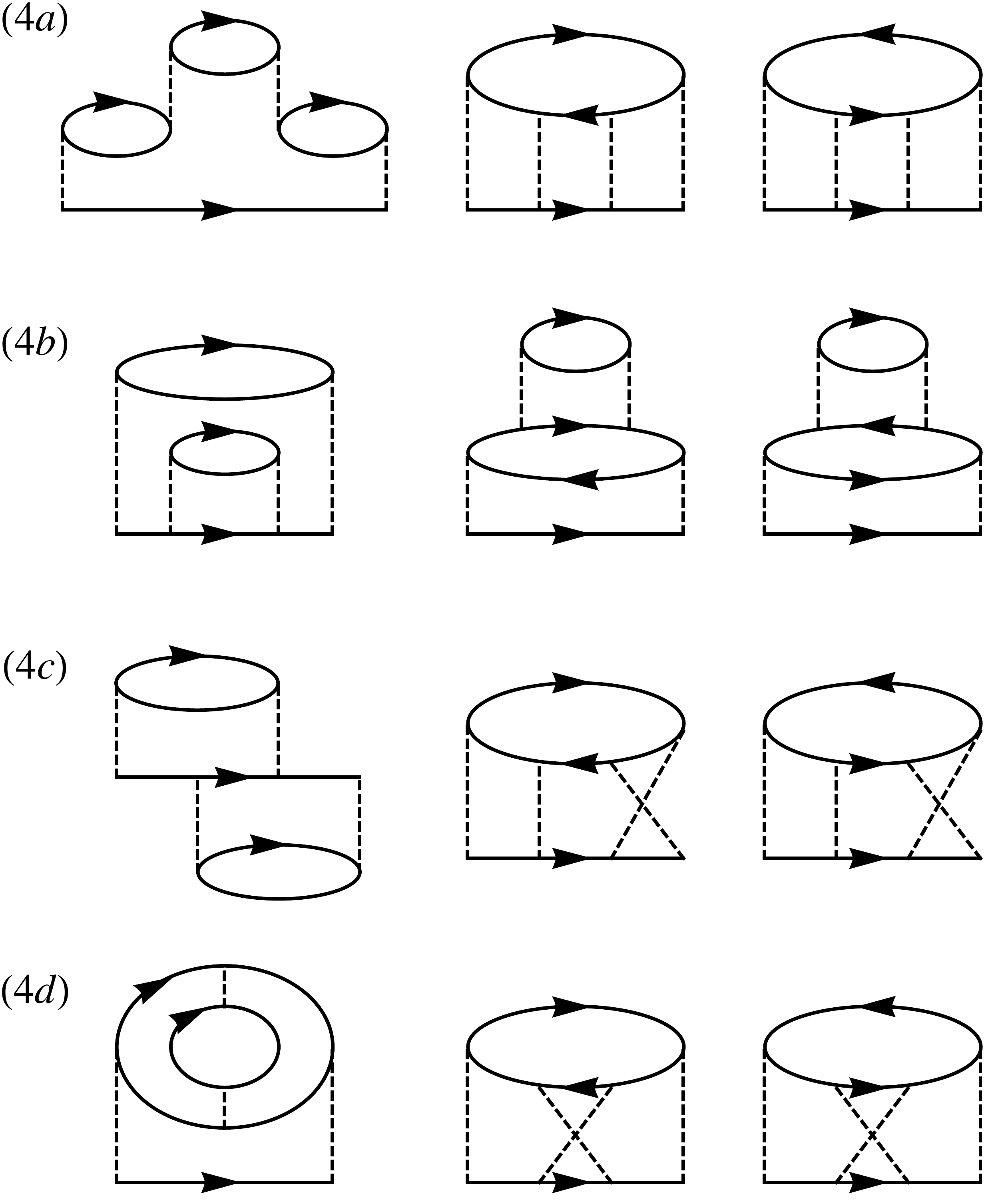}
\caption{The nonvanishing self-energy diagrams at fourth order for the PM phase at half filling.}
\label{fourth-order diagrams}
\end{figure}

For the PM phase at half filling, we consider the fourth-order perturbation theory.
At fourth order, the number of diagrams that we have to consider dramatically increases,
so that we restrict ourselves to the case where the particle-hole symmetry holds.
In this case, the odd-order diagrams disappear. We take $\alpha_\sigma=\frac{1}{2}$ to cancel all the tadpoles and the Hartree term.
What remains are the second-order diagram [Fig.~\ref{third-order diagrams}(2)]
and 12 fourth-order diagrams,\cite{YosidaYamada,Freericks1994,FreericksJarrell1994,GebhardJeckelmannMahlertNishimotoNoack2003}
as shown in Fig.~\ref{fourth-order diagrams}. 

We classify the 12 diagrams in four groups [Figs.~\ref{fourth-order diagrams}(4a)-\ref{fourth-order diagrams}(4d)], each of which contains three diagrams. 
Using the particle-hole symmetry $\mathcal G_{0,\sigma}(t,t')=-\mathcal G_{0,\sigma}(t',t)$, one can show that those three classified in the same group
give exactly the same contribution.\cite{FreericksJarrell1994} Thanks to this fact, it is enough to consider one of the three for each group. In total, 
the number of diagrams to be computed is reduced to four. We represent this simplification by writing the fourth-order self-energy as
\begin{align}
\Sigma^{(4)}(t,t')=3[\Sigma^{(4a)}(t,t')+\Sigma^{(4b)}(t,t')+\Sigma^{(4c)}(t,t')+\Sigma^{(4d)}(t,t')],
\end{align}
where we have omitted the spin label $\sigma$. We can explicitly evaluate each contribution of the self-energy diagrams to obtain 
\begin{align}
&\Sigma^{(4a)}(t,t')
=
U(t)U(t')\mathcal{G}_0(t,t')
\nonumber
\\
&\times
\int_{\mathcal{C}} d\bar{t} \int_{\mathcal{C}} d\bar{t}'\, U(\bar{t})U(\bar{t}')
\left[\mathcal{G}_0(t,\bar{t})\mathcal{G}_0(\bar{t},\bar{t}')\mathcal{G}_0(\bar{t}',t')\right]^2,
\\
&\Sigma^{(4b)}(t,t')
=
U(t)U(t')\left[\mathcal{G}_0(t,t')\right]^2
\nonumber
\\
&\times
\int_{\mathcal{C}} d\bar{t} \int_{\mathcal{C}} d\bar{t}'\, U(\bar{t})U(\bar{t}')
\mathcal{G}_0(t,\bar{t})\mathcal{G}_0(\bar{t}',t')
\left[\mathcal{G}_0(\bar{t},\bar{t}')\right]^3,
\\
&\Sigma^{(4c)}(t,t')
=
U(t)U(t')\int_{\mathcal{C}} d\bar{t} \int_{\mathcal{C}} d\bar{t}'\, U(\bar{t})U(\bar{t}')
\nonumber
\\
&\times
\mathcal{G}_0(t,\bar{t})\mathcal{G}_0(\bar{t},\bar{t}')\mathcal{G}_0(\bar{t}',t')
\left[\mathcal{G}_0(t,\bar{t}')\mathcal{G}_0(\bar{t},t')\right]^2,
\\
&\Sigma^{(4d)}(t,t')
=
-U(t)U(t')\mathcal{G}_0(t,t')
\int_{\mathcal{C}} d\bar{t} \int_{\mathcal{C}} d\bar{t}'\, U(\bar{t})U(\bar{t}')
\nonumber
\\
&\times
\mathcal{G}_0(t,\bar{t})\mathcal{G}_0(\bar{t},t')
\mathcal{G}_0(t,\bar{t}')\mathcal{G}_0(\bar{t}',t')
\left[\mathcal{G}_0(\bar{t},\bar{t}')\right]^2.
\end{align}
Note that they involve double contour integrals. However, for $\Sigma^{(4a)}$ and $\Sigma^{(4b)}$,
we can decouple the integrals by defining the contour functions
\begin{align}
\chi^{(4a)}(t,t')
&=
\int_{\mathcal C} d\bar{t}\, U(\bar{t})[\mathcal G_0(t,\bar{t})\mathcal G_0(\bar{t},t')]^2,
\\
\chi^{(4b)}(t,t')
&=
\int_{\mathcal C} d\bar{t}\, U(\bar{t})[\mathcal G_0(t,\bar{t})]^3\mathcal G_0(\bar{t},t')],
\end{align}
which involve single contour integrals.
With these, $\Sigma^{(4a)}$ and $\Sigma^{(4b)}$ can be rewritten as
\begin{align}
\Sigma^{(4a)}(t,t')
&=
U(t)U(t')\mathcal G_0(t,t') \int_{\mathcal C} d\bar{t}\, [\mathcal G_0(t,\bar{t})]^2\chi^{(4a)}(\bar{t},t'),
\\
\Sigma^{(4b)}(t,t')
&=
U(t)U(t')[\mathcal G_0(t,t')]^2 \int_{\mathcal C} d\bar{t}\, \mathcal G_0(t,\bar{t})\chi^{(4b)}(\bar{t},t'),
\end{align}
which again involves only single integrals. Unfortunately, this kind of reduction is not possible for $\Sigma^{(4c)}$ and $\Sigma^{(4d)}$.
Hence, the computational cost for the fourth-order diagrams is $O(N^4)$, which is one order higher than the calculation 
of the third-order diagrams or solving the DMFT self-consistency. The maximum time $t_{\rm max}$ up to which one can let the system evolve is
therefore quite limited compared to the third-order perturbation theory.

For the fourth-order self-consistent perturbation theory, we only take the two-particle irreducible diagrams among Fig.~\ref{fourth-order diagrams},
which are those grouped in (4a), (4c), and (4d).\cite{Freericks1994} 
The diagrams in (4b) are two-particle reducible, and are not considered in the self-consistent perturbation theory.

\section{Application to equilibrium phases}
\label{application to equilibrium}

To establish the validity of the weak-coupling perturbation theory as an impurity solver for DMFT,
we first apply it to the equilibrium phases of the Hubbard model. In particular, we focus on the PM phase (Sec.~\ref{paramagnetic phase}) away from half filling and 
the AFM phase (Sec.~\ref{antiferromagnetic phase}), where the conventional second-order perturbation theory fails.\cite{GeorgesKotliarKrauthRozenberg1996}
There has been a proposal to improve it for arbitrary filling 
by introducing control parameters in such a way that the perturbation theory recovers the correct strong-coupling limit.\cite{KajueterKotliar1996}
However, it is not known at this point how to generalize this approach to nonequilibrium situations.
Here we explore the different types of perturbation theories that have been overviewed in Sec.~\ref{weak-coupling perturbation theory} and clarify which ones
improve the quality of the approximation compared to previously known results.

\subsection{Paramagnetic phase}
\label{paramagnetic phase}

\begin{figure}[tbp]
\begin{center}
\includegraphics[width=7cm]{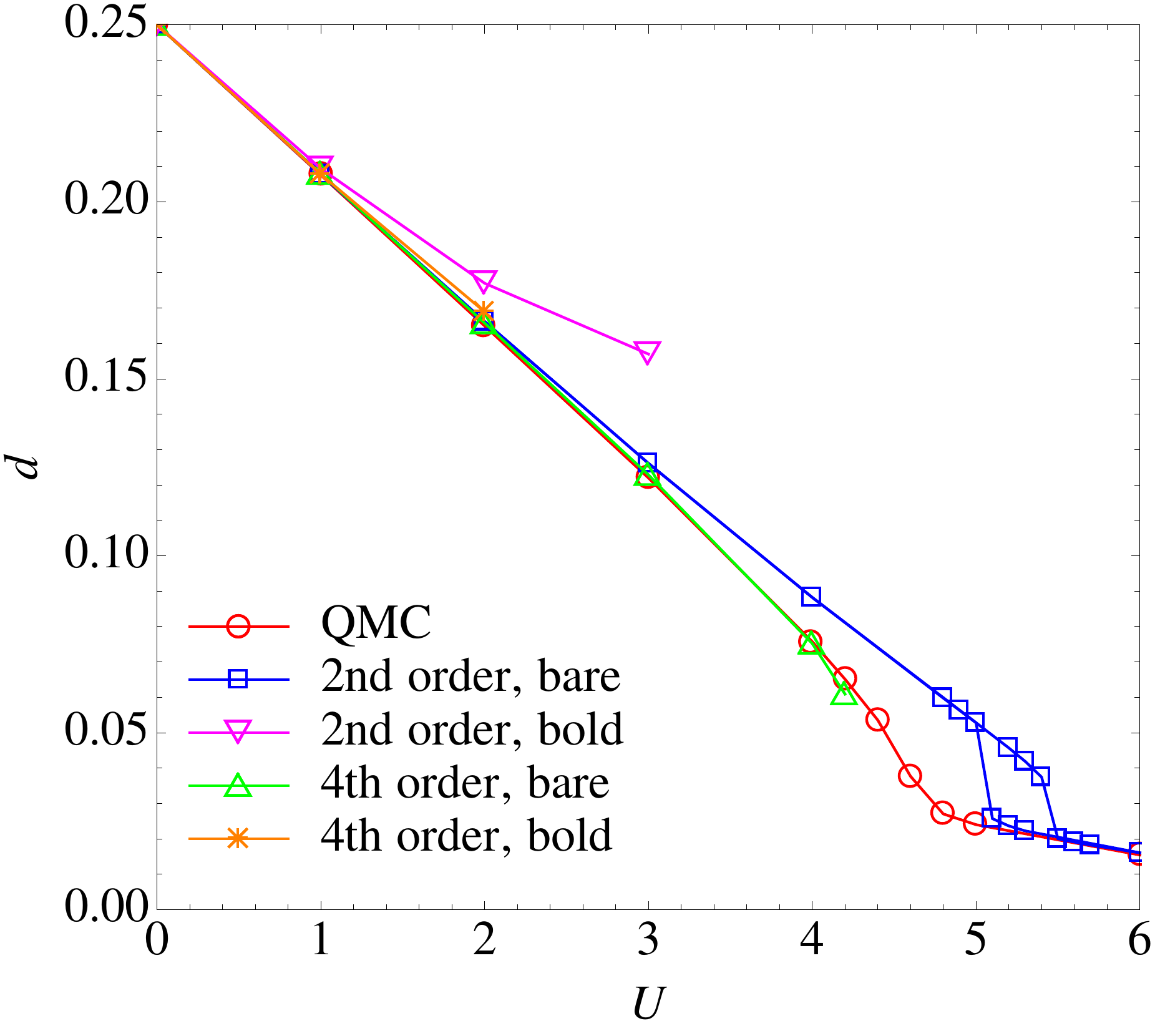}
\caption{(Color online) The double occupancy for the Hubbard model in equilibrium at half filling with $\beta=16$ calculated from DMFT
with various impurity solvers.}
\label{eq double occupancy}
\end{center}
\end{figure}

Let us consider the PM phase of the Hubbard model (\ref{Hubbard}), and first look at the half filled system. In Fig.~\ref{eq double occupancy},
we show the results for the double occupancy,
\begin{align}
d=\langle \hat{n}_\uparrow \hat{n}_\downarrow \rangle=\frac{1}{2U}\sum_\sigma (\Sigma_\sigma\ast G_\sigma)^M(0^-),
\end{align}
given by DMFT with various perturbation expansions. $d$ is a good measure of correlation effects.
As is well known, the second-order perturbation theory with bare diagrams (which is often referred to 
as the IPT) works remarkably well over the entire $U$ regime.\cite{GeorgesKotliarKrauthRozenberg1996} 
In particular, it captures the Mott metal-insulator transition.
Quantitatively, deviations from QMC (exact) start to appear around $U\sim 3$ in the weak-coupling regime.
The fourth-order bare expansion improves the results up to $U\sim 4$, just before the Mott transition occurs ($U\sim 5$).
It quickly fails to converge at $U\gtrsim 4.3$ (convergence is not recovered by mixing the old and new solutions during the DMFT iterations).
The bold diagrams (self-consistent perturbation theory) give worse results than the bare expansions (Fig.~\ref{eq double occupancy}). The second-order bold expansion
deviates from QMC at $U\sim 2$, and it does not converge at $U>3$. The fourth-order bold diagram improves the second-order bold results for $U\lesssim 2$, but
it fails to converge at $U>2$. Hence, at half filling the fourth-order bare expansion gives the best results in the weak-coupling regime ($U\le 4$).

\begin{figure}[tbp]
\begin{center}
\includegraphics[width=7cm]{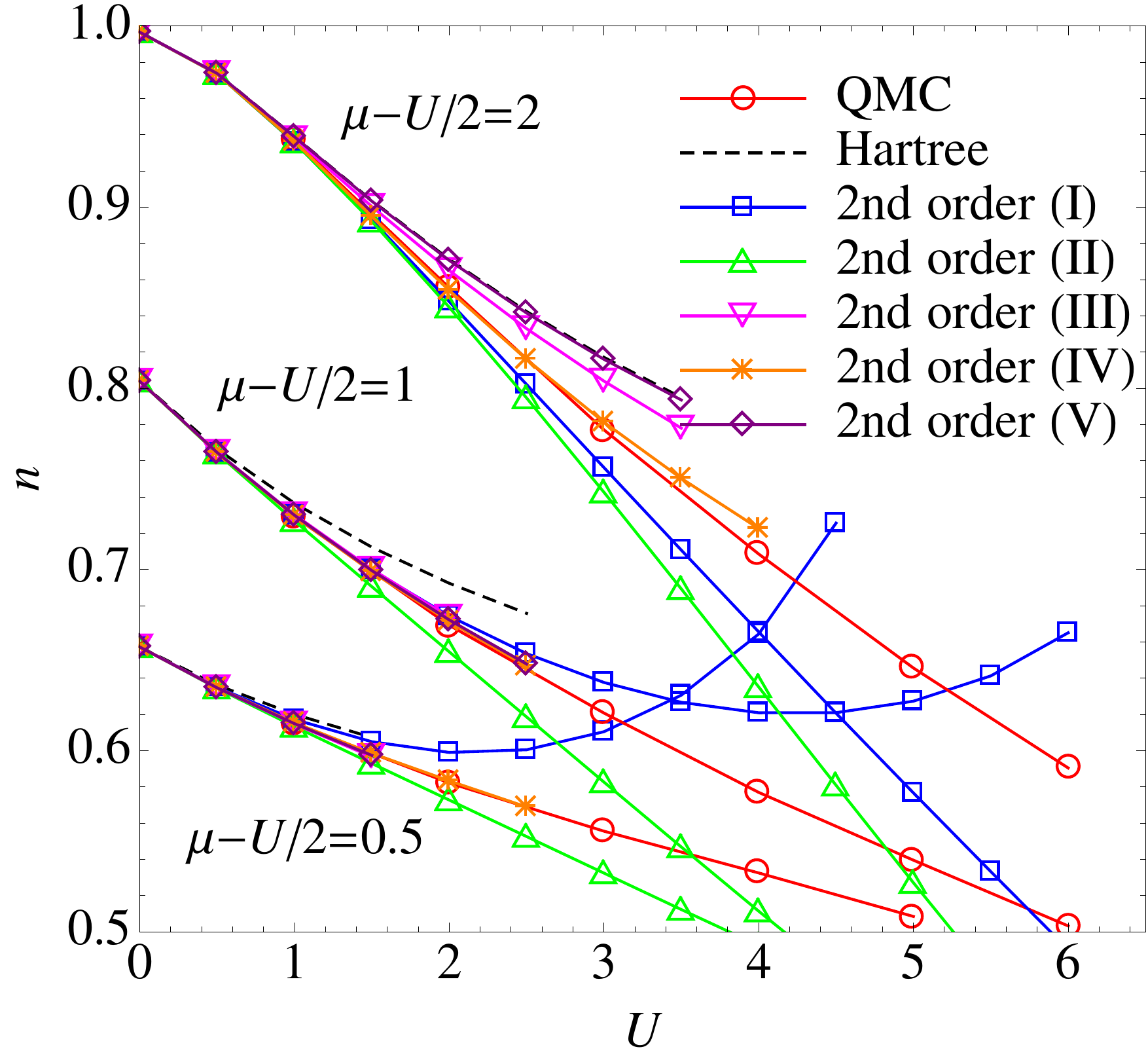}
\caption{(Color online) The density for the Hubbard model in equilibrium with $\beta=16$ calculated by DMFT
with QMC, Hartree approximation, and the second-order perturbation theories. The labels (I)-(V) correspond to the classification in Table~\ref{n-alpha}.}
\label{density second-order}
\end{center}
\end{figure}

\begin{figure}[tbp]
\begin{center}
\includegraphics[width=7cm]{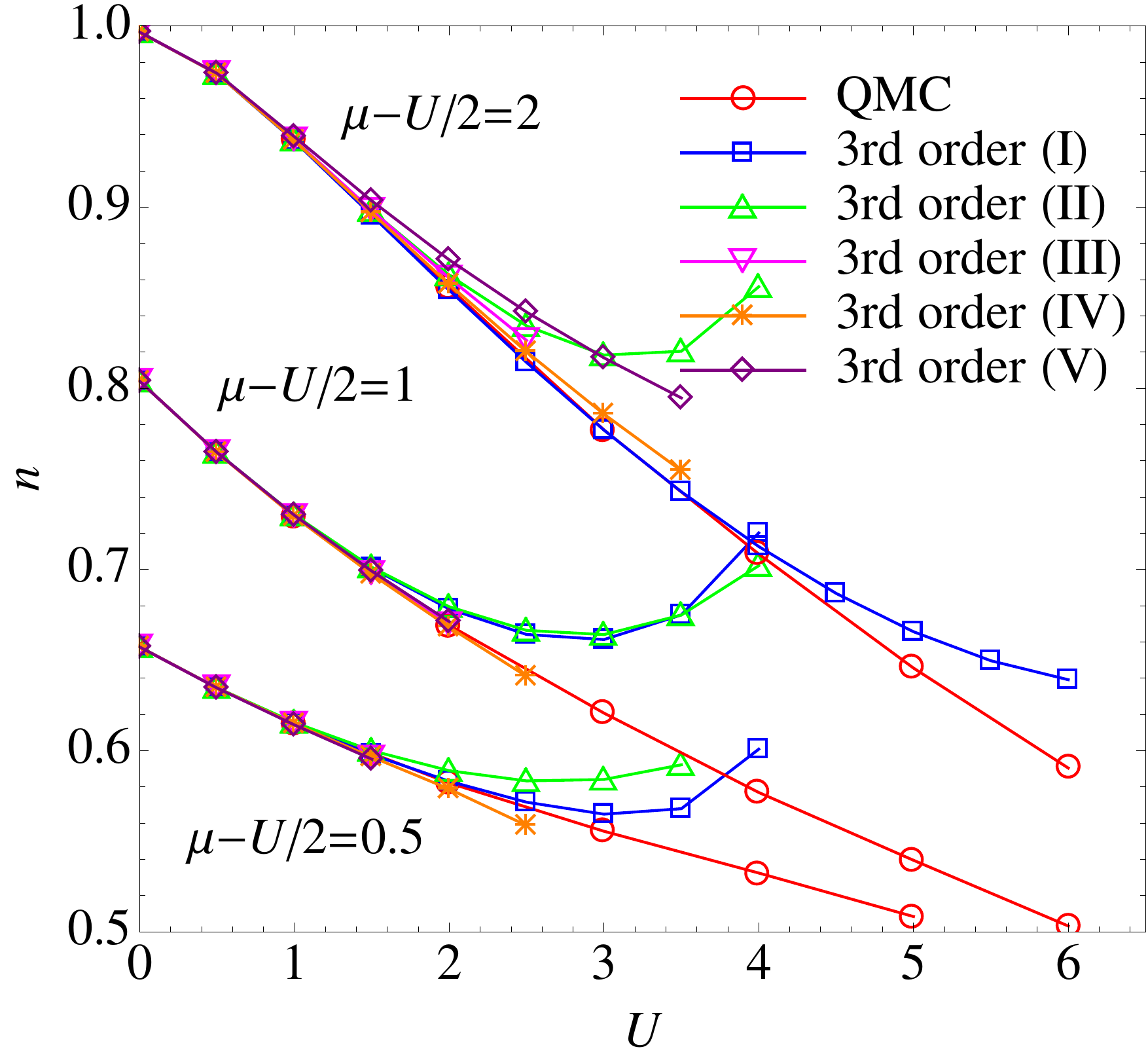}
\caption{(Color online) The density for the Hubbard model in equilibrium with $\beta=16$ calculated by DMFT
with the third-order perturbation theories. The labels (I)-(V) correspond to the classification in Table~\ref{n-alpha}.}
\label{density third-order}
\end{center}
\end{figure}

Away from half filling, we calculate the density per spin, $n=G^M(0^-)$, as a function of $U$ for a fixed chemical potential $\mu-U/2=0.5, 1, 2$.
The results obtained by DMFT with QMC, the Hartree approximation, and the second-order perturbation theories are shown in Fig.~\ref{density second-order},
while the results from the third-order perturbation expansions are shown in Fig.~\ref{density third-order}. 
We consider five types of perturbation expansions (I)-(V) as indicated in Table~\ref{n-alpha}. 
The QMC results indicate that there are Mott transitions in the strong $U$ regime (e.g., $U_c\sim 6$ for $\mu-U/2=1$),
where the density $n$ 
approaches 0.5. 
The Hartree approximation (dashed line in Fig.~\ref{density second-order}), which only includes the Hartree term (\ref{hartree eq}) as the self-energy correction,
deviates from QMC already at relatively small $U$ $(\sim 1)$. Among the various second-order expansions, type (IV) (bare second-order and bold Hartree diagrams with $\alpha_\sigma=n_\sigma$) 
seems to be closest to the QMC result up to $U\sim 2.5$ for $\mu-U/2=0.5, 1$ and $U\sim 4$ for $\mu-U/2=2$. 
However, this approach, as well as types (III) and (V), leads to a convergence problem in
the DMFT calculation as one goes to larger $U$ 
[which is why the lines for types (III)-(V) in Figs.~\ref{density second-order} and \ref{density third-order} are terminated]. On the other hand, type (I) easily converges even for large $U$.

By comparing the second- (Fig.~\ref{density second-order}) and third-order (Fig.~\ref{density third-order}) perturbation theories, we see a systematic improvement
of the results in most cases. In particular, the third-order type (I) becomes better than type (IV) and gets closest to QMC for $\mu-U/2=2$. It agrees with QMC up to $U\sim 4$.
Again type (I) shows
excellent convergence for the entire $U$ range, in contrast to the other approaches. For $\mu-U/2=1$, the results of type (I) are not improved from second
to third order, while other types fail to converge around $U\approx 2.5$. 
Thus, it remains difficult to access the intermediate filling regime ($0.1<|n-0.5|<0.2, U>2$) using these weak-coupling perturbation expansions. 
If one goes far away from half filling (dilute regime), the system effectively behaves 
as a weakly correlated metal, and the perturbative approximations become valid.

Let us remark that it was pointed out earlier by Yosida and Yamada \cite{YosidaYamada} that the bare weak-coupling perturbation theory
is well behaved for the Anderson impurity model if it is expanded around the nonmagnetic Hartree solution.
This corresponds to the expansion of type (IV) ($\alpha_\sigma=n_\sigma$) in our classification (Table.~\ref{n-alpha}). 
Thus, their observation is consistent with our conclusion that type (IV) is as good as type (I) and is better than the other schemes.
The difference is that when type (IV) expansion is applied to DMFT, 
it suffers from a convergence problem in the intermediate-coupling regime. 

\subsection{Antiferromagnetic phase}
\label{antiferromagnetic phase}

\begin{figure}[tbp]
\begin{center}
\includegraphics[width=8cm]{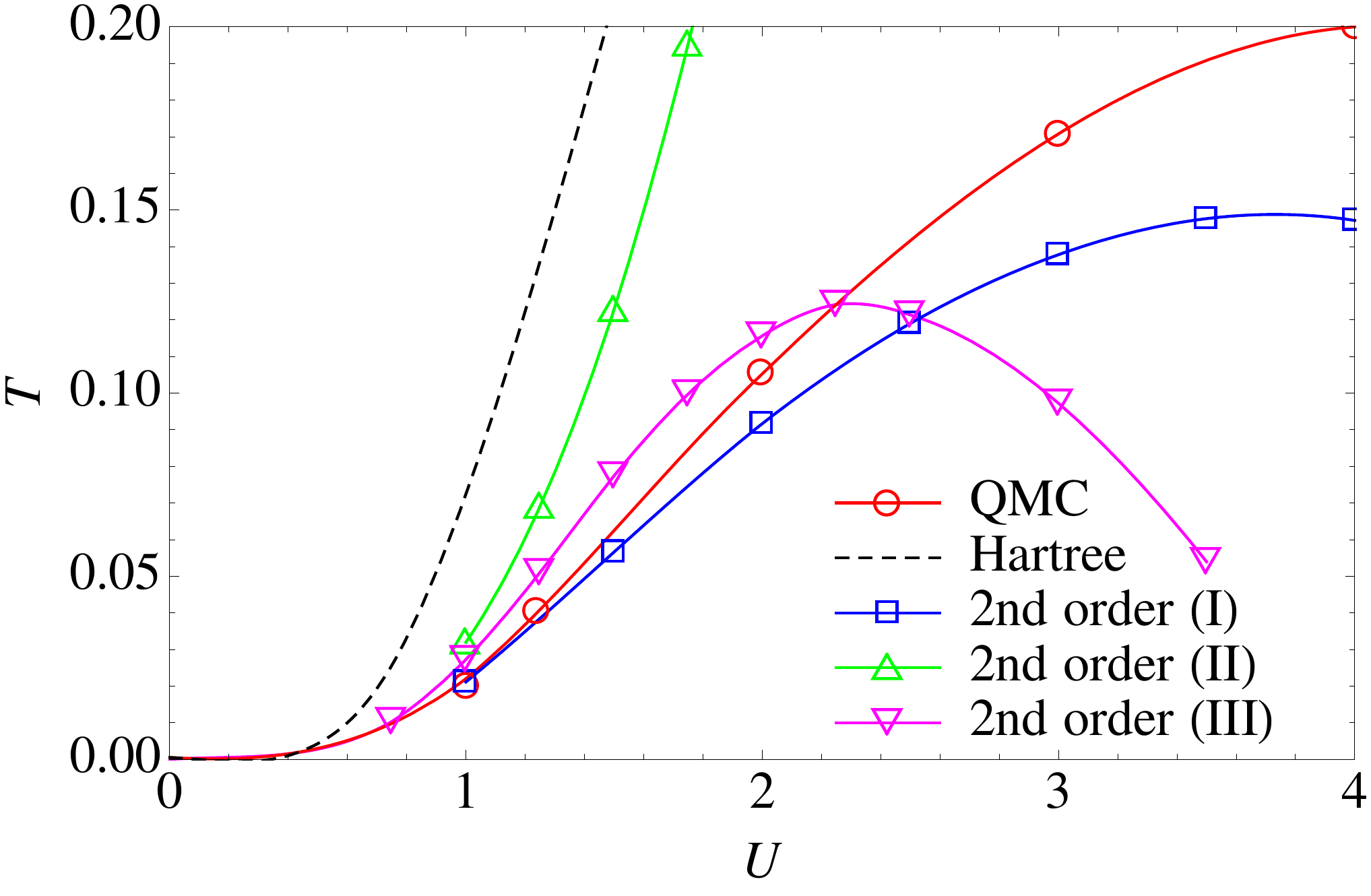}
\caption{(Color online) The AFM transition temperature for the Hubbard model at half filling derived from DMFT with various impurity solvers.
The region below (above) the curves represents the AFM
(PM) phase. The labels (I)-(III) correspond to the classification in Table~\ref{n-alpha}. 
The QMC data are taken from Ref.~\onlinecite{Koga2011}.}
\label{afm phase diagram second-order}
\end{center}
\end{figure}

\begin{figure}[tbp]
\begin{center}
\includegraphics[width=8cm]{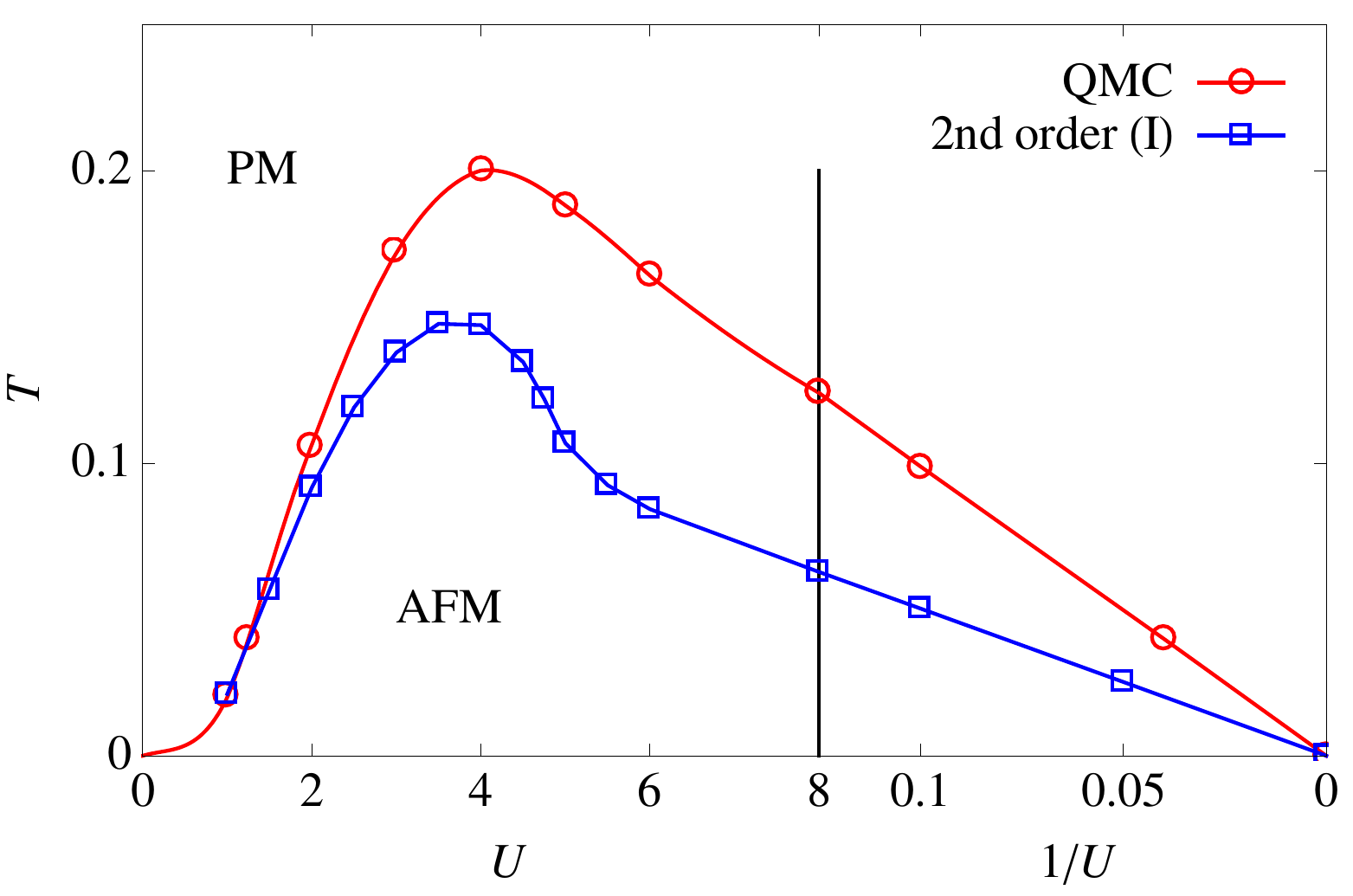}
\caption{(Color online) The AFM phase diagram of the Hubbard model covering the weak-coupling and strong-coupling limits.
The QMC data are taken from Ref.~\onlinecite{Koga2011}.}
\label{global afm phase diagram}
\end{center}
\end{figure}

Next, we test the validity of the perturbative impurity solvers for the equilibrium AFM phase of the Hubbard model at half filling.
We show the AFM phase diagram in the weak-coupling regime obtained from DMFT with QMC, the Hartree approximation, and the second-order perturbation theories
in Fig.~\ref{afm phase diagram second-order}. We also depict the phase diagram covering the entire $U$ range in Fig.~\ref{global afm phase diagram}.
QMC provides the exact critical temperature $T_c$, which in the small-$U$ limit behaves as $T_c\propto v_\ast e^{-1/D(\epsilon_F)U}$
[similar to the BCS formula for the superconducting phase; $D(\epsilon_F)$ is the DOS at the Fermi energy].
$T_c$ takes the maximum value at $U\approx 4$ and slowly decays as $T_c\propto v_\ast^2/U$ in the strong-coupling regime. This is analogous to 
the BCS-BEC crossover for superconductivity which is often discussed in the context of cold-atom systems. Here,
it corresponds to a crossover from the spin density wave in the weak-coupling regime to the AFM Mott insulator with local magnetic moments
in the strong-coupling regime.

The Hartree approximation (dashed curve in Fig.~\ref{afm phase diagram second-order}) correctly reproduces the weak-coupling asymptotic form, $T_c\propto v_\ast e^{-1/D(\epsilon_F)U}$,
but starts to deviate from the QMC result already at $U\sim 0.5$. The second-order perturbation theories of types (I)-(III) give better results than 
the Hartree approximation as shown in Fig.~\ref{afm phase diagram second-order}. However, quantitatively the agreement with QMC is still not so good
for $U>1$. This problem was previously pointed out for the second-order perturbation expansion of type (III).\cite{GeorgesKotliarKrauthRozenberg1996}
We have not plotted $T_c$ estimated from the second-order perturbations of types (IV) and (V), since type (IV) gives a discontinuous (first-order) phase transition
which is not correct for the AFM order, and type (V) yields a pathological discontinuous jump of the magnetization as a function of temperature within the ordered phase
which is physically unreasonable.

\begin{figure}[tbp]
\begin{center}
\includegraphics[width=8cm]{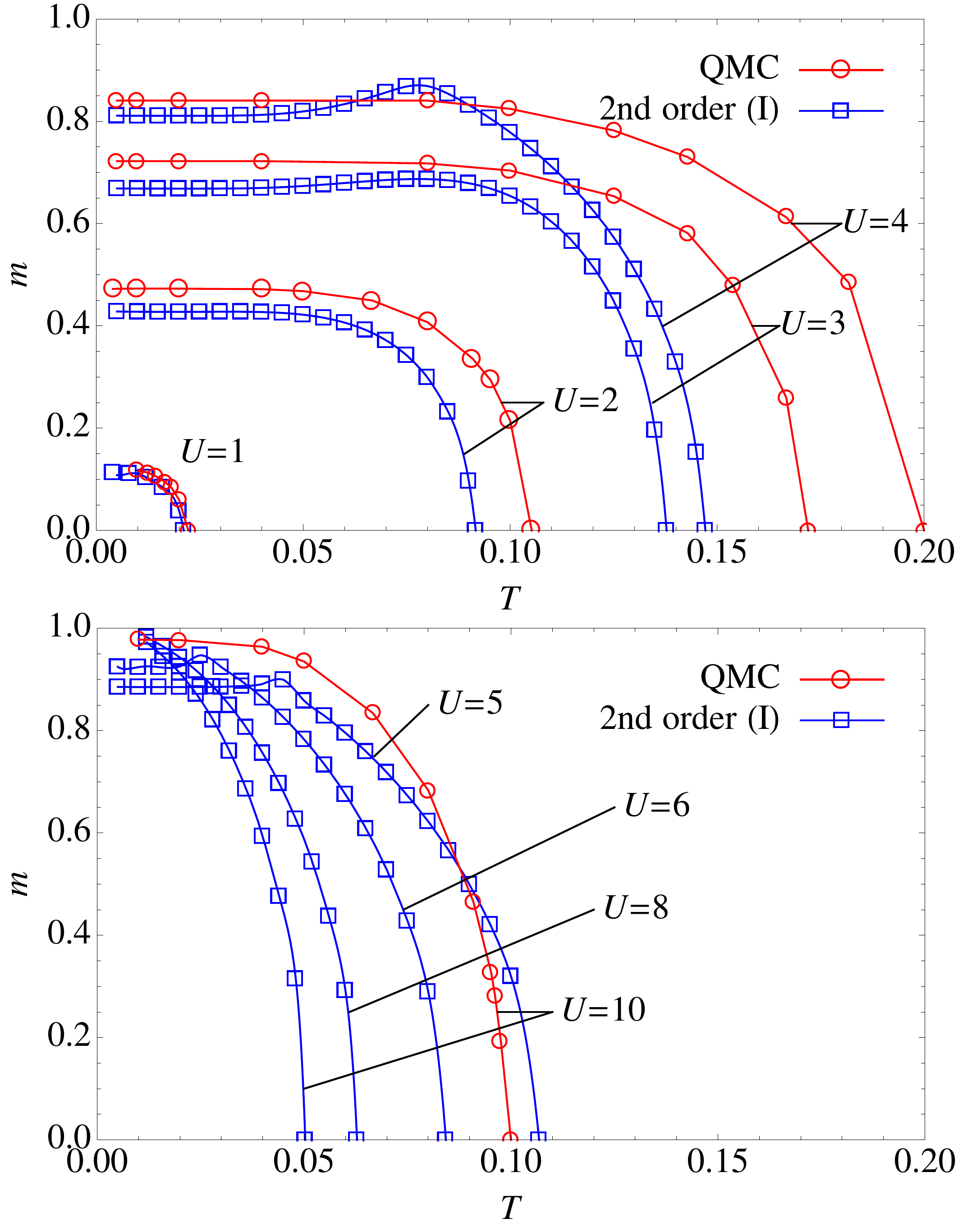}
\caption{(Color online) The staggered magnetization for the Hubbard model at half filling 
evaluated by DMFT with QMC and the second-order perturbation theory of type (I) in the weak-coupling (top panel)
and strong-coupling (bottom panel) regimes.
The QMC data for $U=2, 10$ are taken from Ref.~\onlinecite{Koga2011}.}
\label{magnetization 2nd-order}
\end{center}
\end{figure}
\begin{figure}[tbp]
\begin{center}
\includegraphics[width=8cm]{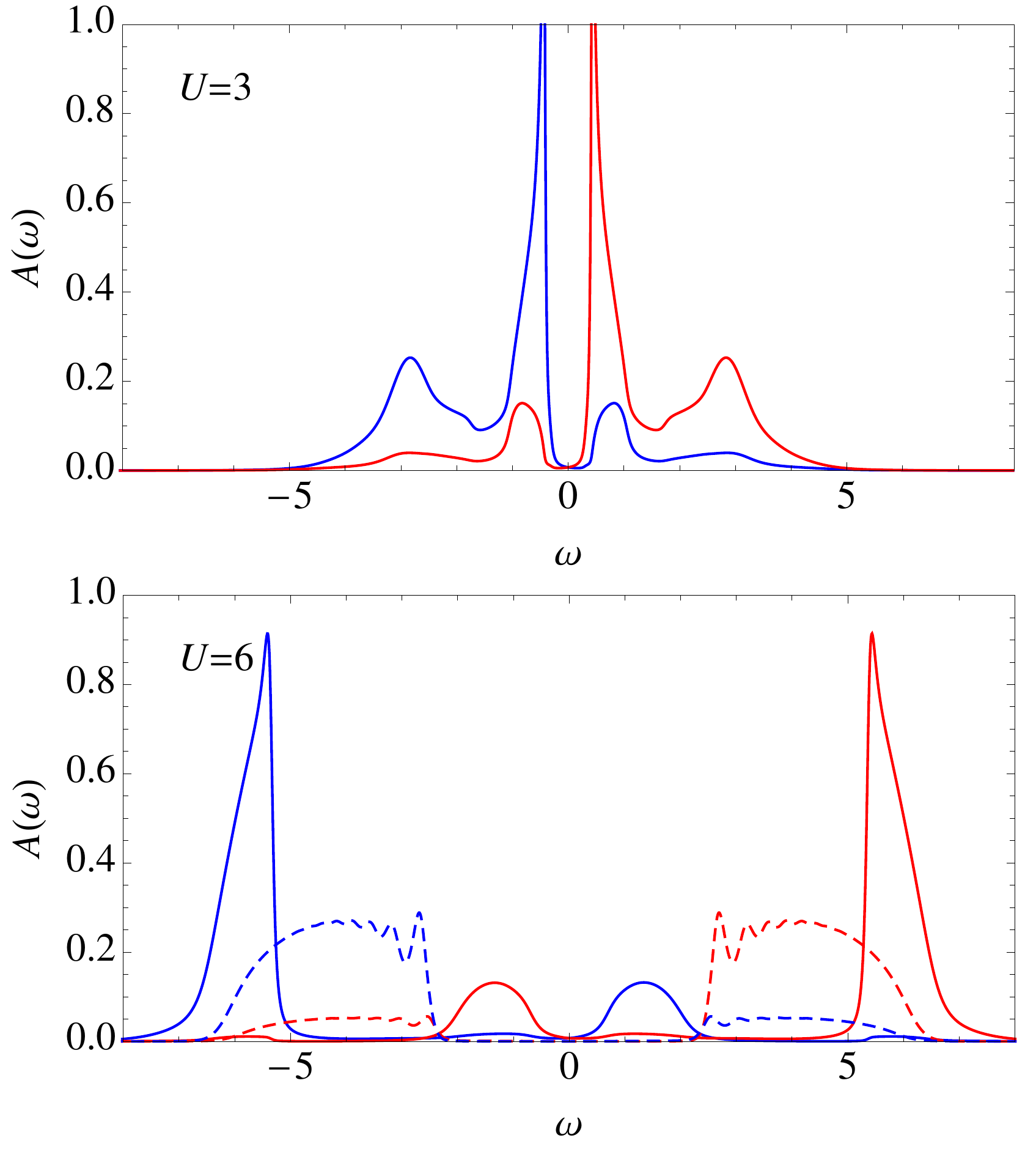}
\caption{(Color online) The spectral function of the majority (minority) spin component [blue (red) solid curve]
for the AFM phase of the Hubbard model at half filling calculated by DMFT with the second-order
perturbation theory of type (I). The dashed lines show the spectral function calculated by DMFT with the noncrossing approximation.
The parameters are $U=3, \beta=20$ (top panel) and $U=6, \beta=16$ (bottom).}
\label{spectral function 2nd-order}
\end{center}
\end{figure}
Type (I) continues to converge in the strong-coupling
regime, in contrast to other second-order approaches that fail to converge at some point. What is special about this weak-coupling expansion is that 
it qualitatively captures the BCS-BEC crossover; i.e., the critical temperature scales appropriately both in the weak- and strong-coupling limits 
(Fig.~\ref{global afm phase diagram}). Quantitatively, the value of $T_c$ given by the type (I) second-order scheme is roughly a factor of 2 lower than the QMC result 
in the large-$U$ regime.
If one restricts the DMFT solution to the PM phase, it is known that the second-order bare-diagram expansion (IPT) reproduces the correct
strong-coupling limit. The AFM critical temperature, on the other hand, depends on the treatment of the Hartree term
(note that even in the PM phase the evaluation of the spin susceptibility may depend on the choice of the Hartree diagram
since it enters in the vertex correction), and only the approach of type (I) among the various methods that we tested survives for large $U$. 
It would be interesting to compare the situation
with the $T$-matrix approximation\cite{KellerMetznerSchollwoeck1999,KellerMetznerSchollwoeck2001} 
that is often adopted in the study of the attractive Hubbard model. It takes account of a series of ladder diagrams for the self-energy and
similarly reproduces the BCS-BEC crossover for $T_c$.

We also plot the staggered magnetization $m=\sum_\sigma \sigma n_{\sigma}=\sum_\sigma \sigma G_\sigma^M(0^-)$ 
for the ordered state evaluated by QMC and the second-order perturbation theory of type (I) in the weak-coupling
(the top panel of Fig.~\ref{magnetization 2nd-order}) and strong-coupling (bottom panel) regimes.
For small $U$, the second-order perturbation theory gives a smooth curve for the magnetization as a function of $T$. As one increases $U$, there emerges a kink in the magnetization curve for $U\ge 4$ (Fig.~\ref{magnetization 2nd-order}),
which is an artifact of the perturbation theory as confirmed by comparison to the QMC results. 
Hence, although $T_c$ behaves reasonably in the large $U$ regime, it is unlikely that the second order perturbation of type (I) correctly describes the strong-coupling state. 

Figure~\ref{spectral function 2nd-order} plots the spectral function $A_\sigma(\omega)=-\frac{1}{\pi}{\rm Im}G_\sigma^R(\omega)$ obtained from the second-order perturbation of type (I). The weak-coupling regime ($U=3$, top panel of Fig.~\ref{spectral function 2nd-order}) shows coherence peaks separated by the AFM energy gap and accompanied by the Hubbard sidebands. However, as one goes to the strong-coupling regime ($U=6$, bottom panel of Fig.~\ref{spectral function 2nd-order}), the coherence peaks are rapidly shifted away from the Fermi energy, 
and two additional bands appear around $\omega=\pm 1.5$.
This is quite different from the result of the noncrossing approximation\cite{WernerTsujiEckstein2012} (dashed lines in the bottom panel of Fig.~\ref{spectral function 2nd-order}), which is supposed to be reliable in the strong-coupling regime, and shows spin-polaron peaks on top of the Mott-Hubbard bands with a large energy gap. Therefore, we conclude that the second-order perturbation of type (I) does not correctly describe the AFM state in the strong-coupling regime.

\begin{figure}[tbp]
\begin{center}
\includegraphics[width=8cm]{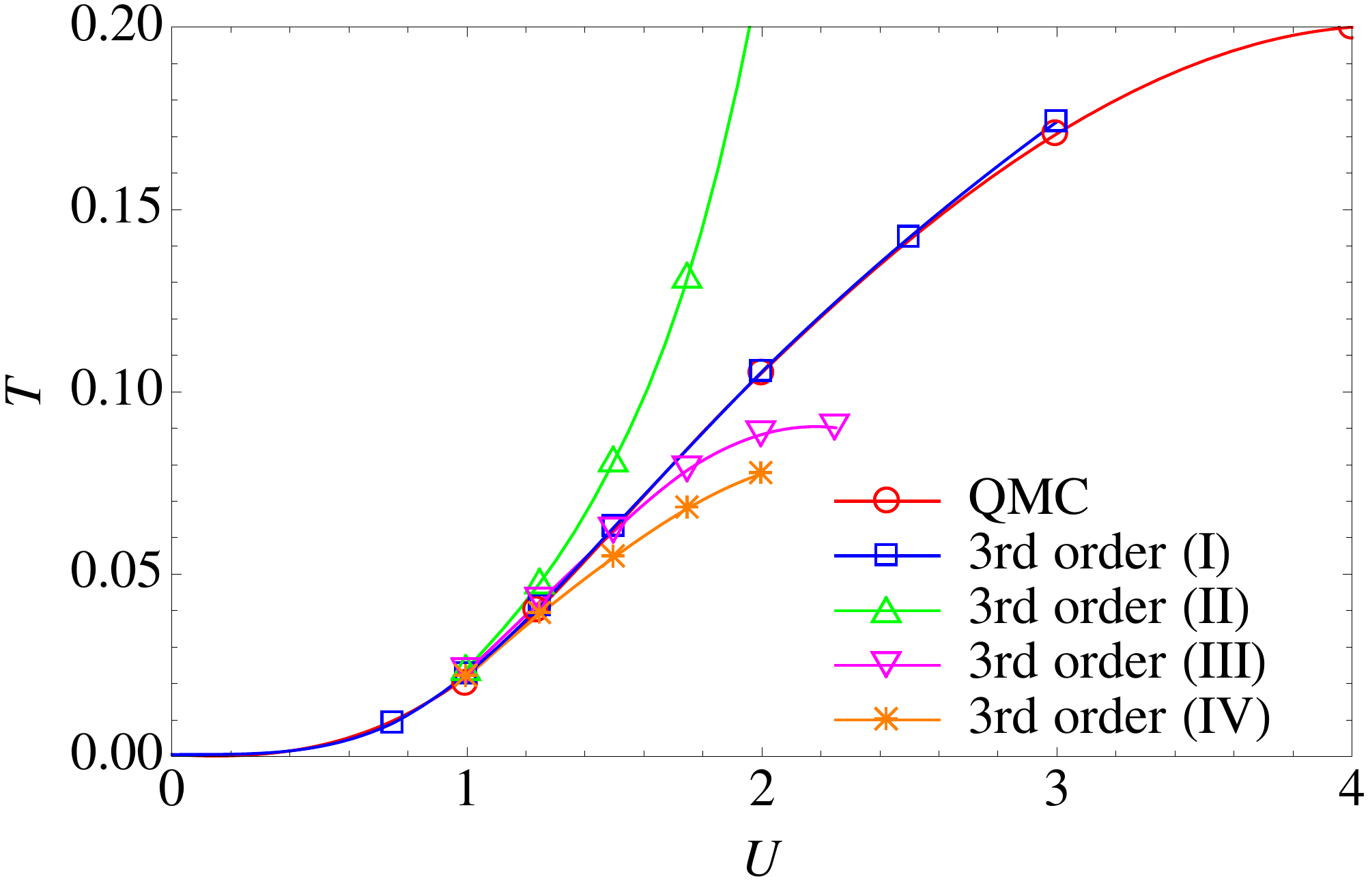}
\caption{(Color online) The AFM transition temperature for the Hubbard model at half filling derived from DMFT with various third-order perturbation impurity solvers. The region below (above) the curves represents the AFM
(PM) phase.
The labels (I)-(IV) correspond to the classification in Table~\ref{n-alpha}.}
\label{phase diagram 3rd-order}
\end{center}
\end{figure}
\begin{figure}[tbp]
\begin{center}
\includegraphics[width=8cm]{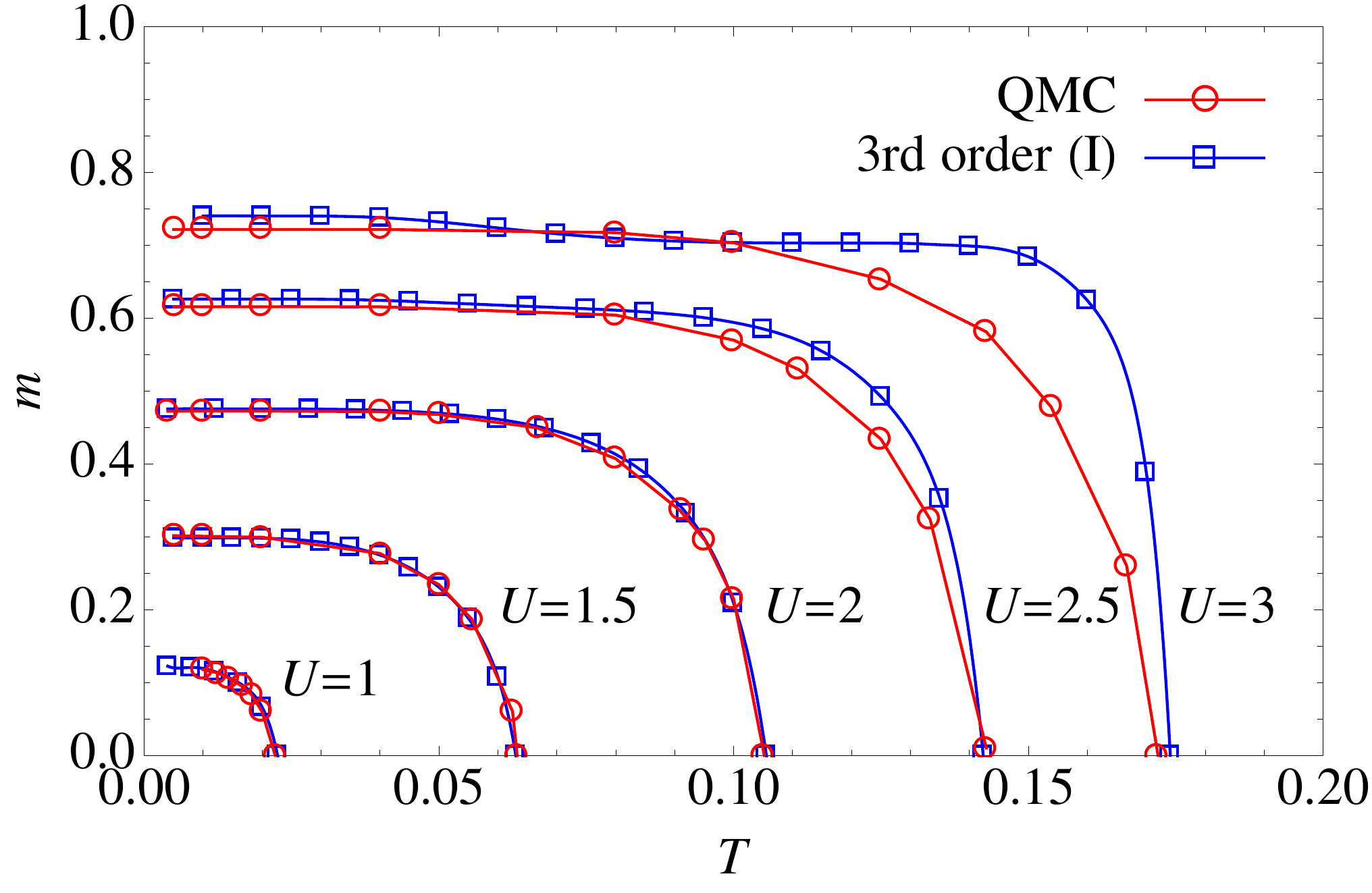}
\caption{(Color online) The staggered magnetization of the Hubbard model at half filling evaluated by DMFT with QMC and the third-order perturbation theory of type (I).}
\label{magnetization 3rd-order}
\end{center}
\end{figure}
The phase diagram derived from DMFT using various third-order perturbation theories is shown in Fig.~\ref{phase diagram 3rd-order}.
Again we do not draw the $T_c$ curve for type (V), since it has a discontinuous jump of the magnetization as a function of temperature in the ordered state.
We find that the third-order perturbation of type (I) (all the diagrams including the Hartree term are bare) reproduces $T_c$ very accurately up to $U=3$. 
A comparable accuracy cannot be obtained with the other third-order expansions. To establish the validity of this approach, we calculate the staggered magnetization
below $T_c$, which is illustrated in Fig.~\ref{magnetization 3rd-order}. By comparing the results with those of QMC, we can see that
the type (I) third-order approach predicts not only accurate $T_c$ but also correct magnetizations for $U\le 2.5$.\cite{TsujiEcksteinWerner2012}
When $U$ becomes larger than 2.5, deviations from the exact QMC results start to appear, and the curvature of the magnetization curve gets steeper.
Thus, the third-order perturbation of type (I) is the method of choice for studying the AFM phase in the weak-coupling regime ($U<3$).
This is again consistent with the observation of Yosida and Yamada \cite{YosidaYamada} that the bare weak-coupling expansion
works well if expanded around the nonmagnetic Hartree solution (i.e., $\alpha_\sigma=\frac{1}{2}$).

\section{Interaction quench in the paramagnetic phase of the Hubbard model}
\label{quench paramagnetic}

Having examined the performance of the weak-coupling perturbation theories for the equilibrium state of the Hubbard model,
we move on to studying the validity of the perturbative methods for nonequilibrium problems. In this section, we focus on
the interaction quench problem for the PM phase of the Hubbard model; i.e., we consider the Hamiltonian (\ref{Hubbard}) with the interaction parameter
abruptly varied as
\begin{align}
U(t)=
\begin{cases}
U_i & t=0^- \\
U_f & t>0
\end{cases}.
\end{align}
The interaction quench problem for the Hubbard model has been previously studied 
using the flow equation and unitary perturbation theory,\cite{MoeckelKehrein2008,MoeckelKehrein2010}
nonequilibrium DMFT,\cite{EcksteinKollarWerner2009, EcksteinKollarWerner2010}
time-dependent Gutzwiller variational method,\cite{SchiroFabrizio2010,SchiroFabrizio2011}
generalized Gibbs ensemble,\cite{KollarWolfEckstein2011}
equation-of-motion approach,\cite{HamerlaUhrig2013,HamerlaUhrig2013b}
and quantum kinetic equation.\cite{StarkKollar2013}
In the weak-coupling regime,
the physics of ``prethermalization''\cite{BergesBorsanyiWetterich2004} has been discussed.

Here we take the parameters, $U_i=0$ (noninteracting initial state) and the initial temperature $T=0$, to allow a systematic comparison with these previous results.
We use the second-order and fourth-order perturbation theories for the half-filling case in Sec.~\ref{quench half filling}
and the third-order perturbation theory for calculations away from half filling in Sec.~\ref{quench away from half filling}.
We restrict ourselves to the PM solution of the nonequilibrium DMFT equations throughout this section.

\subsection{Half filling}
\label{quench half filling}

\begin{figure}[tbp]
\begin{center}
\includegraphics[width=8cm]{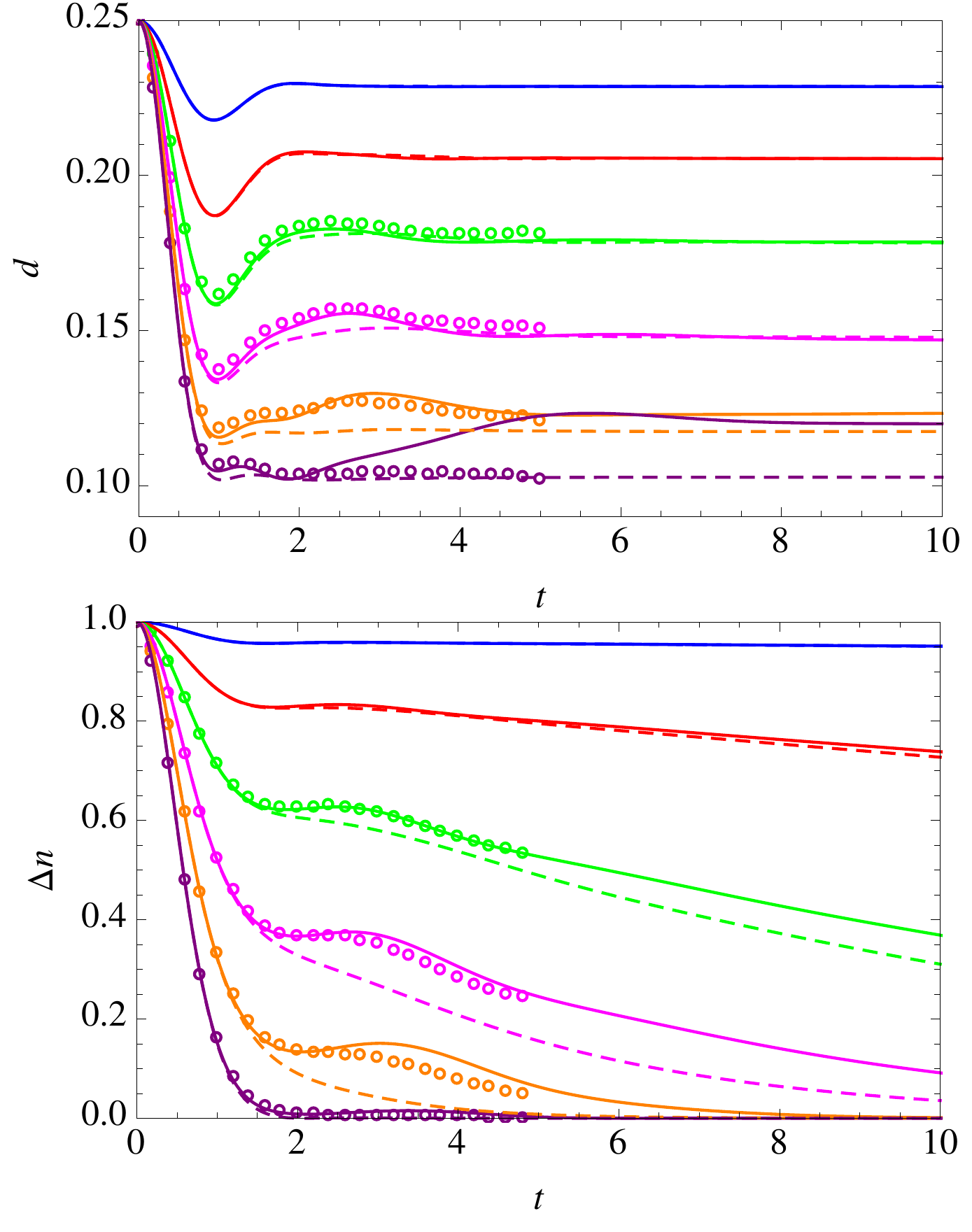}
\caption{(Color online) Time evolution of the double occupancy (top panel) and the jump of the momentum distribution function (bottom panel)
after the interaction quenches $U=0\to 0.5, 1, \dots, 3$ (from top to bottom) in the PM phase 
of the Hubbard model at half filling calculated by the nonequilibrium DMFT 
with QMC (circles, taken from Ref.~\onlinecite{EcksteinKollarWerner2009}), 
the bare second-order (dashed curves), and bare fourth-order (solid curves) perturbation theories.}
\label{double occupancy 4th-order}
\end{center}
\end{figure}

To study the relaxation behavior of the Hubbard model after the interaction quench,
we calculate the time evolution of the double occupancy $d(t)=\langle n_\uparrow(t)n_\downarrow(t)\rangle$
within nonequilibrium DMFT via the formula,
\begin{align}
d(t)=-\frac{i}{2U}\sum_\sigma(\Sigma_\sigma\ast G_\sigma)^<(t,t),
\end{align}
which can be derived from the equation of motion. Initially the system is noninteracting, so that $d(0^-)=\langle n_\uparrow\rangle\langle n_\downarrow\rangle
=1/2\times 1/2=1/4$ at half filling.
The results for $d(t)$ obtained with different impurity solvers (QMC, bare second-order, and bare fourth-order perturbation theory)
are plotted for $U_f\le 3$ in the top panel of Fig.~\ref{double occupancy 4th-order}. As one can see, the results of the second and fourth order agree very well with 
the QMC results up to $U_f= 1.5$. After the quench, the double occupancy quickly 
relaxes to an almost constant value, which is quite close to the thermal value of the final state.
\cite{EcksteinKollarWerner2009}
At $U_f=2$ and $2.5$, the difference between the second and fourth order becomes larger, and the latter reproduces the correct result
of the double occupancy with an irregular hump around $t\sim 2.5$. The second-order perturbation predicts an overdamping of the double occupancy without a hump.

\begin{figure}[tbp]
\begin{center}
\includegraphics[width=8cm]{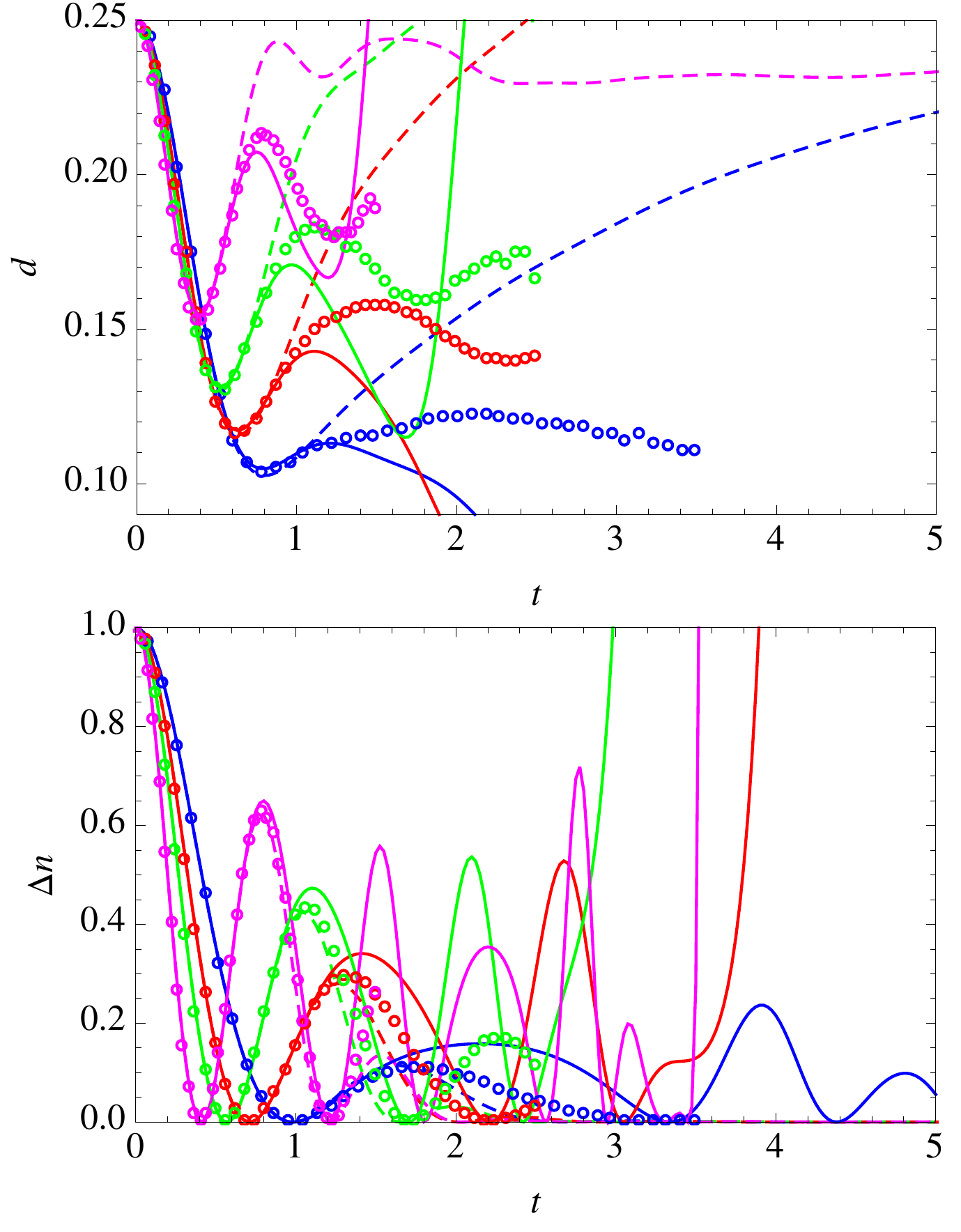}
\caption{(Color online) Time evolution of the double occupancy (top panel) and the jump of the momentum distribution function (bottom)
after the interaction quenches $U=0\to 4, 5, 6, 8$ (from bottom to top in the first minima of $d$, and from top to bottom
in the initial decrease of $\Delta n$) in the PM phase of the Hubbard model at half filling calculated by the nonequilibrium DMFT 
with QMC (circles, taken from Ref.~\onlinecite{EcksteinKollarWerner2009}),
the bare second-order (dashed curves), and bare fourth-order (solid curves) perturbation theories.}
\label{double occupancy 4th-order strong-coupling}
\end{center}
\end{figure}

The results for the double occupancy in the intermediate- and strong-coupling regimes ($4\le U_f\le 8$)
are shown in the top panel of Fig.~\ref{double occupancy 4th-order strong-coupling}.
As we increase the interaction strength, the perturbation theories quickly deviate from the QMC results,
and an agreement is found only on very short time scales.
The fourth-order expansion thus fails to improve the second-order results in this regime, and seems to be numerically more unstable than
the second order. We discuss later that this is related to pathological violation of energy conservation.

We also compute the momentum distribution,
\begin{align}
n_{\bm k}(t)=-iG_{\bm k}^<(t,t). 
\end{align}
The distribution of the initial noninteracting state at $T=0$ is $n_{\bm k}=\theta(\mu-\epsilon_{\bm k})$,
which has a discontinuous jump $\Delta n=1$ at the Fermi energy. This jump does not immediately disappear after the quench, but survives for some period.\cite{EcksteinKollarWerner2009}
It is a measure of how close or far the system is from the thermalized state. If the system fully thermalizes after the quench, the jump $\Delta n$ should vanish since
a thermal state at nonzero temperature has a smooth distribution.
In the bottom panel of Fig.~\ref{double occupancy 4th-order}, we show the time evolution of $\Delta n$ obtained by the nonequilibrium DMFT with QMC, bare second-order and bare fourth-order perturbations for $U_f\le 3$.
The QMC results suggest that, in contrast to the double occupancy, $\Delta n$ does not directly relax to the thermal value ($\Delta n=0$), but is trapped at some intermediate value for some time.
Although local quantities such as the double occupancy look thermalized at this moment (prethermalization), the distribution is clearly nonthermal.
\cite{MoeckelKehrein2008}
After prethermalization, $\Delta n$ slowly relaxes to zero. The time scale of this relaxation is much longer than that for the double occupancy.

The difference between the second- and fourth-order perturbations for $\Delta n$ is clearer than in the case of the double occupancy. It already becomes evident at $U_f=1.5$.
Before $\Delta n$ approaches the plateau ($t<1.5$), both methods give almost the same results. 
However, the second-order results do not show a clear plateaulike structure in the prethermalization regime ($1.5\lesssim t\lesssim 3$), 
while the fourth order does. One can see in the bottom panel of Fig.~\ref{double occupancy 4th-order} that the results of the fourth-order perturbation theory 
for $\Delta n$ agree fairly well with those of QMC for $U_f\le 3$.

We also show the results for $\Delta n$ in the intermediate- and strong-coupling regimes ($4\le U_f\le 8$)
in the bottom panel of Fig.~\ref{double occupancy 4th-order strong-coupling}.
It has been known that the behavior of $\Delta n$ qualitatively changes from the weak- to the strong-coupling regime
near $U_c^{\rm dyn}=3.2$ (dynamical transition).
\cite{EcksteinKollarWerner2009,SchiroFabrizio2010} On the weak-coupling side $\Delta n$ monotonically decays without touching zero, while
on the strong-coupling side $\Delta n$ oscillates between zero and nonzero values.
One can see that the bare perturbation theories (both second and fourth order) correctly reproduce the short-time dynamics
of these collapse-and-revival oscillations in $\Delta n$
with period $2\pi/U$. Especially, they capture the sharp qualitative change of the short-time behavior of $\Delta n$. In the second-order perturbation theory
the oscillations of $\Delta n$ always damp faster than those given by the fourth-order calculation.
Similarly to the double occupancy, the $\Delta n$ obtained from the perturbation theories fail to reproduce the QMC results after 
the second- and fourth-order calculations start to deviate with each other.

\begin{figure}[tbp]
\begin{center}
\includegraphics[width=8cm]{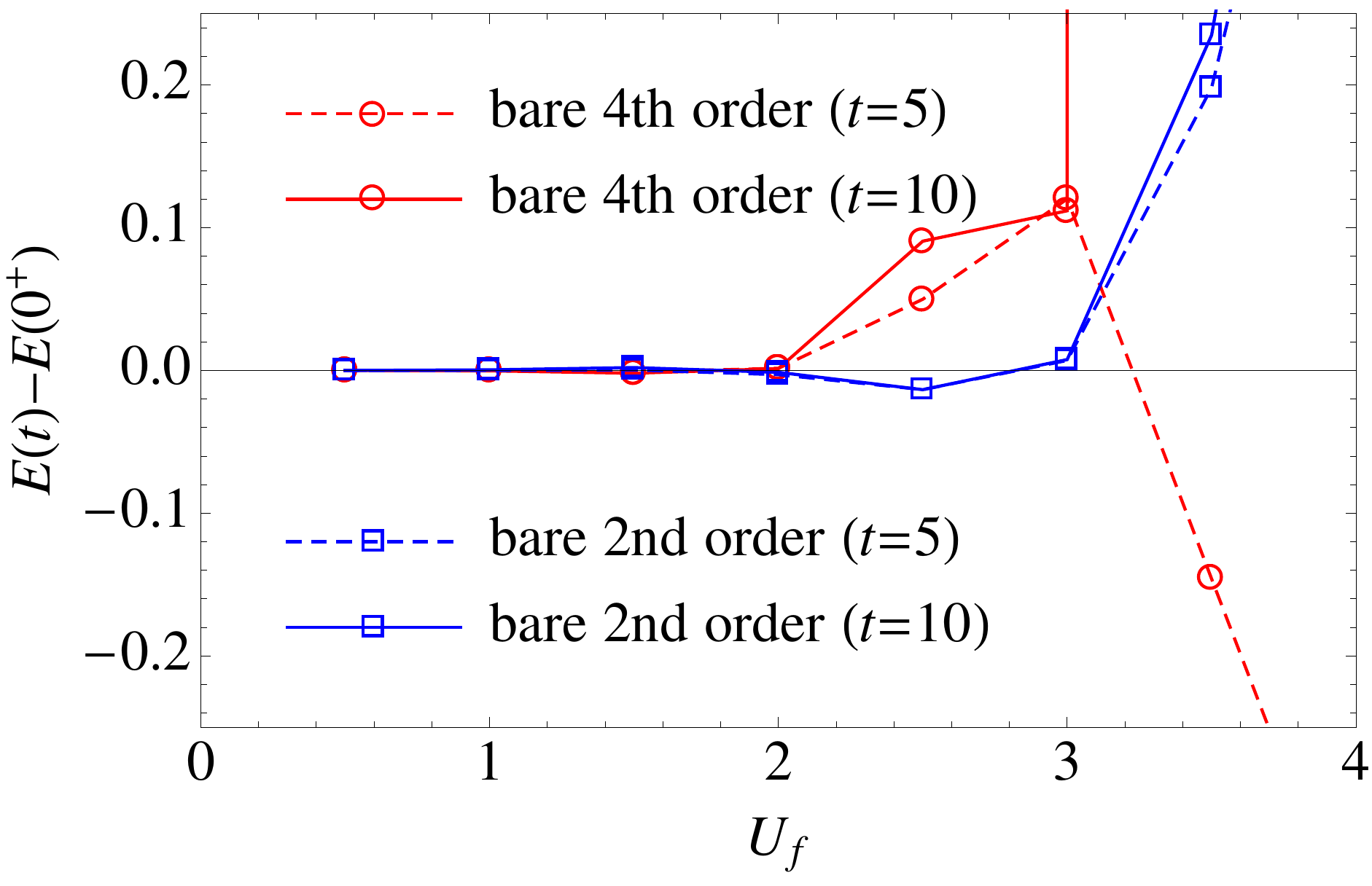}
\caption{(Color online) The drift of the total energy $E(t)-E(0^+)$ (measured at $t=5,10$) in the simulation of the interaction quench $U=0\to U_f$ using the bare second-order and bare fourth-order perturbation solvers.}
\label{drift of total energy}
\end{center}
\end{figure}

The quality of the bare-diagram perturbation theory can be judged by looking at the evolution of the total energy (Fig.~\ref{drift of total energy}).
The total energy of the Hubbard model is given by
\begin{align}
E(t)=\sum_{\bm k\sigma}\epsilon_{\bm k}n_{\bm k\sigma}(t,t) +U(t)\left(d(t)-\frac{1}{4}\right).
\end{align}
For the semicircular DOS in the PM and AFM phases, the kinetic-energy term can be rewritten in terms of the local Green's functions as
\begin{align}
\sum_{\bm k\sigma}\epsilon_{\bm k}n_{\bm k\sigma}(t,t)
&=
-i\sum_\sigma (\Delta_\sigma\ast G_\sigma)^<(t,t)
\nonumber
\\
&=
-iv_\ast^2
\sum_\sigma (G_{\bar{\sigma}}\ast G_\sigma)^<(t,t).
\end{align}
Since the bare-diagram expansions are not a conserving approximation, it is not guaranteed that the total energy is conserved after the quench, even though the Hamiltonian becomes time independent.
To make a systematic comparison, we consider the difference between $E(t>0)$ and $E(0^+)$, which should be zero if the total energy is indeed conserved.
As one can see in Fig.~\ref{drift of total energy}, the total energy is nicely conserved up to $U_f=2$. However, when $U_f$ exceeds $2$, the conservation of the total energy
is suddenly violated for both the second-order and the fourth-order perturbation theories 
[the drift in total energy is smaller in the second-order perturbation]. 
In particular, the fourth-order expansion does not extend the interaction region in which the total energy is conserved.
If one compares the energy drift at $t=5$ and $t=10$ in Fig.~\ref{drift of total energy}, one sees that the drift is ``saturated'' for $U_f\le 3$ at the second order
and for $U_f\le 2$ at the fourth order; i.e., the drift only occurs on a certain short time scale and the difference to the correct total energy does not grow anymore thereafter.
In other words, the bare perturbation theory ``does not accumulate'' numerical errors as time evolves for $U_f\le 2$.
From this, we can conclude that the bare perturbation theories remain reliable up to long times for these $U_f$, which is a big advantage of this approach.

The breakdown of the total-energy conservation generically implies a 
deviation of the results for $d$ or $\Delta n$ from QMC in the regime $U_f>2$ (Figs.~\ref{double occupancy 4th-order} and \ref{double occupancy 4th-order strong-coupling}). 
Let us remark that this does not necessarily mean that the bare perturbation theory always fails to describe 
the dynamics of the Hubbard model with $U>2$. It all depends on how the system is perturbed (interaction quench, slow ramp, electric-field excitation, etc.), the initial state (noninteracting or interacting), and other details of the problem. The general tendency is that the total energy is conserved when the excitation energy is small and/or the interaction strength is weak.

\begin{figure}[tbp]
\begin{center}
\includegraphics[width=8cm]{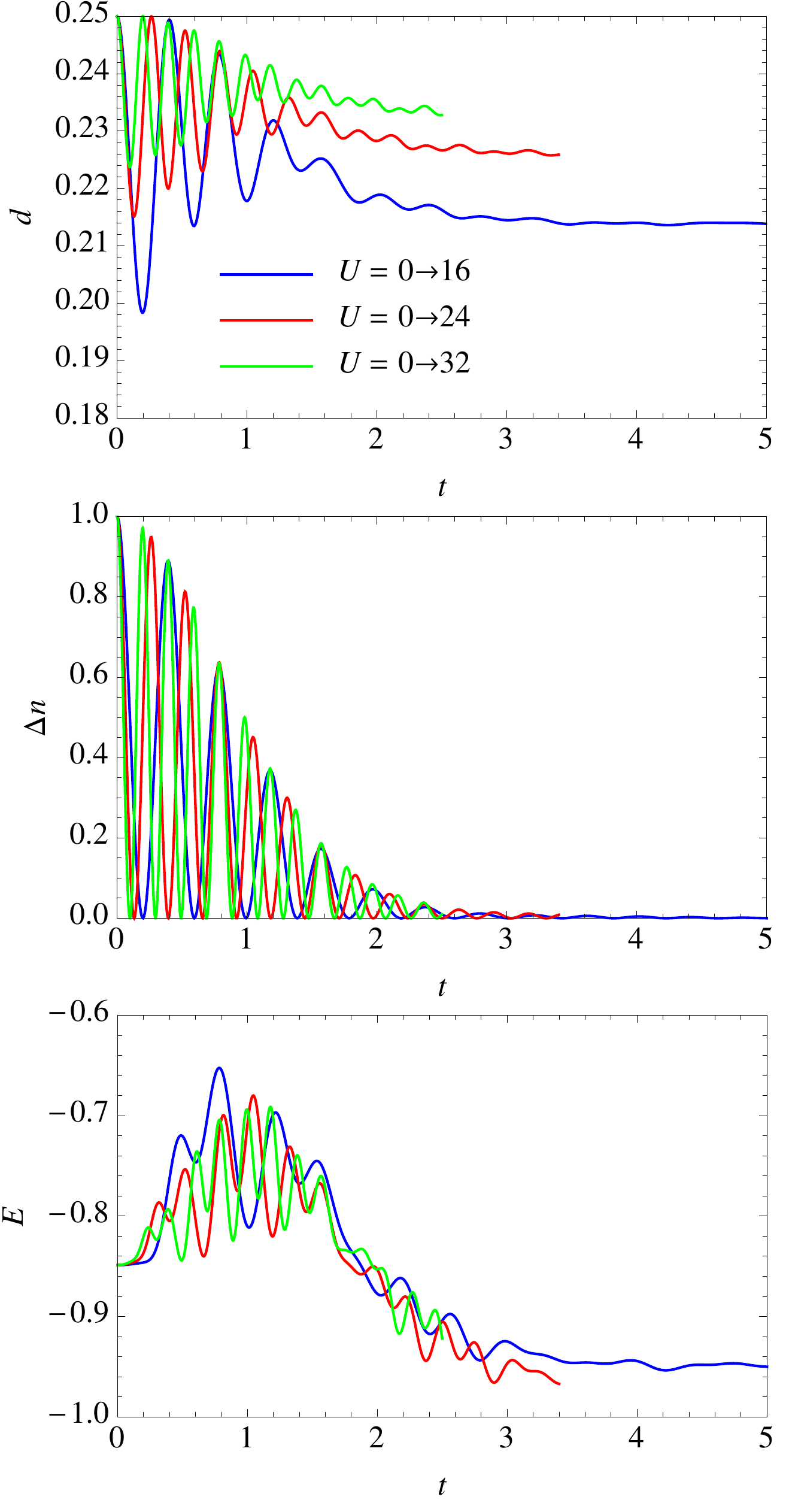}
\caption{(Color online) Time evolution of the double occupancy (top), the jump of the momentum distribution function (middle), and the total energy (bottom)
after the interaction  quenches $U=0\to 16, 24, 32$ (from long- to short-time data) in the PM phase of the Hubbard model at half filling
calculated by nonequilibrium DMFT with the bare second-order perturbation theory.}
\label{very large U}
\end{center}
\end{figure}
If we further increase $U_f$, the second-order bare perturbation theory again starts to work reasonably well. In Fig.~\ref{very large U}, we plot
$d$, $\Delta n$, and $E$ for quenches $U=0\to 16, 24, 32$. The simulation is numerically stable within the accessible time range, and the observables do not diverge as time grows.
The results nicely show the coherent collapse-and-revival oscillations of period $2\pi/U$, which are characteristic of the atomic limit.
We also observe that the envelope curve of rapidly oscillating $\Delta n$ is ``universal''; i.e., it is almost invariant against large enough $U$.
In contrast, the fourth-order bare perturbation theory fails to produce physically reasonable results for these $U$.
The seeming success of the second-order bare perturbation theory (IPT) for very large $U$ appears to be related to the fact that IPT reproduces the correct atomic limit 
of the Hubbard model in equilibrium.\cite{GeorgesKotliarKrauthRozenberg1996}
However, it is {\it a priori} not obvious that IPT also describes the correct nonequilibrium dynamics near the atomic limit, since
the dynamics here starts from the noninteracting state which is very far from the atomic limit,
and errors can accumulate in the strong-coupling regime
as the system time evolves. Indeed, if one looks at the total energy $E$ (bottom panel of Fig.~\ref{very large U}),
there is a non-negligible energy drift whose magnitude ($\sim 10\%$ of the absolute value of $E$) is roughly independent of $U$.
Despite the violation of energy conservation, IPT seems to work surprisingly well out of equilibrium near the atomic limit.

\begin{figure}[tbp]
\begin{center}
\includegraphics[width=8cm]{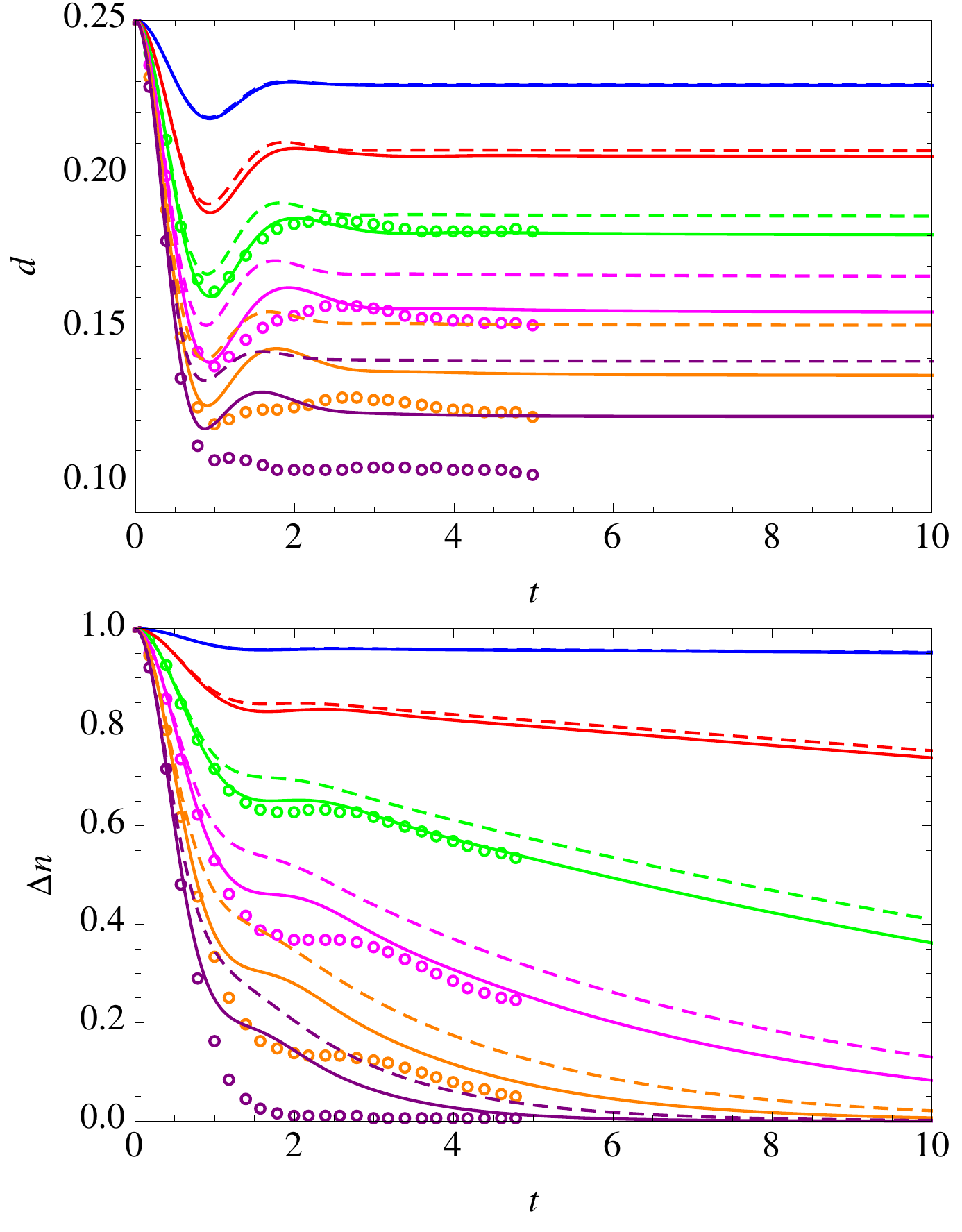}
\caption{(Color online) Time evolution of the double occupancy (top panel) and the jump of the momentum distribution function (bottom)
after the interaction quenches $U=0\to 0.5, 1, \dots, 3$ (from top to bottom) in the PM phase 
of the Hubbard model at half filling calculated by the nonequilibrium DMFT 
with QMC (circules, taken from Ref.~\onlinecite{EcksteinKollarWerner2009}), 
the bold second-order (dashed curves), and the bold fourth-order (solid curves) self-consistent perturbation theories.}
\label{double occupancy self-consistent}
\end{center}
\end{figure}

\begin{figure}[tbp]
\begin{center}
\includegraphics[width=8cm]{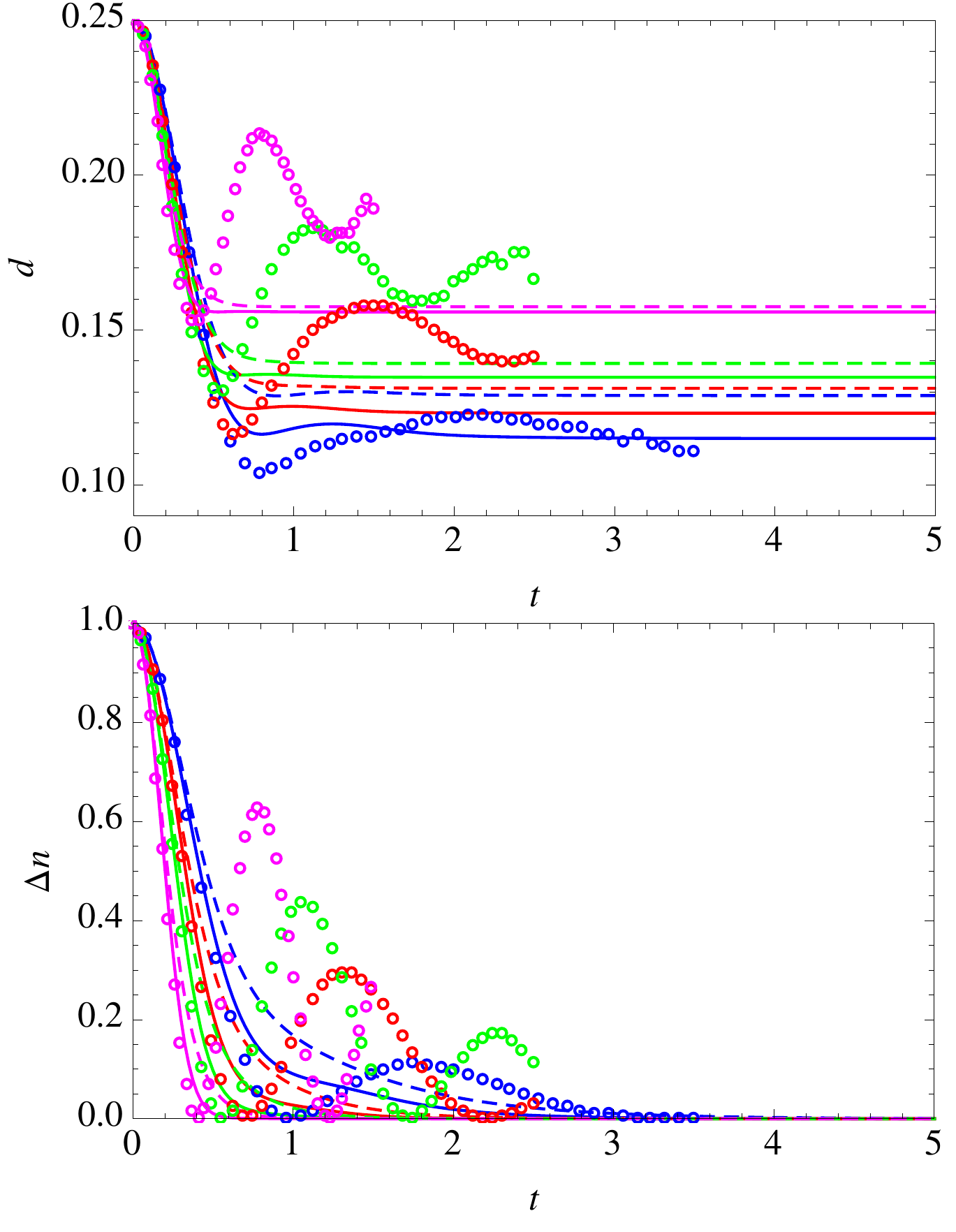}
\caption{(Color online) Time evolution of the double occupancy (top panel) and the jump of the momentum distribution function (bottom)
after the interaction quenches $U=4, 5, 6, 8$ (from bottom to top in the long-time limit of $d$, and from top to bottom in $\Delta n$)
in the PM phase of the Hubbard model at half filling calculated by the nonequilibrium DMFT 
with QMC (circules, taken from Ref.~\onlinecite{EcksteinKollarWerner2009}), 
the bold second-order (dashed curves), and the bold fourth-order (solid curves) self-consistent perturbation theories.}
\label{double occupancy self-consistent strong-coupling}
\end{center}
\end{figure}

We have also tested the second-order and fourth-order self-consistent perturbation theories (bold-diagram expansions). 
The results for the double occupancy and the jump in the momentum distribution for $U_f\le 3$ ($U_f\ge 4$)
are shown in the top and bottom panels of Fig.~\ref{double occupancy self-consistent} (Fig.~\ref{double occupancy self-consistent strong-coupling}), respectively. 
By comparing with QMC, one can see that the self-consistent perturbation theories are not particularly good.
Although we have a slight improvement from the second-order to the fourth-order expansion, a deviation from the QMC results still remains, even in the short-time dynamics. The detailed dynamics of $d$ and $\Delta n$ in the transient and long-time regimes
is not correctly reproduced. The damping of the double occupancy is too strong in both the weak-coupling and the strong-coupling regimes 
(see also Ref.~\onlinecite{EcksteinKollarWerner2009}). 
This may be due to a too-tight self-consistency condition, i.e., the self-consistency within the perturbation theory and the DMFT self-consistency.
[Flaws in the (second-order) self-consistent perturbation theory, when applied to the equilibrium DMFT,\cite{Muller-Hartmann1989b}
were pointed out already in Ref.~\onlinecite{GeorgesKotliar1992}. In particular, it was noted that it does not reproduce the high-energy features (Hubbard sidebands)
of the spectral function.]
For $\Delta n$, the height of the prethermalization plateau is not correctly reproduced for $U_f\le 3$. On the strong-coupling side,
$\Delta n$ relaxes monotonically without showing any oscillation. This evidences that the self-consistent perturbation theory cannot describe
the dynamical transition found in the interaction-quenched Hubbard model at half filling.
\cite{EcksteinKollarWerner2009,SchiroFabrizio2010}
Hence, even though the self-consistent perturbation theory is a conserving approximation, it is not the impurity solver of choice for nonequilibrium DMFT.

\subsection{Away from half filling}
\label{quench away from half filling}

When the filling is shifted away from half filling, the particle-hole symmetry is lost, and odd-order diagrams start to contribute in the calculation.
Here we consider the interaction quench problem for the PM phase of the Hubbard model at quarter filling, i.e., $n_\sigma=1/4$,
and apply the second-order and third-order perturbation theories. We adopt the type (I) and type (IV) approaches in the classification of Table~\ref{n-alpha}, i.e., 
the bare-diagram expansions having the bare tadpole diagram with $\alpha_\sigma=1/2$ and bold tadpole with $\alpha_\sigma=n_\sigma$, 
since they have been shown to be relatively good approximations away from half filling in Sec.~\ref{paramagnetic phase}.

\begin{figure*}[tbp]
\begin{center}
\includegraphics[width=16cm]{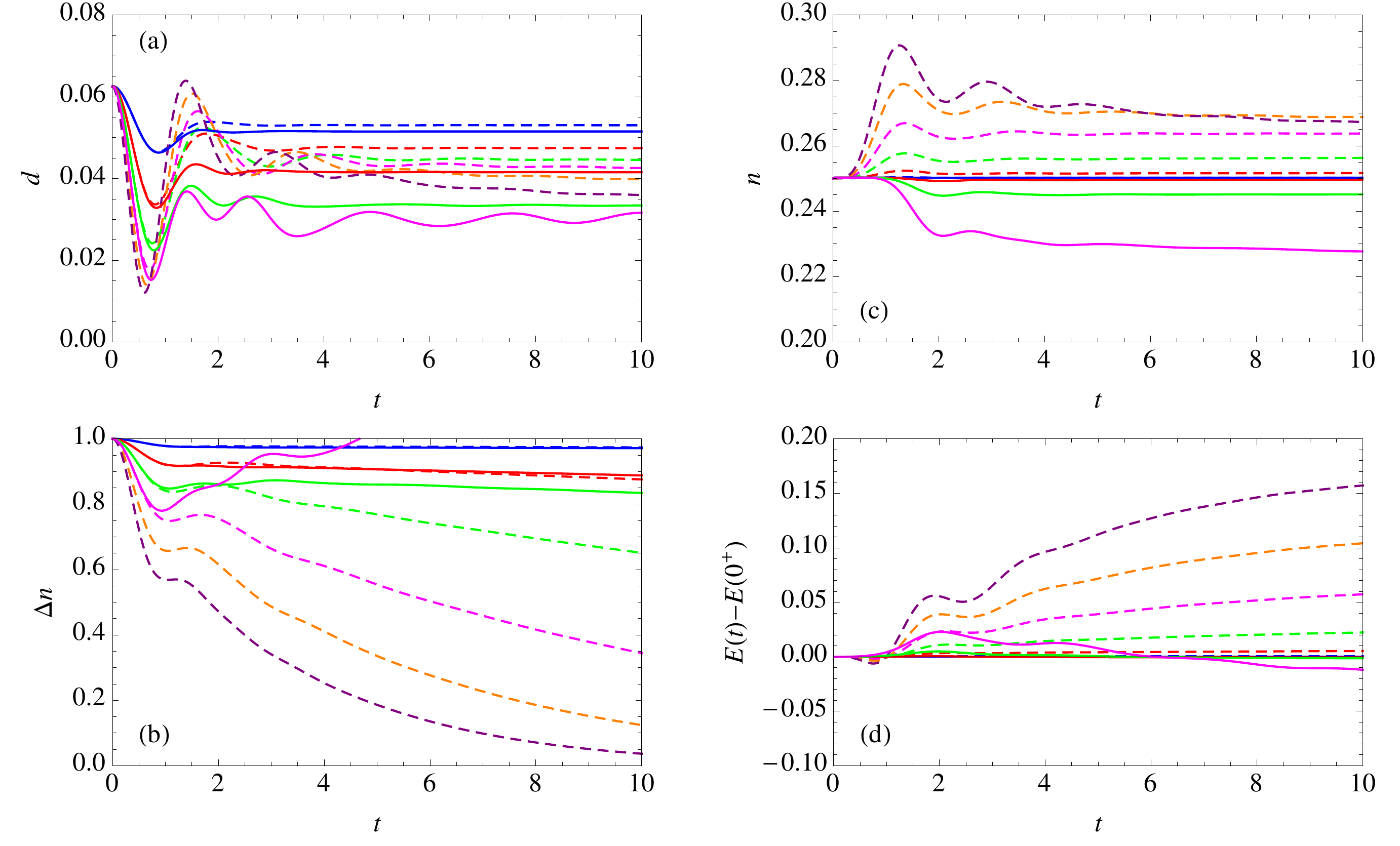}
\caption{(Color online) Time evolution of the double occupancy (a), the jump of the momentum distribution function (b), 
the density (c), and the total-energy drift (d) after interaction quenches in the PM phase of the Hubbard model at quarter filling
calculated by the nonequilibrium DMFT with the type (I) second-order (dashed curves, $U=0\to 0.5, 1, \dots, 3$), and type (I) third-order 
(solid curves, $U=0\to 0.5, 1, 1.5, 2$) perturbation theories.}
\label{double occupancy 3rd-order}
\end{center}
\end{figure*}
\begin{figure*}[tbp]
\begin{center}
\includegraphics[width=16cm]{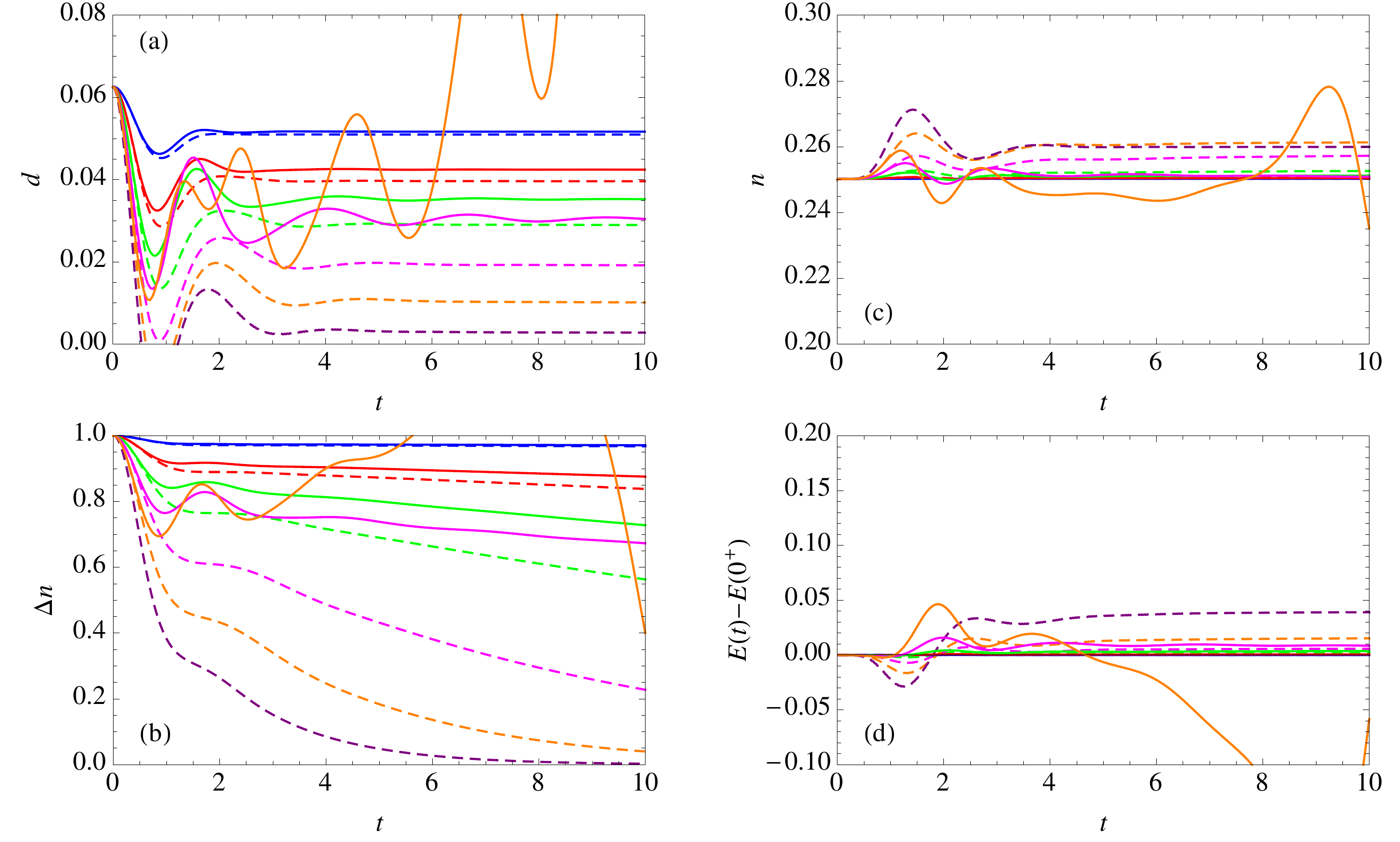}
\caption{(Color online) Time evolution of the double occupancy (a), the jump of the momentum distribution function (b), 
the density (c), and the total-energy drift (d) after interaction quenches in the PM phase of the Hubbard model at quarter filling
calculated by the nonequilibrium DMFT with the type (IV) second-order (dashed curves, $U=0\to 0.5, 1, \dots, 3$), and type (IV) third-order 
(solid curves, $U=0\to 0.5, 1, \dots, 2.5$) perturbation theories.}
\label{double occupancy 3rd-order type IV}
\end{center}
\end{figure*}
We plot the results produced by the type (I) and type (IV) expansions in Figs.~\ref{double occupancy 3rd-order} and \ref{double occupancy 3rd-order type IV}, respectively.
In Fig.~\ref{double occupancy 3rd-order}(a), we show the time evolution of the double occupancy calculated by the type (I) scheme. 
The noninteracting initial state has $d(0)=1/4 \times 1/4=1/16=0.0625$.
The second-order and third-order perturbations give quantitatively different evolutions after the quench. When $U_f$ is small enough ($U_f\le 1$), the double occupancy quickly damps to a thermal value.
As one increases $U_f$, an enhanced oscillation starts to appear in both the second-order and the third-order calculations. 
In Fig.~\ref{double occupancy 3rd-order}(b), we plot the time evolution
of the jump $\Delta n$ at the Fermi energy in the momentum distribution. Initially, the system has a Fermi distribution with $T=0$, so that $\Delta n(0)=1$. The second-order calculations (dashed curves in the bottom panel of Fig.~\ref{double occupancy 3rd-order})
show that after a rapid decrease, $\Delta n$ stabilizes at an intermediate value for a certain time and then slowly decays to zero. This behavior (prethermalization) is quite similar to what we have seen in the case of half filling.
If we use the third-order perturbation theories, however, the results differ from those of the second order for $U_f\ge 1.5$. In particular, at $U_f=2$ the jump $\Delta n$ starts to oscillate and finally exceeds $1$,
implying that the third-order calculation gives physically unreasonable results.

To examine the validity of the perturbation theories, we show the density and total energy as a function of time in Figs.~\ref{double occupancy 3rd-order}(c) and \ref{double occupancy 3rd-order}(d).
They should be conserved throughout the time evolution. The results suggest that the total energy and density ($n=1/4$) are reasonably conserved when $U_f\le 1$ in both the second-order and the third-order perturbations.
Only in this parameter regime, the simulation is reliable. This limitation is more severe than in the half-filling case, where the total energy is sufficiently conserved up to $U_f=2$.

We also investigated the type (IV) approach in Table~\ref{n-alpha} and show the results in Fig.~\ref{double occupancy 3rd-order type IV}.
The behavior of $d$ and $\Delta n$ [Figs.~\ref{double occupancy 3rd-order type IV}(a) and \ref{double occupancy 3rd-order type IV}(b)] looks qualitatively similar to the result of the type (I) expansion,
while there are quantitative differences such as the value of $d$ after relaxation and the plateau height for $\Delta n$ in the prethermalization regime.
The simulation with the third-order expansion of type (IV) becomes particularly unstable at $U_f=2.5$, showing rapid oscillations in $d$ and an irregular evolution in $\Delta n$.
If one looks at the density and total energy given by the type (IV) perturbation [Figs.~\ref{double occupancy 3rd-order type IV}(c) and \ref{double occupancy 3rd-order type IV}(d)], 
the conserving nature is somewhat improved with respect to the type (I) calculation. The conservation starts to break down earlier in $U_f$ in the third-order expansion compared to
the second-order one. Thus, we do not see a systematic improvement away from half filling by proceeding to higher-order perturbation expansions. It should be noted that also the weak-coupling QMC method can only reach times which are about a factor of two shorter than in the case of half-filling, because the odd-order diagrams contribute to the sign problem. 
Hence, the development of a useful impurity solver for nonequilibrium DMFT calculations away from half filling in the weak-coupling regime remains an open issue.

\section{Dynamical symmetry breaking induced by an interaction ramp in the Hubbard model}
\label{ramp symmetry breaking}

So far, we have considered the interaction quench dynamics of the Hubbard model without any long-range order. Since the formalism of the nonequilibrium DMFT has been generalized to the AFM phase 
in Sec.~\ref{nonequilibrium dmft afm}, we can apply the perturbative impurity solvers to the dynamics of such an ordered state.

In this section, we study dynamical symmetry breaking in the Hubbard model induced by an interaction ramp by means of the nonequilibrium DMFT with the third-order perturbation theory of type (I) (Table~\ref{n-alpha}).
This impurity solver correctly reproduced the AFM phase diagram (Fig.~\ref{phase diagram 3rd-order}) and the magnetization (Fig.~\ref{magnetization 3rd-order}) in the weak-coupling regime.
We begin with the PM initial state in thermal equilibrium, and then change the interaction parameter continuously (ramp) as
\begin{align}
U(t)=
\begin{cases}
U_i+(U_f-U_i)t/t_{\rm ramp} & 0\le t\le t_{\rm ramp}, \\
U_f & t> t_{\rm ramp},
\end{cases}
\end{align}
where $t_{\rm ramp}$ is the ramp time, to go across the phase transition line in the phase digram (Fig.~\ref{phase diagram 3rd-order}). 
We consider an interaction ramp ($t_{\rm ramp}>0$) rather than a quench ($t_{\rm ramp}=0$) to reduce the increase of the energy, 
but it turns out that the results do not significantly depend on $t_{\rm ramp}$.

In order to trigger the symmetry breaking,
we introduce a tiny staggered magnetic field $h$ in the initial state. We assume that the seed field $h$ is uniform in space, so that the order parameter (staggered magnetization $m$) grows uniformly.
From a large-scale point of view, this assumption is probably not appropriate, since the direction of symmetry breaking is random at each position, which leads to domain structures and topological defects in between
(Kibble-Zurek scenario \cite{Kibble1976, Zurek1985}). However, our interest here lies in the fast microscopic dynamics of the order parameter, where our set up can be justified. For convenience, we ramp off the seed field in the following way:
\begin{align}
h(t)=
\begin{cases}
h(1-t/t_{\rm ramp}) & 0\le t\le t_{\rm ramp}, \\
0 & t>t_{\rm ramp}.
\end{cases}
\end{align}

\subsection{Nonequilibrium DMFT results}
\begin{figure}[tbp]
\begin{center}
\includegraphics[width=7cm]{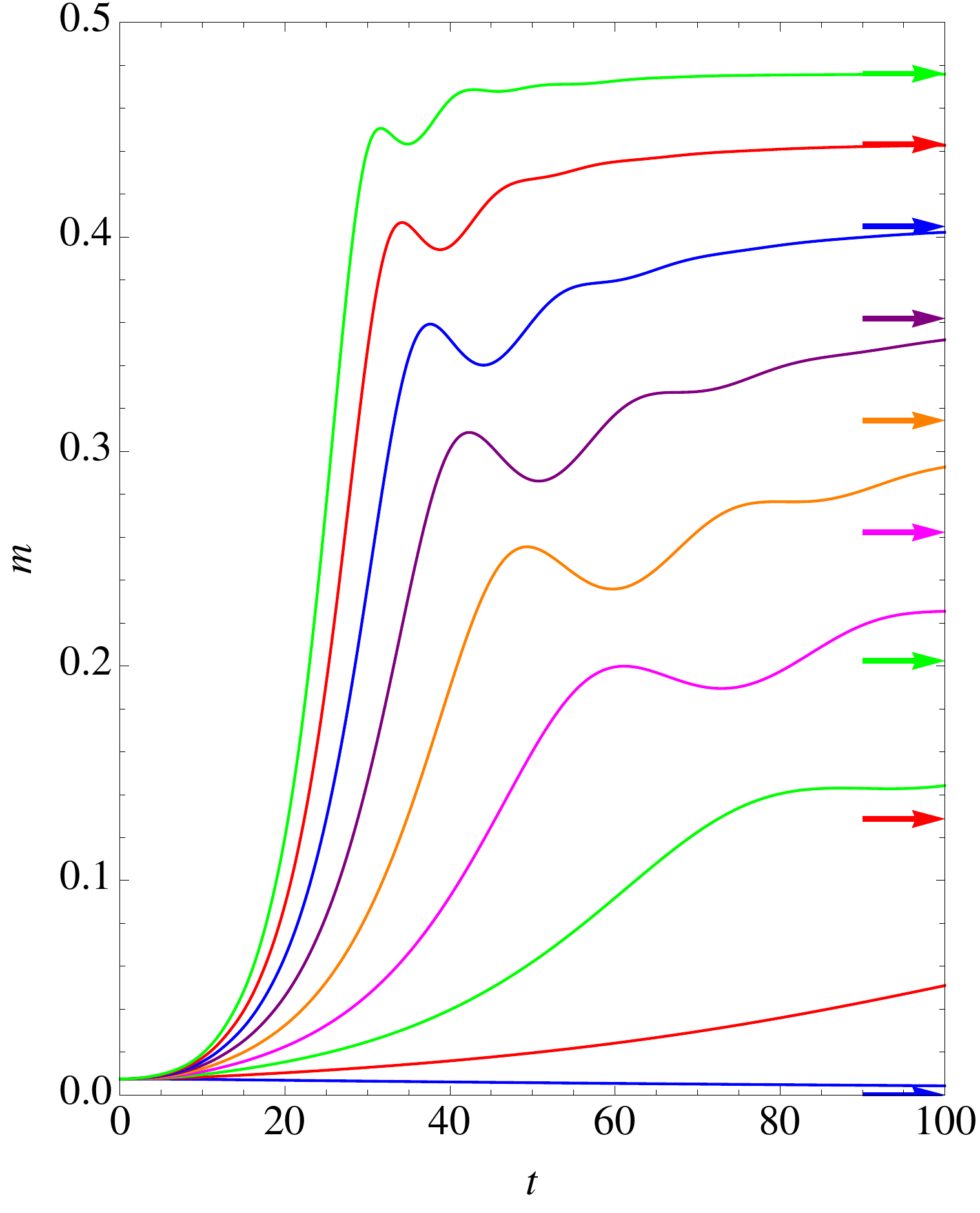}
\caption{(Color online) Time evolution of the staggered magnetization after the interaction ramps $U=1.75\to 1.8, 1.9, \dots, 2.6$ (from bottom to top) in the Hubbard model.
$\beta=11$, $h=10^{-4}$, and $t_{\rm ramp}=10$. The arrows indicate the corresponding thermal values reached in the long-time limit.}
\label{symmetry breaking}
\end{center}
\end{figure}
\begin{figure}[tbp]
\begin{center}
\includegraphics[width=8cm]{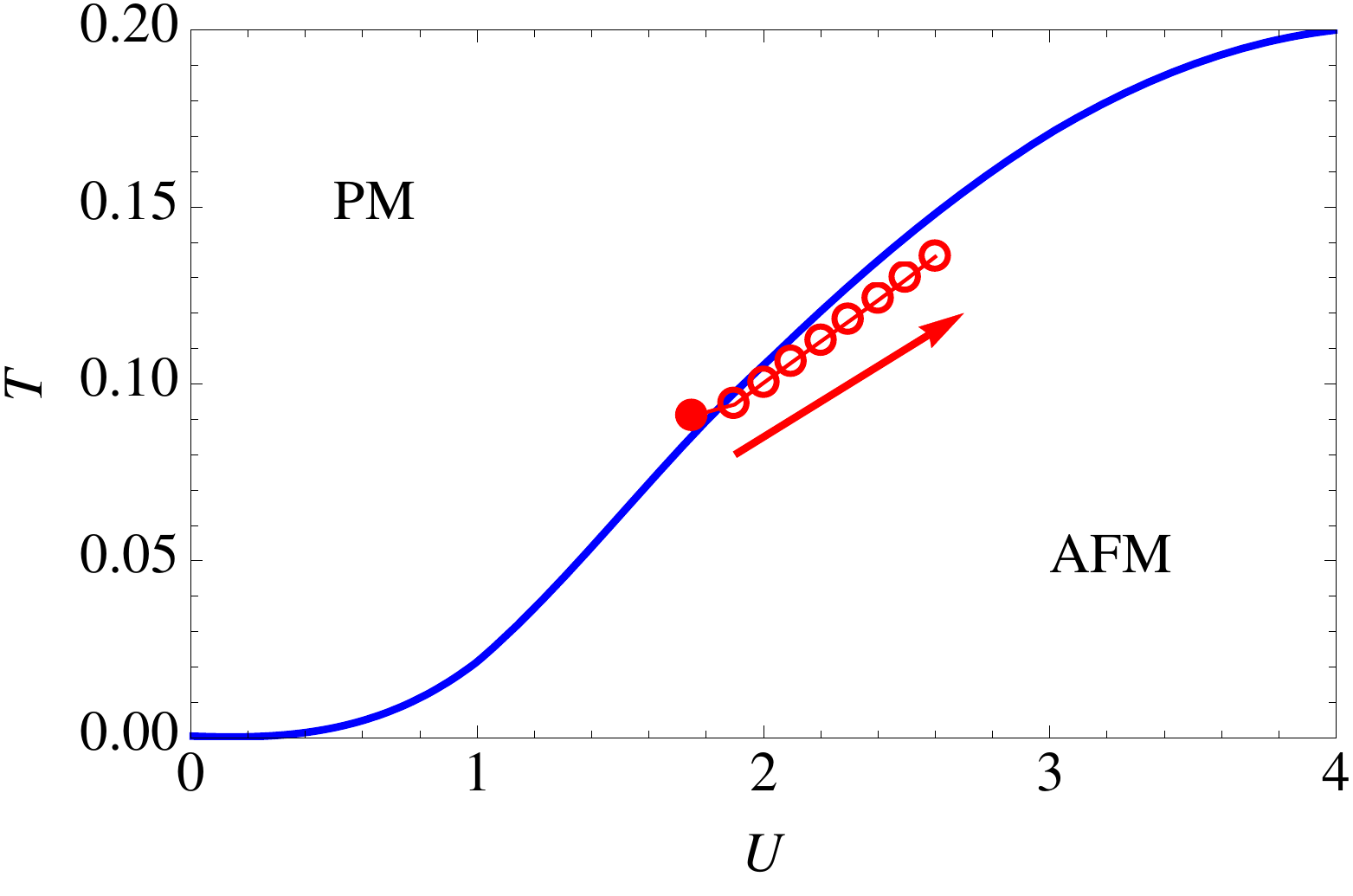}
\caption{(Color online) The initial (solid circle, $U_i=1.75$) and final (open circles, $U_f=1.9, 2, \dots, 2.6$ from left to right) thermal states in the simulation of the dynamical symmetry breaking in Fig.~\ref{symmetry breaking}. The labels ``PM'' and ``AFM'' indicate the paramagnetic and antiferromagnetic phases,
respectively.}
\label{effective temperature}
\end{center}
\end{figure}
In Fig.~\ref{symmetry breaking}, we show the evolution of the staggered magnetization after the interaction ramp obtained by the nonequilibrium DMFT.
The parameters are chosen such that $U_i=1.75$, $\beta=11$, $h=10^{-4}$, and $t_{\rm ramp}=10$.
The initial state is in the PM phase and is quite close to the AFM phase boundary (solid red circle in Fig.~\ref{effective temperature}). We fix the initial state and systematically change $U_f$
to perform a series of interaction-ramp simulations. The initial magnetization is very small but finite due to the presence of the staggered magnetic field $h$.
After the interaction ramp, the PM state becomes unstable, and the order parameter starts to grow exponentially ($m\propto e^{t/\tau_i}$ with $\tau_i$ the initial growth rate). 
It is followed by an amplitude oscillation and a gradual relaxation toward the final state. Here the oscillation is not as coherent as in the case of a ramp out of the symmetry-broken phase,\cite{TsujiEcksteinWerner2012} and one can see a softening of the amplitude mode in Fig.~\ref{symmetry breaking}.

\begin{figure}[tbp]
\begin{center}
\includegraphics[width=8cm]{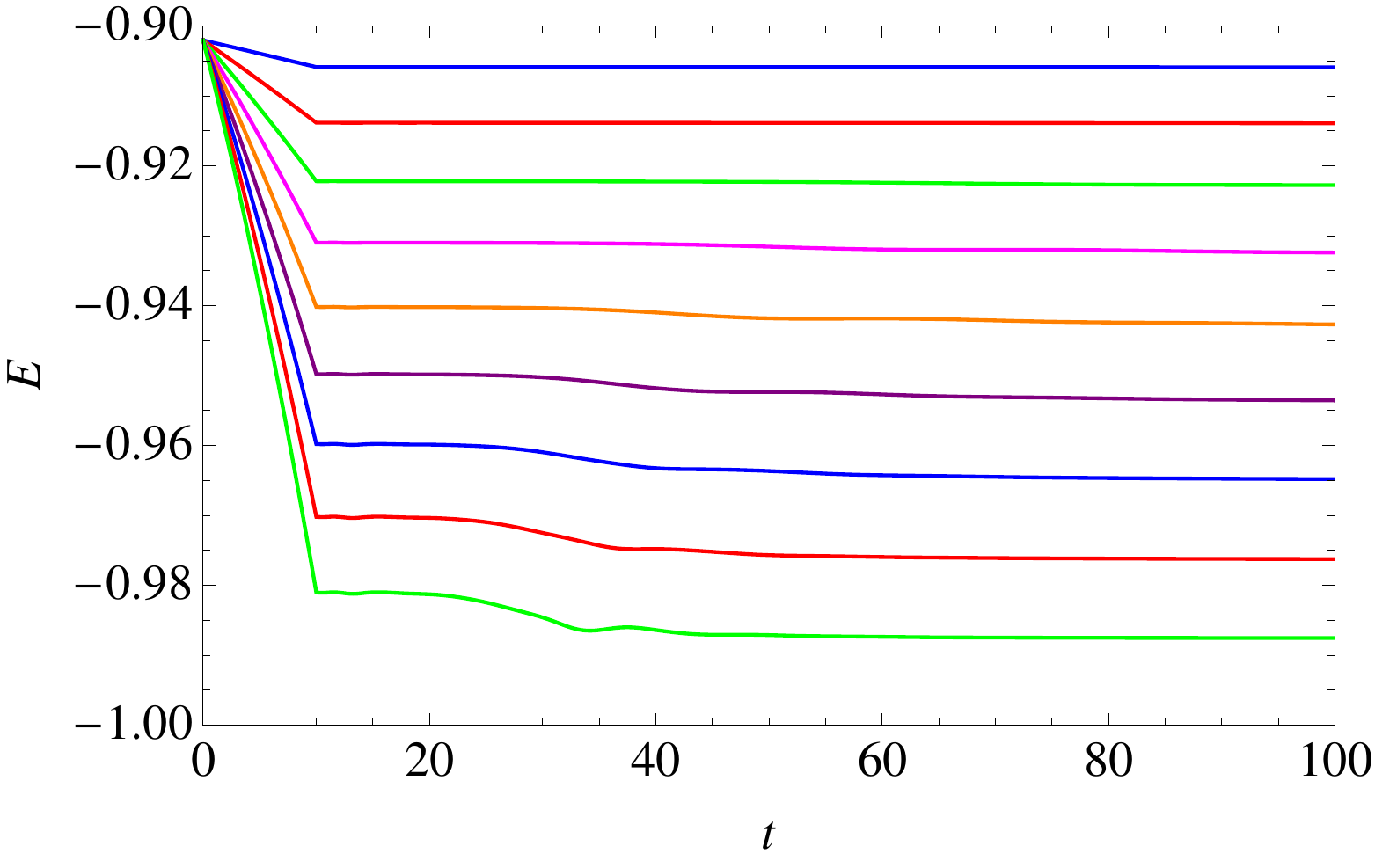}
\caption{(Color online) Time evolution of the total energy for the interaction ramps that correspond to Fig.~\ref{symmetry breaking}.}
\label{total energy symmetry breaking}
\end{center}
\end{figure}

In the long-time limit, the system finally thermalizes in the nonintegrable Hubbard model.
We can estimate the final temperature by searching for the equilibrium thermal state with effective temperature $T_{\rm eff}$ that has the same total energy as the time-evolving state,
since the total energy should be conserved after the interaction ramp. In Fig.~\ref{total energy symmetry breaking}, we plot the total energy for the interaction ramps that correspond to Fig.~\ref{symmetry breaking}.
For $U_f\lesssim 2.1$, the total energy is nicely conserved after the interaction ramps. As $U_f$ is further increased, there emerges a small energy drift during the symmetry breaking ($10\le t\lesssim 40$).
After the symmetry breaking, the conservation of the total energy is recovered. Thus, we have a slight inaccuracy in the simulation of the interaction ramps for larger $U_f$. We use the final value of the total energy,
$E_{\rm tot}(t=100)$, to extract the effective temperature of the thermal states reached in the long-time limit. In Fig.~\ref{effective temperature}, we indicate the final thermalized states in the phase diagram by open circles.
As we increase $U_f$, $T_{\rm eff}$ increases in the vicinity of the phase boundary. Similarly to the case of the dynamical phase transition out of the AFM phase,\cite{TsujiEcksteinWerner2012}
it seems to trace more or less the constant entropy curve,\cite{WernerParcolletGeorgesHassan2005}
although the interaction ramps that we consider here are not at all adiabatic processes.

The arrows in Fig.~\ref{symmetry breaking} indicate the thermal value of the order parameter ($m_{\rm th}$) that is realized in the long-time limit. We notice that there is a large deviation between the transient magnetization and the thermal values $m_{\rm th}$ for $1.9\le U_f\le 2.2$. Especially, the center of the oscillation of the amplitude mode is different from the long-time limit $m_{\rm th}$, so that the evolution of the order parameter is a superposition of a damped oscillation and a slow drift. This reminds us of the behavior of the order parameter seen in the dynamical phase transition 
from the AFM to PM phase induced by an interaction ramp,\cite{TsujiEcksteinWerner2012}
where $m$ does not decay immediately after the ramp but is ``trapped'' to a nonthermal value for a long time. It has been shown for that case that on a relatively short time scale the order-parameter dynamics is governed by the presence of a ``nonthermal critical point,'' in the vicinity of which the period of the amplitude mode diverges.

One may define two time scales that characterize the trapping of the nonthermal fixed point, namely the ``approach time'' to and the ``escape time''
from the nonthermal fixed point. The escape time is determined by $U_f$.
In the $U_f \to 0$ limit it diverges to infinity, while for $U_f \to 3$ it becomes quite
short (Fig.~\ref{symmetry breaking}) since thermalization is accelerated.
This is consistent with the previous observation of fast thermalization in the PM
phase of the Hubbard model.\cite{EcksteinKollarWerner2009}
On the other hand, the characterization of the approach time is unclear because
it depends on the definition. If one considers the initial exponential
growth of the order parameter as part of the nonthermal fixed-point behavior, then the
approach time is very short and does not significantly depend on $U_i$ and $U_f$.
Indeed, as we see in Sec.~\ref{Hartree approximation},
this exponential growth exists in the Hartree solution, which
characterizes the nonthermal fixed point. The weak dependence of the approach time on $U$
is consistent with Refs.~\onlinecite{MoeckelKehrein2008} and \onlinecite{EcksteinKollarWerner2009}.
However, if one interprets the approach time as the time necessary for
the order parameter to enter the coherently oscillating regime, then it is roughly determined by $\tau_i$.

\begin{figure}[tbp]
\begin{center}
\includegraphics[width=8cm]{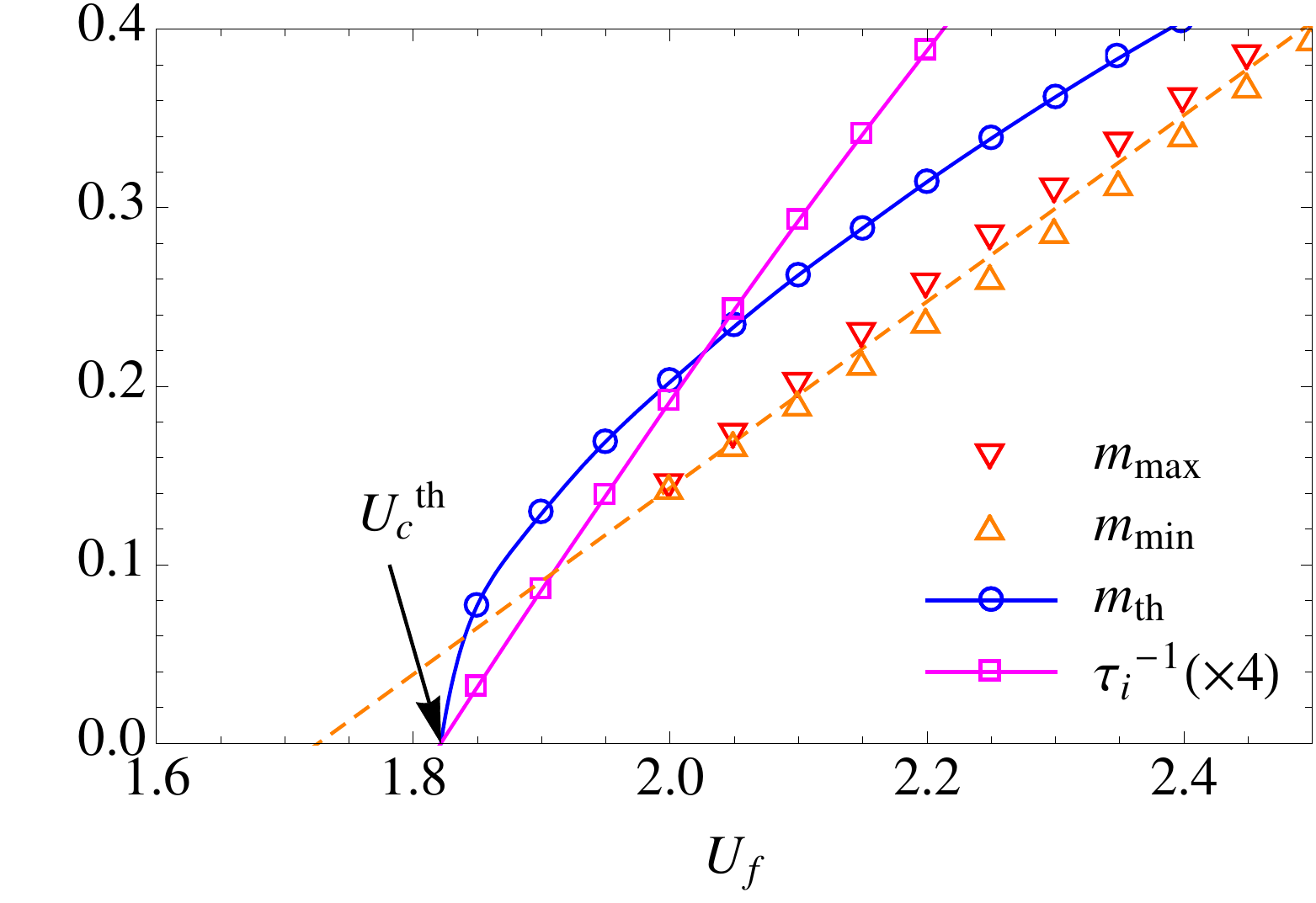}
\caption{(Color online) Various quantities that characterize the critical behavior of the dynamical symmetry breaking with $U_i=1.75$: 
$m_{\rm max}$ and ($m_{\rm min}$) is the maximum (minimum) of the first cycle of the oscillation in $m(t)$,
$m_{\rm th}$ is the thermal value taken in the long-time limit, and $\tau_i$ is the rate of the initial exponential growth.
The dashed line is an extrapolation of the middle points of $m_{\rm max}$ and $m_{\rm min}$.}
\label{critical behavior}
\end{center}
\end{figure}

To analyze the critical behavior of the dynamical symmetry breaking near the phase transition point, we plot several relevant quantities in Fig.~\ref{critical behavior}.
$m_{\rm th}$ (the thermal value reached in the long-time limit) vanishes at the thermal critical point ($U_f=U_c^{\rm th}$) as 
\begin{align}
m_{\rm th}\sim (U_f-U_c^{\rm th})^{\frac{1}{2}}.
\label{beta exponent}
\end{align}
This is consistent with the mean-field prediction $m_{\rm th}\sim (U_f-U_c^{\rm th})^\beta$ with the mean-field critical exponent $\beta=\frac{1}{2}$. 
$\tau_i$ is the time constant of the initial exponential growth ($m\propto e^{t/\tau_i}$),
which diverges as
\begin{align}
\tau_i\sim (U_f-U_c^{\rm th})^{-1}.
\end{align}
Note that these exponents are universal; i.e., 
they do not depend on details of the problem (the initial condition, perturbation of the system, etc.).
We also measured the maximum of the first peak ($m_{\rm max}$) and the minimum of the first dip ($m_{\rm min}$) of the amplitude oscillation, and we plot these quantities in Fig.~\ref{critical behavior}.
$m_{\rm max}$ and $m_{\rm min}$ characterize the ``trapping'' of the order parameter in the transient regime. They behave differently from $m_{\rm th}$:
$m_{\rm max}$ and $m_{\rm min}$ are always smaller than $m_{\rm th}$. The middle point $m_{\rm nth}\equiv (m_{\rm max}+m_{\rm min})/2$ (nonthermal magnetization) seems to depend linearly on $U_f$, which must be contrasted with 
the square-root dependence for $m_{\rm th}$ (\ref{beta exponent}). 
We see in Sec.~\ref{Hartree approximation} that this linear scaling can be justified in the weak-correlation limit.
The linear extrapolation of the middle points (dashed line in Fig.~\ref{critical behavior}) implies that the trapped order parameter
vanishes at a certain point $U_f=U_\ast^{\rm nth}$, which is different from the thermal critical point ($U_f=U_c^{\rm th}$), as 
\begin{align}
m_{\rm nth}\sim (U_f-U_\ast^{\rm nth})^1.
\label{nonthermal criticality}
\end{align}
As we see later in Sec.~\ref{Hartree approximation}, this nonthermal critical behavior becomes ``exact'' in the small $U$ regime 
(where the Hartree approximation is applicable) with $U_\ast^{\rm nth}$ identical to $U_c^{\rm th}$.
There are several possible interpretations of the behavior (\ref{nonthermal criticality}) for larger $U$: One is that the nonthermal critical point is
shifted from $U_f=U_c^{\rm th}$ to $U_\ast^{\rm nth}$ due to correlation effects. Another interpretation is that the nonthermal critical point
still exists at $U_f=U_c^{\rm th}$, but $m_{\rm max}$ and $m_{\rm min}$ are lifted up due to thermalization toward $m_{\rm th}$.
In any case, it is likely that the qualitative features of the nonthermal critical point survive to some extent in the moderate $U$ regime,
so that it affects the order-parameter dynamics during the dynamical symmetry breaking. 

\begin{figure}[tbp]
\begin{center}
\includegraphics[width=8cm]{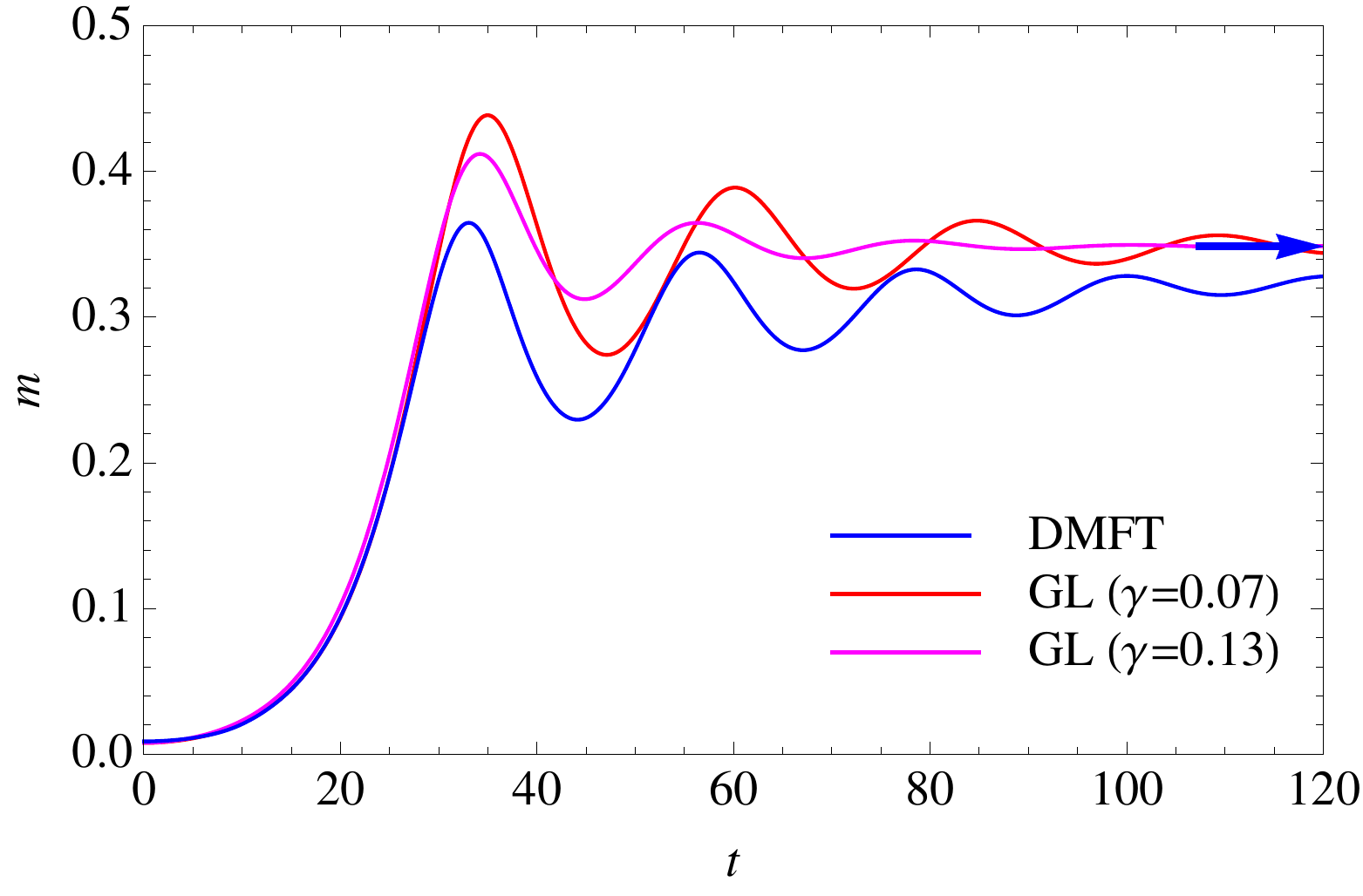}
\caption{(Color online) Comparison between the DMFT result of $m(t)$ for the quench $U=1.25\to 2$, $\beta=22$
with the phenomenological Ginzburg-Landau theory. The arrow indicates the thermal value of $m$
reached in the long-time limit.}
\label{magnetization U=1.25}
\end{center}
\end{figure}

\begin{figure}[tbp]
\begin{center}
\includegraphics[width=8cm]{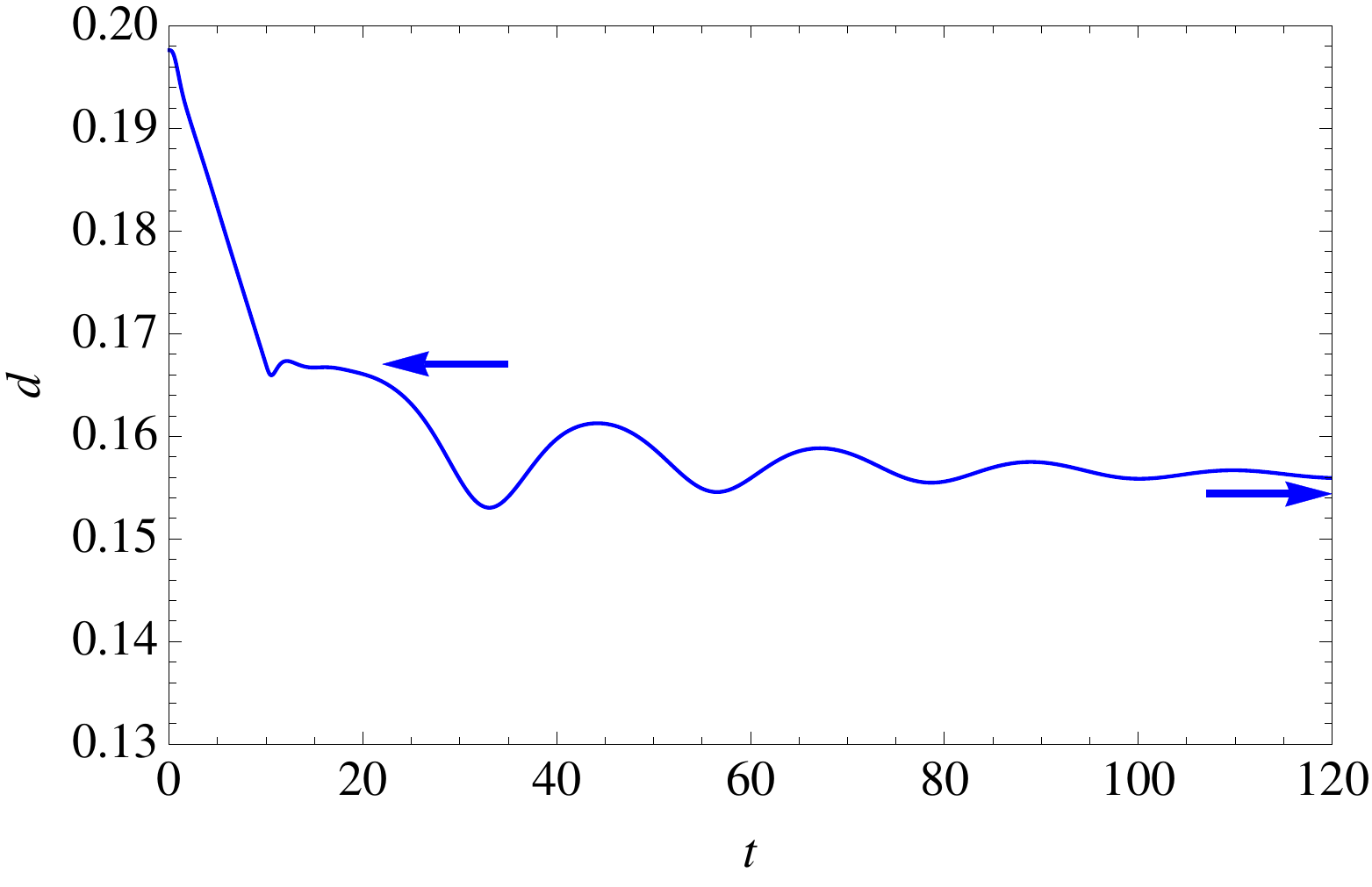}
\caption{(Color online) The time evolution of the double occupancy for the quench $U=1.25\to 2$, $\beta=22$. 
The arrow on the left indicates the thermal value of $m$ for the PM phase, while
the one on the right shows the value in the AFM phase.}
\label{double occupancy U=1.25}
\end{center}
\end{figure}

If we start with a smaller $U_i$, the amplitude mode induced by the interaction ramp becomes more coherent. In Fig.~\ref{magnetization U=1.25}, we show the time evolution of $m$
for $U_i=1.25$, $U_f=2$, $\beta=22$, $h=10^{-4}$, and $t_{\rm ramp}=10$ (blue curve). We can clearly see many oscillation cycles. Again the oscillation center slowly drifts to the thermal value $m_{\rm th}$ (arrow in Fig.~\ref{magnetization U=1.25}).
Figure \ref{double occupancy U=1.25} illustrates the corresponding evolution of the double occupancy. After the ramp ($t=10$), the double occupancy quickly approaches
the thermalized value within the PM phase indicated by the arrow on the left in Fig.~\ref{double occupancy U=1.25}. This suggests that the system prethermalizes within the PM phase before the dynamical symmetry breaking occurs.
After the order parameter $m$ starts to grow, the double occupancy also oscillates along with the amplitude oscillation of $m$. In the same way as $m$, the double occupancy slowly approaches the thermal value
for the AFM phase (the arrow on the right in Fig.~\ref{double occupancy U=1.25}).

The dynamics of the order parameter is reflected in the time-resolved spectral function $A_\sigma(\omega,t)$, which is defined by the retarded Green's function,
\begin{align}
A_\sigma(\omega,t)=-\frac{1}{\pi}{\rm Im}\int_0^{\infty} d\bar{t}\,e^{i\omega\bar{t}}
G_\sigma^R(t+\bar{t}/2,t-\bar{t}/2).
\label{def spectral function}
\end{align}
This function represents the single-particle spectrum at time $t$.
Since the range of the time arguments is limited ($\le t_{\rm max}$), we have to introduce a cutoff in the semi-infinite integral in Eq.~(\ref{def spectral function}). As a result, the energy resolution $\Delta\omega$ is restricted (energy-time uncertainty). Here $\Delta\omega\sim 2\pi/t_{\rm max}\sim 0.06$, which is fine enough
to resolve the AFM energy gap.

\begin{figure}[tbp]
\begin{center}
\includegraphics[width=8cm]{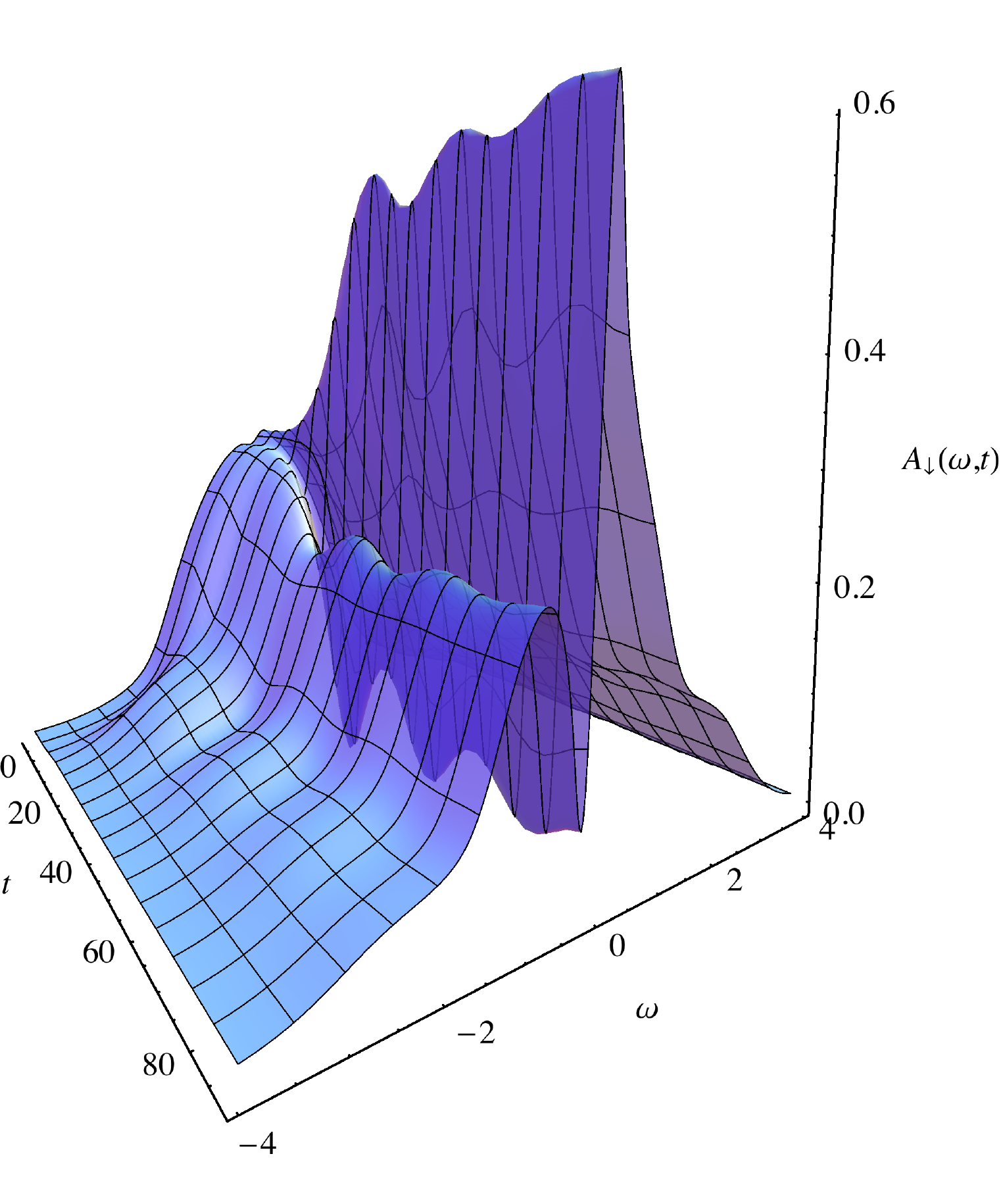}
\caption{(Color online) The time-resolved spectral function $A(\omega,t)$ of the minority spin component
for the quench $U=1.25\to 2$, $\beta=22$.}
\label{spectral function}
\end{center}
\end{figure}
In Fig.~\ref{spectral function}, we depict $A_\sigma(\omega,t)$ for the interaction ramp ($U=1.25\to 2$) which corresponds to the blue magnetization curve in Fig.~\ref{magnetization U=1.25}.
At first, the system is noninteracting, so that $A_\sigma(\omega)=\sqrt{4v_\ast^2-\omega^2}/(2\pi v_\ast^2)$ (noninteracting DOS). After the interaction ramp, an energy gap 
is dynamically generated at the Fermi energy ($\omega=0$) in the spectral function. 
Once the gap has opened, the magnitude of the gap [the distance between the coherence peaks in $A(\omega)$] stays nearly constant in time. On the other hand,
there is a coherently oscillating spectral-weight transfer between the lower ($\omega<0$) and higher ($\omega>0$) energy region, consistent with the time-evolution of the order parameter $m$.
The drift of the oscillation center is also reflected in $A(\omega, t)$. Therefore, the spectral function captures the characteristic behavior of the order-parameter dynamics.
Experimentally, it is not easy to observe the time evolution of the staggered magnetization directly. However, the change of the spectral function can be detected by
time-resolved photoemission spectroscopy and pump-probe optical spectroscopy. 
These techniques thus provide a way of tracking the evolution of the staggered magnetization.

\subsection{Comparison to the phenomenological Ginzburg-Landau equation}
\label{Ginzburg-Landau}

To analyze the behavior of the order parameter after the interaction ramp,
we compare the nonequilibrium DMFT results with the phenomenological Ginzburg-Landau (GL) equation.\cite{Schmid1966,AbrahamsTsuneto1966,SadeMelo1993}
The GL equation has been widely used to describe the order-parameter dynamics in superconductors and other ordered phases.
It is justified when the quasiparticle energy relaxation time is much longer than the time scale of the order-parameter dynamics.\cite{Barankov2004}

Here we adopt a phenomenological description, assuming that the motion of the order parameter is governed by the free-energy potential of the final thermal state [$F_{\rm th}(m)$]
after the interaction ramp; i.e., the initial free energy is suddenly quenched to the final one (sudden approximation).
Our equation reads
\begin{align}
\partial_t^2 m+\gamma\partial_t m=-\frac{\partial F_{\rm th}(m)}{\partial m}=-2a_{\rm th}m-4bm^3,
\label{GL equation}
\end{align}
where 
$\gamma$ is a ``friction'' constant, and the free energy of the final thermal state is expanded as $F_{\rm th}(m)=a_{\rm th}m^2+bm^4$.
To distinguish the coefficient $a$ of the thermal free energy from that for the nonthermal potential that will be defined later,
we put the subscript ``th''.
We can freely rescale both sides of Eq.~(\ref{GL equation}), so that we choose the coefficient of $\partial_t^2 m$ to be unity.
By taking the final free energy, we can guarantee
that the order parameter converges to the thermal value of the final state in the long-time limit.
Of course, the transient state right after the interaction ramp is far from equilibrium, so one cannot expect that the whole dynamics is reproduced by this sudden approximation.
Here we use the phenomenological approach to demonstrate to what extent the order parameter behaves differently from the conventional GL picture.

In equilibrium, the order parameter takes the thermal value
\begin{align}
m_{\rm th}=\sqrt{-\frac{a_{\rm th}}{2b}}.
\label{mth}
\end{align}
Initially, the order parameter grows exponentially, $m\propto e^{t/\tau_i}$.
Since the order parameter is small at the initial stage, one can neglect the second term on the right-hand side of Eq.~(\ref{GL equation}).
Substituting $m\propto e^{t/\tau_i}$ in Eq.~(\ref{GL equation}), one obtains the relation
\begin{align}
\tau_i^{-2}+\gamma\tau_i^{-1}=-2a_{\rm th}.
\label{tau_i}
\end{align}
$m_{\rm th}$ and $\tau_i$ can be directly measured.
If we fix one parameter (say $\gamma$), we can identify the other parameters $a_{\rm th}$ and $b$ using Eqs.~(\ref{mth}) and (\ref{tau_i}).

In Fig.~\ref{magnetization U=1.25}, we plot the solution of the time-dependent GL Eq.~(\ref{GL equation}) for $\gamma=0.07$ and $\gamma=0.13$
on top of the nonequilibrium DMFT result. We have agreement in the initial exponential growth and the final value, whereas the transient dynamics 
of the GL calculations looks quite different from the DMFT result. The GL equation cannot describe the trapping effect of the order parameter, i.e., 
the center of the amplitude oscillation is fixed to $m_{\rm th}$ from the beginning. The amplitude, damping rate, and phase shift of the oscillation are not correctly captured by the GL equation, no matter how the value of the free parameter $\gamma$ is chosen.
If we try to fit the frequency of the amplitude mode ($\gamma=0.13$), the damping is too strong. If we try to fit the amplitude of the oscillation ($\gamma=0.07$), we have a phase mismatch. 
Furthermore, the GL equation does not capture the softening of the amplitude mode. Thus, we conclude that the DMFT order-parameter dynamics which shows a softening amplitude mode and a trapping by a nonthermal critical point is out of the adiabatic regime, so
the GL description is not applicable.

\subsection{Comparison to the time-dependent Hartree approximation}
\label{Hartree approximation}

Finally, let us compare the nonequilibrium DMFT results with the Hartree approximation, which may be valid in the opposite limit,  
where the order parameter changes fast compared to the quasiparticle scattering time in the weak-coupling regime. 
In the Hartree approximation, one takes the tadpole diagram (Fig.~\ref{hartree}) as the self-energy, 
\begin{align}
\Sigma_{a\sigma}(t,t')
&=
U(t)n_{a\bar{\sigma}}\delta_{\mathcal C}(t,t').
\end{align}
In the AFM phase the local density is
\begin{align}
n_\sigma^A(t)&=\bar{n}+\frac{1}{2}\sigma m(t),
\\
n_\sigma^B(t)&=\bar{n}-\frac{1}{2}\sigma m(t),
\end{align}
where $\bar{n}$ is the average density per spin, and $\sigma=\uparrow, \downarrow=\pm$. 
At half filling, $\bar{n}=\frac{1}{2}$.

As shown in the Supplemental Material of Ref.~\onlinecite{TsujiEcksteinWerner2012}, the Dyson equation (\ref{lattice Dyson}) and its conjugate equation can be written in the form of a Bloch equation for spin precession,
\begin{align}
\partial_t \bm f_{\bm k}(t)
&=
\bm b_{\bm k}(t)\times \bm f_{\bm k}(t).
\label{Bloch equation}
\end{align}
Here we use a vector representation $\bm f_{\bm k}=(f_{\bm k}^x, f_{\bm k}^y, f_{\bm k}^z)$ for the momentum distributions, analogous to Anderson's pseudospin representation for superconductivity.\cite{Anderson1958}
The components are defined by
\begin{align}
f_{\bm k}^x(t)
&=
\frac{1}{2}\sum_\sigma [n_{\bm k\sigma}^{BA}(t)+n_{\bm k\sigma}^{AB}(t)],
\\
f_{\bm k}^y(t)
&=
\frac{i}{2}\sum_\sigma \sigma[n_{\bm k\sigma}^{BA}(t)-n_{\bm k\sigma}^{AB}(t)],
\\
f_{\bm k}^z(t)
&=
\frac{1}{2}\sum_\sigma \sigma[n_{\bm k\sigma}^{AA}(t)-n_{\bm k\sigma}^{BB}(t)],
\end{align}
where $n_{\bm k\sigma}^{ab}(t)\equiv -iG_{\bm k\sigma}^{ab<}(t,t)$ is the momentum distribution function for the $a, b=A, B$ sublattice.
$n_{\bm k\sigma}^{AA}(t)+n_{\bm k\sigma}^{BB}(t)$ is a constant of motion (time independent).
The effective magnetic field $\bm b_{\bm k}(t)$ in Eq.~(\ref{Bloch equation}) is given by
\begin{align}
\bm b_{\bm k}(t)=(-2\epsilon_{\bm k},0,U(t)m(t)).
\end{align}
The order parameter is self-consistently determined by the condition
\begin{align}
m(t)=\sum_{\bm k} f_{\bm k}^z(t).
\label{mean field}
\end{align}
With this, the equation of motion for $\bm f_{\bm k}(t)$ is closed. Note that Eq.~(\ref{Bloch equation}) holds for arbitrary filling $\bar{n}$.

Let us first look at the equilibrium solution in the Hartree approximation.
By solving the static Dyson equation, we obtain the momentum distributions,
\begin{align}
&
\begin{pmatrix}
n_{\bm k\sigma}^{AA} & n_{\bm k\sigma}^{AB} \\
n_{\bm k\sigma}^{BA} & n_{\bm k\sigma}^{BB}
\end{pmatrix}
=
T\sum_{n} e^{i\omega_n 0^+}
\nonumber
\\
&\times
\begin{pmatrix}
i\omega_n+\mu-U(\bar{n}+\frac{1}{2}\bar{\sigma} m) & -\epsilon_{\bm k} \\
-\epsilon_{\bm k} & i\omega_n+\mu-U(\bar{n}-\frac{1}{2}\bar{\sigma} m)
\end{pmatrix}^{-1}.
\end{align}
From this, we can explicitly calculate $f_{\bm k}^z$,
\begin{align}
f_{\bm k}^z
&=
-T\sum_n e^{i\omega_n 0^+}\frac{Um}{(i\omega_n+\mu-U\bar{n})^2-\left(\frac{Um}{2}\right)^2-\epsilon_{\bm k}^2}
\nonumber
\\
&=
-\frac{Um}{2\sqrt{\left(\frac{Um}{2}\right)^2+\epsilon_{\bm k}^2}}
\textstyle\bigg[f\left(\sqrt{\left(\frac{Um}{2}\right)^2+\epsilon_{\bm k}^2}-\mu+U\bar{n}\right)
\nonumber
\\
&\quad
\textstyle-f\left(-\sqrt{\left(\frac{Um}{2}\right)^2+\epsilon_{\bm k}^2}-\mu+U\bar{n}\right)\bigg],
\end{align}
with $f(\epsilon)=1/(e^{\beta\epsilon}+1)$ the Fermi distribution function.
Substituting this result into the mean-field condition (\ref{mean field}), we get, near the thermal phase transition point 
(at $U=U_c^{\rm th}$ where $m$ is infinitesimal), 
the equilibrium mean-field equation
\begin{align}
1=-U_c^{\rm th} \sum_{\bm k}\frac{f(\epsilon_{\bm k}-\mu+U_c^{\rm th}\bar{n})-f(-\epsilon_{\bm k}-\mu+U_c^{\rm th}\bar{n})}
{2\epsilon_{\bm k}}.
\label{static mean field}
\end{align}
It is equivalent to the result of the random phase approximation,
$1=U_c^{\rm th}\chi_0(\bm q)$, where $\chi_0(\bm q)=-T\sum_{n\bm k} G_{\bm k}(i\omega_n)G_{\bm k+\bm q}(i\omega_n)$,
and $\bm q=(\pi,\pi,\dots)$.
Note that Eq.~(\ref{static mean field}) determines the transition temperature for the Hartree approximation 
as shown in Fig.~\ref{afm phase diagram second-order}.

Now we move on to analyze the time evolution based on Eq.~(\ref{Bloch equation}).
This type of equation is equivalent\cite{TsujiEcksteinWerner2012} to the time-dependent BCS equation for superconductivity, which is known to be exactly solvable.
\cite{Barankov2004,Yuzbashyan2005,WarnerLeggett2005,Barankov2006,Yuzbashyan2006}
Following the heuristic argument in Ref.~\onlinecite{Barankov2004}, 
we can construct an analytic solution of Eq.~(\ref{Bloch equation}) for the AFM dynamical symmetry breaking induced by an interaction quench from the PM initial state
($U=U_i\to U_f$).

First, we propose an ansatz for Eq.~(\ref{Bloch equation}),
\begin{align}
f_{\bm k}^x(t)&=f_0(\bm k)+f_1(\bm k)m(t)^2,
\\
f_{\bm k}^y(t)&=f_2(\bm k)\partial_t m(t),
\\
f_{\bm k}^z(t)&=f_3(\bm k)m(t),
\end{align}
where $f_i(\bm k)$ ($i=0,1,2,3$) is an arbitrary time-independent function. The mean-field condition (\ref{mean field}) becomes
\begin{align}
\sum_{\bm k} f_3(\bm k)=1.
\label{f3 self-consistency}
\end{align}
We substitute this ansatz into Eq.~(\ref{Bloch equation}) and obtain
\begin{align}
2f_1(\bm k)m(t)\partial_t m(t)&=-U(t)m(t)f_2(\bm k)\partial_t m(t),
\label{fx}
\\
f_2(\bm k)\partial_t^2m(t)&=2\epsilon_{\bm k}f_3(\bm k)m(t)
\label{fy}
\nonumber
\\
&\quad +U(t)m(t)[f_0(\bm k)+f_1(\bm k)m(t)^2],
\\
f_3(\bm k)\partial_t m(t)&=-2\epsilon_{\bm k}f_2(\bm k)\partial_t m(t).
\label{fz}
\end{align}
From Eqs.~(\ref{fx}) and (\ref{fz}), we have
\begin{align}
2f_1(\bm k)&=-U_f f_2(\bm k),
\\
f_3(\bm k)&=-2\epsilon_{\bm k}f_2(\bm k),
\end{align}
with $U(t>0)=U_f$.
Substituting these equations into Eq.~(\ref{fy}), we get
\begin{align}
f_2(\bm k)\partial_t^2m(t)
&=
[-(2\epsilon_{\bm k})^2f_2(\bm k)+U_f f_0(\bm k)]m(t)-\frac{U_f^2}{2}f_2(\bm k)m(t)^3.
\end{align}
Let us assume that $f_2(\bm k)\neq 0$. Then, we can divide the above equation by $f_2(\bm k)$ to obtain
\begin{align}
\partial_t^2m(t)&=\left[-(2\epsilon_{\bm k})^2+U_f\frac{f_0(\bm k)}{f_2(\bm k)}\right]m(t)-\frac{U_f^2}{2}m(t)^3.
\end{align}
This should hold for arbitrary $\bm k$, which suggests that the coefficient of $m(t)$ on the right-hand side must be independent of $\bm k$.
This motivates us to set it to a $\bm k$-independent constant,
\begin{align}
-(2\epsilon_{\bm k})^2+U_f\frac{f_0(\bm k)}{f_2(\bm k)}\equiv a_{\rm nth},
\end{align}
or
\begin{align}
f_2(\bm k)=U_f\frac{f_0(\bm k)}{(2\epsilon_{\bm k})^2+a_{\rm nth}}.
\end{align}
If such a constant $a_{\rm nth}(>0)$ exists, $m(t)$ satisfies the following ``GL-like'' equation,
\begin{align}
\partial_t^2 m(t)&=-\frac{\partial F_{\rm nth}(m)}{\partial m},
\label{GL like}
\\
F_{\rm nth}(m)&=-\frac{1}{2}a_{\rm nth}m^2+\frac{U_f^2}{8}m^4,
\end{align}
where $F_{\rm nth}(m)$ is a nonthermal ``free-energy'' potential. Note that $F_{\rm nth}(m)\neq F_{\rm th}(m)$. In particular, terms with orders higher than four are absent in $F_{\rm nth}(m)$.
Now the mean-field condition (\ref{f3 self-consistency}) becomes
\begin{align}
-U_f\sum_{\bm k}\frac{2\epsilon_{\bm k}}{(2\epsilon_{\bm k})^2+a_{\rm nth}}f_0(\bm k)=1.
\label{condition for a}
\end{align}
$f_0(\bm k)$ is determined from the initial condition. Let us assume that the initial magnetization $m(0)$ induced by the seed magnetic field is very small
and $\partial_tm(t)=0$, which leads to
\begin{align}
f_{\bm k}^x(0)&=
f(\epsilon_{\bm k}-\mu_i+U_i\bar{n})-f(-\epsilon_{\bm k}-\mu_i+U_i\bar{n})+O(m(0)^2),
\label{initial condition1}
\\
f_{\bm k}^y(0)&=0,
\label{initial condition2}
\\
f_{\bm k}^z(0)&=O(m(0)),
\label{initial condition3}
\end{align}
with $\mu_i$ the chemical potential of the initial state.
Here we have used the noninteracting equilibrium distribution function,
\begin{align}
n_{\bm k\sigma}^{BA}=n_{\bm k\sigma}^{AB}
&=
T\sum_n e^{i\omega_n 0^+}\frac{\epsilon_{\bm k}}{(i\omega_n+\mu_i-U_i\bar{n})^2-\epsilon_{\bm k}^2}
\nonumber
\\
&=
\frac{1}{2}[f(\epsilon_{\bm k}-\mu_i+U_i\bar{n})-f(-\epsilon_{\bm k}-\mu_i+U_i\bar{n})].
\end{align}
Thus, we obtain
\begin{align}
f_0(\bm k)=f(\epsilon_{\bm k}-\mu_i+U_i\bar{n})-f(-\epsilon_{\bm k}-\mu_i+U_i\bar{n}).
\end{align}
Now we have determined all the components of $\bm f_{\bm k}(t)$, which are consistent with the equation of motion (\ref{Bloch equation}),
the self-consistency condition (\ref{mean field}), and the initial condition (\ref{initial condition1})-(\ref{initial condition3}).

If no constant $a_{\rm nth}$ exists which satisfies Eq.~(\ref{condition for a}), then no symmetry breaking occurs.
Hence, the existence of a solution for $a_{\rm nth}$ in Eq.~(\ref{condition for a}) is a prerequisite for dynamical symmetry breaking. 
As we see below, this corresponds to the condition for a symmetry-broken solution in equilibrium; that is, the dynamical symmetry breaking occurs
if and only if $U_f$ exceeds $U_c^{\rm th}$ determined by Eq.~(\ref{static mean field}) with the initial temperature.

\begin{figure}[tbp]
\begin{center}
\includegraphics[width=8cm]{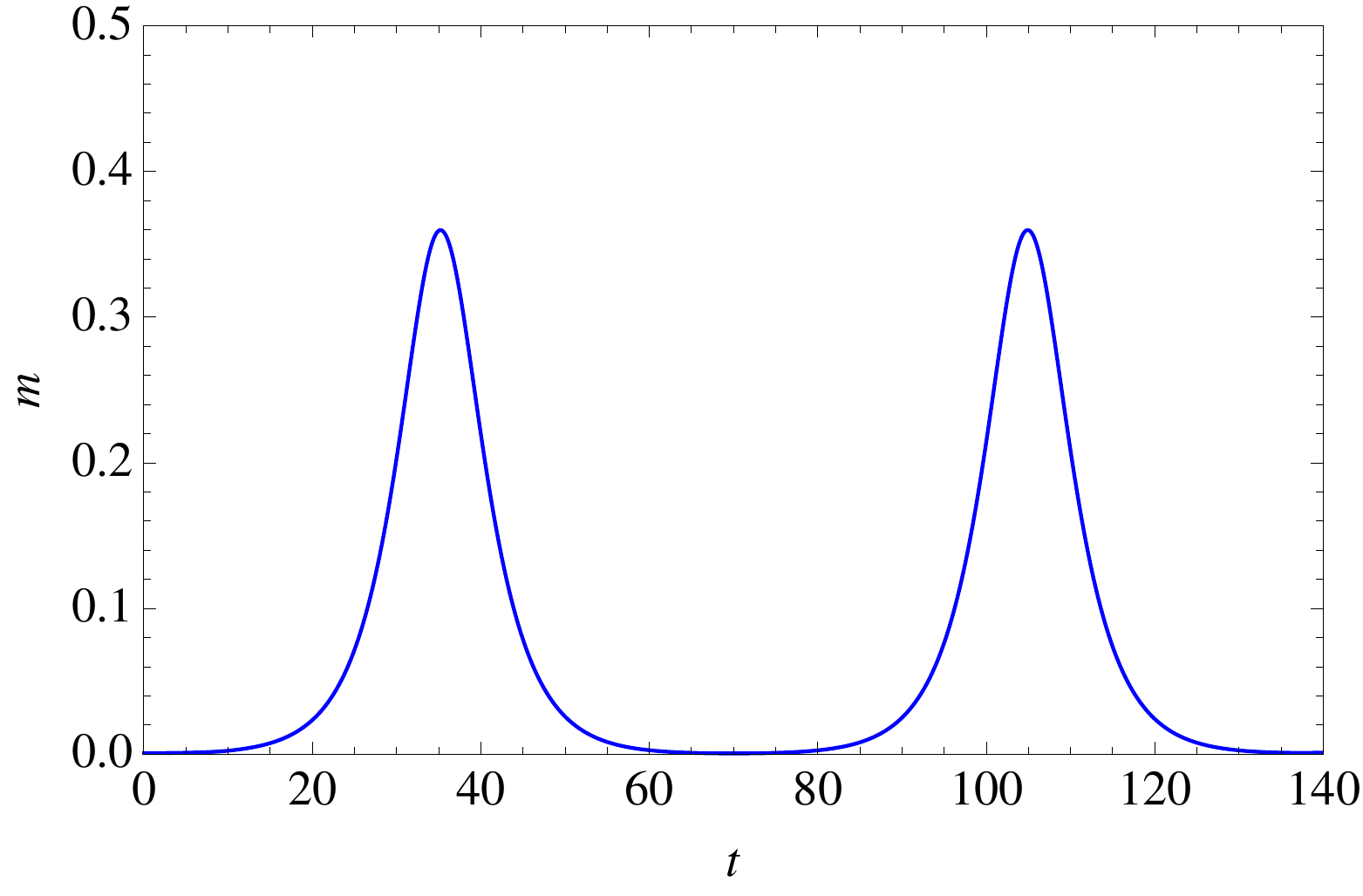}
\caption{(Color online) The time evolution of $m$ for the interaction ramp $U=0\to 1.25$ in the Hubbard model at half filling with $\beta=40$, $h=10^{-4}$, and $t_{\rm ramp}=10$ obtained from the Hartree approximation.}
\label{soliton}
\end{center}
\end{figure}
In Fig.~\ref{soliton}, we plot as an example the solution for the interaction ramp $U=0\to 1.25$ given by the Hartree approximation.
If the ramp is performed fast enough compared to the development of the order parameter, it can be considered as a quench.
The interaction ramp generates an exponential growth of the order parameter, after which it goes through a maximum and returns back to the initial value.
The curve of $m(t)$ in Fig.~\ref{soliton} looks like a soliton. In fact, Eq.~(\ref{GL like}) allows for an analytical soliton solution,
\begin{align}
m(t)=\frac{2\sqrt{a_{\rm nth}}}{U_f\cosh(\sqrt{a_{\rm nth}}t)},
\end{align}
for the initial condition that $m(-\infty)$ is infinitesimal. When the initial value is nonzero, the solution corresponds to a train of solitons as shown in Fig.~\ref{soliton}.
The period of the soliton train depends on the initial condition and is hence nonuniversal, while the maximum of $m(t)$, $m_{\rm max}=2\sqrt{a_{\rm nth}}/U_f$, does not.
As we see below, this $m_{\rm max}$ exhibits a universal behavior, which obeys a scaling law different from that for the conventional GL theory.

The nature of the Hartree solution is quite distinct from the DMFT results (Figs.~\ref{symmetry breaking} and \ref{magnetization U=1.25}):
The Hartree approximation gives a permanently oscillating $m$, whereas the DMFT solution indicates that the amplitude oscillation damps, and $m$ eventually converges to the thermal value $m_{\rm th}$.
The difference is apparently coming from the lack of scattering processes in the Hartree approximation. In other words, it is due to the ``integrability" of the Hartree equation.
However, there seem to exist common universal features in both results. For example, the universality of $m_{\rm max}$ in the Hartree approximation somehow survives even after we take account of
correlations in the nonequilibrium DMFT. We examine this point later.

Let us first take a closer look at the coefficient $a_{\rm nth}$ of the quadratic term in $F_{\rm nth}(m)$, since it controls the phase transition. 
$a_{\rm nth}$ is implicitly determined by Eq.~(\ref{condition for a}), through which $a_{\rm nth}$ can be regarded as a function of $U_f$. Assuming that $a_{\rm nth}=a_{\rm nth}(U_f)$ is a reversible function,
we write $U_f=U_f(a_{\rm nth})$. We now prove that in the vicinity of $a_{\rm nth}=0$, $U_f$ varies as
\begin{align}
U_f(a_{\rm nth})=U_c^{\rm nth}+c\sqrt{a_{\rm nth}}+O(a_{\rm nth}^1);
\label{a expansion}
\end{align}
i.e., $U_f$ has a square-root dependence on $a_{\rm nth}$ ($c$ is an arbitrary constant).
$U_c^{\rm nth}\equiv \lim_{a_{\rm nth}\to 0}U_f(a_{\rm nth})$ can be interpreted as the critical interaction strength;
i.e., the dynamical symmetry breaking is generated when $U_f>U_c^{\rm nth}$.

To identify $U_c^{\rm nth}$, we consider the limit $a_{\rm nth}\to 0$ in Eq.~(\ref{condition for a}).
Substituting $a_{\rm nth}=0$ in Eq.~(\ref{condition for a}) gives 
\begin{align}
-U_c^{\rm nth}\sum_{\bm k} \frac{f(\epsilon_{\bm k}-\mu_i+U_i\bar{n})-f(-\epsilon_{\bm k}-\mu_i+U_i\bar{n})}{2\epsilon_{\bm k}}=1,
\label{U_f(0)}
\end{align}
which is equivalent to the static mean-field Eq.~(\ref{static mean field})
if we identify $\mu_i-U_i\bar{n}$ in Eq.~(\ref{U_f(0)}) with $\mu-U_c^{\rm th}\bar{n}$ in Eq.~(\ref{static mean field})
and $U_c^{\rm nth}$ in Eq.~(\ref{U_f(0)}) with $U_c^{\rm th}$ in Eq.~(\ref{static mean field}). This identification is allowed 
if the particle number is conserved (the Hartree approximation is conserving).
Hence, the relation $U_c^{\rm nth}=U_c^{\rm th}\equiv U_c$ holds exactly for arbitrary filling within the Hartree approximation.

To see the $a_{\rm nth}$ dependence of $U_f$, we take the derivative of Eq.~(\ref{condition for a}) with respect to $a_{\rm nth}$,
\begin{align}
\frac{dU_f^{-1}}{da_{\rm nth}}=\sum_{\bm k}\frac{2\epsilon_{\bm k}}{[(2\epsilon_{\bm k})^2+a_{\rm nth}]^2}f_0(\bm k).
\end{align}
This leads to 
\begin{align}
\frac{dU_f}{d\sqrt{a_{\rm nth}}}
&=
-2U_f^2\sqrt{a_{\rm nth}}\frac{dU_f^{-1}}{da_{\rm nth}}
\nonumber
\\
&=
-2U_f^2\int d\epsilon\,D(\epsilon)\frac{(2\epsilon)^2\sqrt{a_{\rm nth}}}{[(2\epsilon)^2+a_{\rm nth}]^2}
\nonumber
\\
&\quad
\times \frac{f(\epsilon-\mu_i+U_i\bar{n})-f(-\epsilon-\mu_i+U_i\bar{n})}{2\epsilon}.
\label{dU/dsqrt{a}}
\end{align}
Let us consider the quantity
\begin{align}
\frac{(2\epsilon)^2\sqrt{a_{\rm nth}}}{[(2\epsilon)^2+a_{\rm nth}]^2}.
\end{align}
The integral of this quantity does not depend on $a_{\rm nth}$,
\begin{align}
\int_{-\infty}^{\infty} d\epsilon\, \frac{(2\epsilon)^2\sqrt{a_{\rm nth}}}{[(2\epsilon)^2+a_{\rm nth}]^2}
&=
\int_{-\infty}^{\infty} d\epsilon\, \frac{(2\epsilon)^2}{[(2\epsilon)^2+1]^2}
=
\frac{\pi}{4},
\end{align}
and
\begin{align}
\lim_{a_{\rm nth}\to 0} \frac{(2\epsilon)^2\sqrt{a_{\rm nth}}}{[(2\epsilon)^2+a_{\rm nth}]^2}=0
\quad
(\epsilon\neq 0),
\end{align}
which means
\begin{align}
\lim_{a_{\rm nth}\to 0}\frac{(2\epsilon)^2\sqrt{a_{\rm nth}}}{[(2\epsilon)^2+a_{\rm nth}]^2}
&=
\frac{\pi}{4}\delta(\epsilon).
\end{align}
By taking the limit $a_{\rm nth}\to 0$ in Eq.~(\ref{dU/dsqrt{a}}), we have
\begin{align}
\lim_{a_{\rm nth}\to 0}\frac{dU_f}{d\sqrt{a_{\rm nth}}}
&=
-2U_c^2\int d\epsilon\, D(\epsilon)\frac{\pi}{4}\delta(\epsilon) 
\nonumber
\\
&\quad\times
\frac{f(\epsilon-\mu_i+U_i\bar{n})-f(-\epsilon-\mu_i+U_i\bar{n})}{2\epsilon}
\nonumber
\\
&=
-\frac{\pi}{2}U_c^2 D(0)f'(\mu_i-U_i\bar{n})\equiv c,
\end{align}
which is finite as long as $f'(\mu_i-U_i\bar{n})$ is finite. As a result, one obtains the expansion (\ref{a expansion}).
[Zero temperature is an exception, since $f'(\mu_i-U_i\bar{n})$ diverges or vanishes. At half filling there is a logarithmic correction
$U_f(a_{\rm nth})\sim -c'(\ln a_{\rm nth})^{-1}$, while away from half filling it has a linear dependence 
$U_f(a_{\rm nth})\sim U_c+c'' a_{\rm nth}$ around $a_{\rm nth}=0$.]

The result (\ref{a expansion}) implies
\begin{align}
a_{\rm nth}\propto (U_f-U_c)^2 \quad (U_f\ge U_c),
\label{a scaling}
\end{align}
which strikingly contrasts with the behavior of the conventional GL free energy $F_{\rm th}(m)$, having $a_{\rm th}\propto (U_f-U_c)^1$.
The scaling (\ref{a scaling}) is natural from the point of view of the power counting, since $a_{\rm nth}$ has the dimension of (energy)${}^2$.
Putting $a_{\rm nth}=a_0(U_f-U_c)^2$ ($a_0$ is a dimensionless constant), the nonthermal potential becomes
\begin{align}
F_{\rm nth}(m)=-\frac{1}{2}a_0(U_f-U_c)^2m^2+\frac{U_f^2}{8}m^4.
\end{align}
The scaling law (\ref{a scaling}) is ``universal'', i.e., the exponent does not depend on details of the problem 
($\beta$, $\mu_i$, $U_i$, $U_f$, and other parameters).
It defines a new universality class that characterizes the nonequilibrium dynamical symmetry breaking. For example, the maximum of the magnetization curve, $m_{\rm max}$, or
the middle point $m_{\rm nth}=(m_{\rm max}+m_{\rm min})/2=m_{\rm max}/2$, scales as
\begin{align}
m_{\rm nth}\propto m_{\rm max}
\propto (U_f-U_c)^\beta
\label{new scaling}
\end{align}
with
\begin{align}
\beta=1.
\label{new beta}
\end{align}
By comparing this with the thermal scaling (\ref{beta exponent}), we notice that 
$m_{\rm th}$ becomes much bigger than $m_{\rm nth}$ when $U_c^{\rm nth}=U_c^{\rm th}$
(which is the case in the Hartree approximation).
That is, in the vicinity of the critical point the magnitudes of the thermal and nonthermal order parameters are very different.
This leads us to the following scenario. When one goes beyond the Hartree approximation by including correlation effects,
$m$ approaches $m_{\rm th}$ in the long-time limit. However, if there exists a ``nonthermal critical point,'' which may govern the transient order-parameter dynamics,
$m$ is trapped for some duration around $m_{\rm nth}$, which can deviate strongly from the final value ($m_{\rm th}$).

\begin{figure}[tbp]
\begin{center}
\includegraphics[width=8cm]{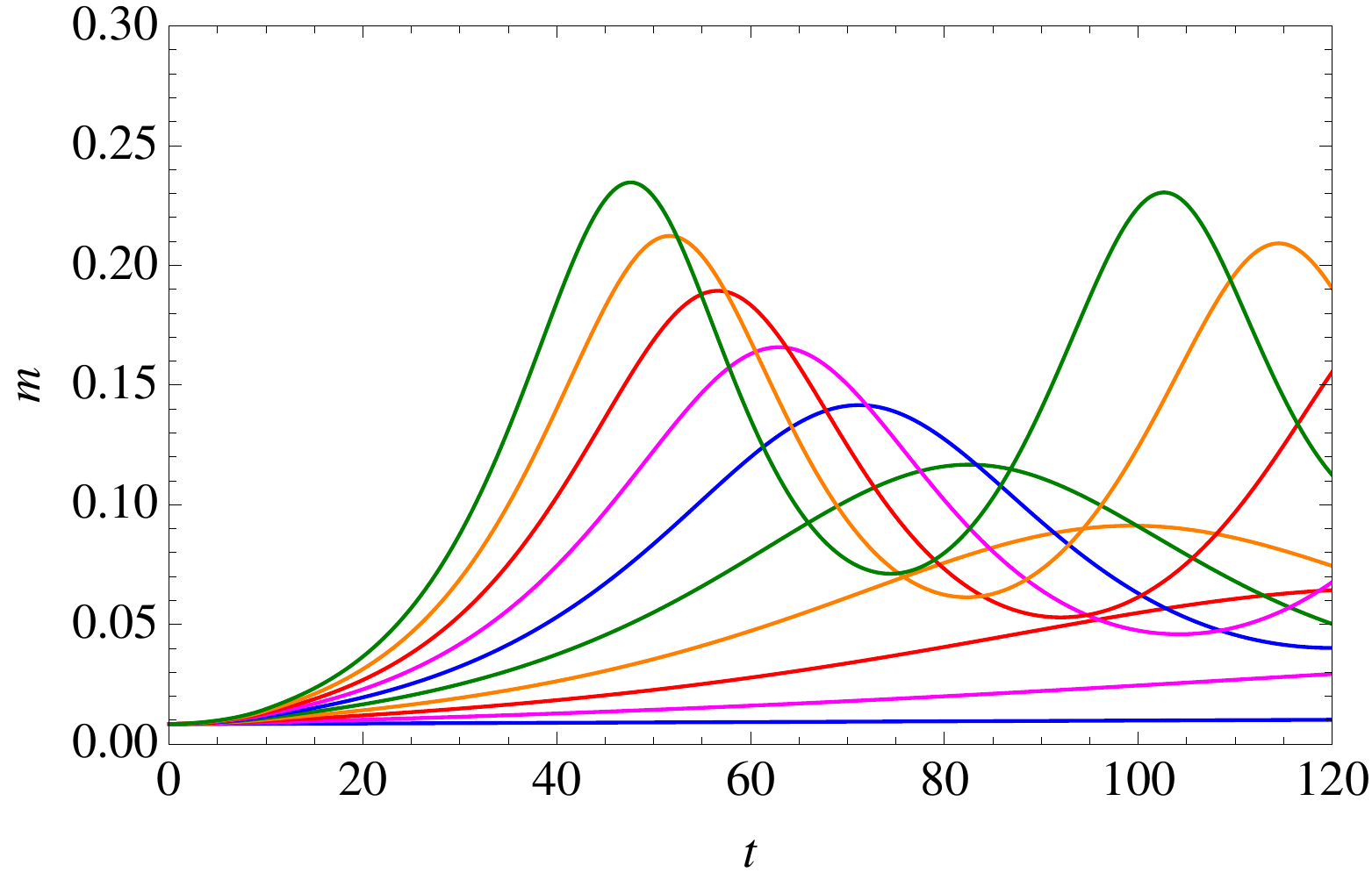}
\caption{(Color online) Time evolution of $m$ for interaction ramps $U=1.0\to 1.05, 1.1, \dots, 1.5$ in the Hubbard model at half filling with $\beta=40$, $h=10^{-4}$, and $t_{\rm ramp}=10$.}
\label{magnetization U=1}
\end{center}
\end{figure}
\begin{figure}[tbp]
\begin{center}
\includegraphics[width=8cm]{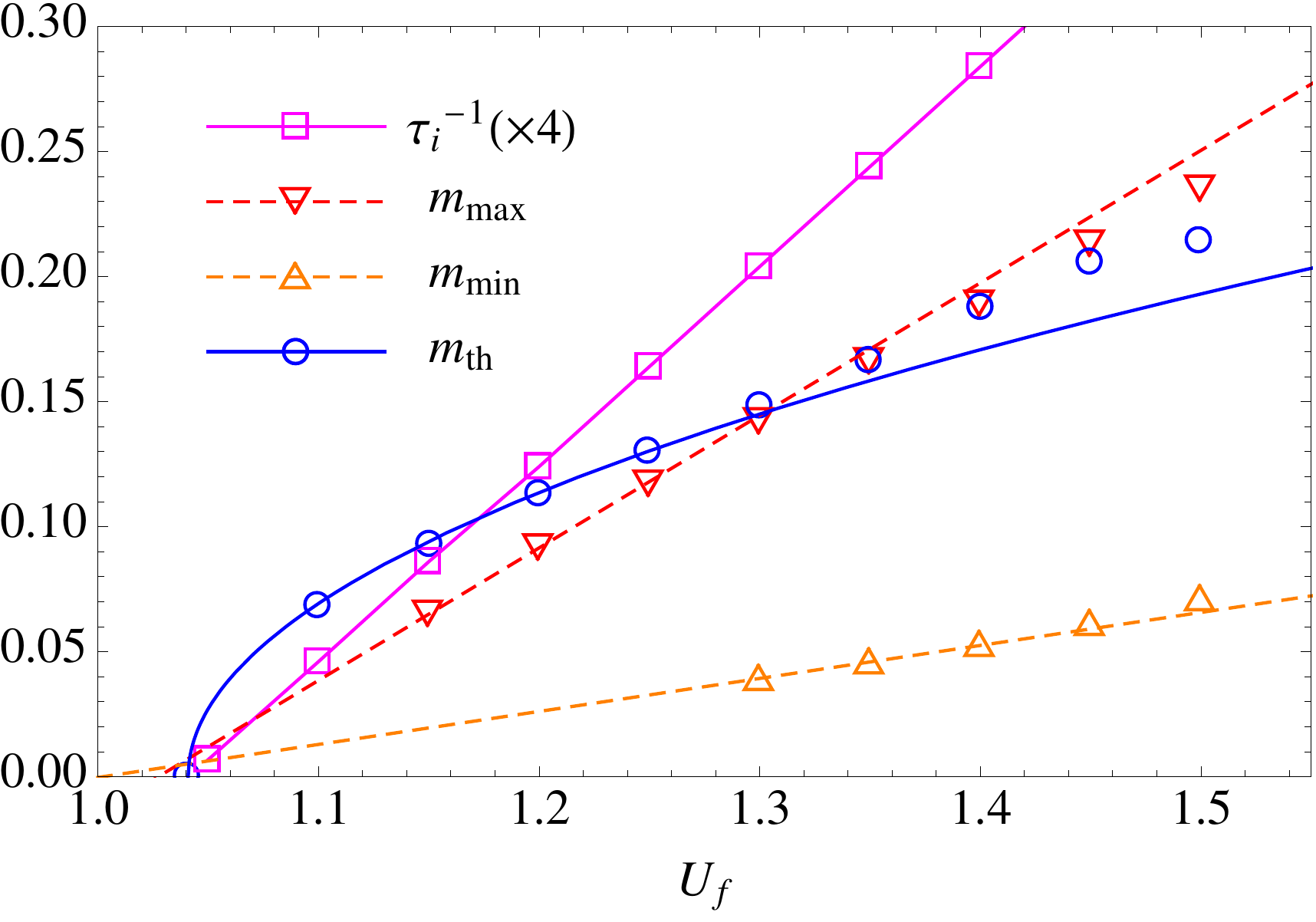}
\caption{(Color online) The initial exponential growth rate $\tau_i$ and the maximum (minimum) value of $m$ at the first peak (dip) of the oscillations
for the interaction ramps in Fig.~\ref{magnetization U=1}.}
\label{critical U=1}
\end{center}
\end{figure}
To confirm the validity of this scenario, we solve a nearly integrable system, i.e., the case with smaller $U_i$ and $U_f$ than in Fig.~\ref{symmetry breaking} or \ref{magnetization U=1.25}, using the nonequilibrium DMFT
with the third-order perturbation of type (I) (beyond the Hartree approximation). In Fig.~\ref{magnetization U=1}, we plot the time evolution of $m$ for interaction ramps $U_i=1\to 1.05, 1.1, \dots, 1.5$.
As we decrease $U_i$ and $U_f$, the magnetization curves approach the form of the Hartree solution (Fig.~\ref{soliton}). Due to the presence of the higher-order diagrams,
the soliton wave becomes incoherent, and the oscillation center slowly drifts upwards. We measured the time constant of the initial exponential growth, $\tau_i$, and the maximum and minimum magnetization,
$m_{\rm max}$ and $m_{\rm min}$, at the first peak and dip of the oscillation, which are plotted in Fig.~\ref{critical U=1}. The dashed lines in Fig.~\ref{critical U=1} are linear extrapolations of $m_{\rm max}$ and $m_{\rm min}$.
One can see that both $m_{\rm max}$ and $m_{\rm min}$ linearly approach the critical point, validating the nonthermal scaling (\ref{new scaling}) and (\ref{new beta}) beyond the Hartree approximation.
Thus, the transient dynamics of the order parameter is governed by the nonthermal critical point.
The extrapolated $m_{\rm max}$ and $m_{\rm min}$ (or the middle point $m_{\rm nth}$) reach zero at $U_f=U_\ast^{\rm nth}$,
which is quite close to the thermal critical point $U_f=U_c^{\rm th}$ where $\tau_i$ diverges.  
This means that even in calculations which treat correlation effects beyond the Hartree approximation, $U_\ast^{\rm nth}$ stays in the close vicinity of $U_c^{\rm th}$, at least in the weak-coupling regime.  

\section{Summary}

We have studied the reliability of impurity solvers based on different variants of the weak-coupling perturbation theory in the context of nonequilibrium DMFT,
focusing on interaction quenches and ramps in a Hubbard model with semicircular DOS of bandwidth 4.
For the PM phase of the Hubbard model at half filling, we have tested the perturbative solvers up to fourth order, and
showed that the non-self-consistent fourth-order calculation (bare-diagram expansion of the self-energy) can systematically improve the result of the non-self-consistent
second-order expansion (IPT) in the weak-coupling regime ($U<2$), 
but suddenly fails to conserve the total energy after an interaction quench from $U=0$ to $U>2$.
The fourth-order scheme thus fails at approximately the same interaction strength as IPT, which means that going to higher orders in perturbation theory is not a viable strategy
for accessing the challenging intermediate coupling regime. While the self-consistent perturbations are conserving, they give a much stronger damping to the thermal state,
a behavior which is inconsistent with the result of unbiased QMC calculations.

For the PM phase away from half filling and the AFM phase of the Hubbard model, 
the perturbation theory has been implemented up to third order. We discussed important details of the weak-coupling expansions,
including the choice of bare and bold diagrams, symmetrization of the interaction term, and the treatment of the Hartree diagrams.
We considered five types of expansions, as classified in Table~\ref{n-alpha}. Types (III)-(V) lead to a convergence problem in the DMFT self-consistency
at moderate and strong interactions, while the type (I) and type (II) schemes always show a nice convergence. Far away from half filling ($|n_\sigma-0.5|\ge 0.25$),
the third-order expansions of types (I) and (IV) give accurate results, whereas the intermediate filling regime ($0.1<|n_\sigma-0.5|<0.25$) remains hard
to treat by the perturbation expansions. As one approaches the half-filling regime further, the accuracy improves again.
For nonequilibrium dynamics away from half filling, the applicability of the weak-coupling perturbation solvers  remains quite limited ($U\le 1$).

The weak-coupling perturbation theories however turn out to be very useful for the study of the AFM phase near half filling.
The third-order expansion of type (I) is the most accurate in the weak $U$ regime ($U<3$), giving reliable results for $T_c$ and the staggered magnetization.
The second-order perturbation of type (I) reproduces the qualitatively correct weak-to-strong-coupling crossover for $T_c$, but it fails to correctly describe the magnetization curve and spectral function in the strong-coupling regime.

We used the nonequilibrium DMFT with the type (I) third-order perturbation theory in an investigation of  
the dynamical AFM symmetry breaking in the Hubbard model induced by an interaction ramp.
The results show that the order parameter starts to grow exponentially, followed by an amplitude oscillation around
some nonthermal value, which may be viewed as a ``trapping'' phenomenon related to the existence of a nonthermal critical point.
This amplitude mode is superimposed on top of a slow drift toward the thermal value, and it manifests itself as an oscillating spectral weight transfer 
in the spectral function. 

The phenomenological GL theory cannot consistently explain the dynamics
of the order parameter, no matter how the fitting parameter (friction constant) is chosen. We compared the results
with the time-dependent Hartree approximation, which allows for analytical soliton solutions. Writing the equation of motion for the order parameter
in the form of a GL-like equation, we showed that the nonthermal potential $F_{\rm nth}$ (corresponding to this Hartree solution)
belongs to a characteristic universality class (with critical exponent $\beta=1$), which is different from the GL universality ($\beta=\frac{1}{2}$).
Due to this fact, the order-parameter dynamics shows a ``universality transition'' in the critical regime,
that is, the order parameter crosses over from the nonthermal value $\propto (U_f-U_c)^1$,
around which the order parameter is transiently trapped, to the thermal value $\propto (U_f-U_c)^{\frac{1}{2}}$,
which is reached in the long-time limit. This qualitative behavior survives to some extent even when $U$ is increased into a regime in which 
the Hartree approximation breaks down. 

There are a couple of issues that are left for future investigations. One is the development of a nonequilibrium impurity solver for calculations away from half filling 
in the weak-coupling regime, which safely preserves the conservation laws. The bare third-order expansions of types (I) and (IV) seem to be valid in the overdoped regime ($|n_\sigma-0.5|>0.25$),
while treating the underdoped regime ($0.1<|n_\sigma-0.5|<0.25$) remains challenging. The weak-coupling techniques that have been studied in this paper
can be applied not only to the AFM phase but also to other symmetry-broken ordered phases such as superconductivity (SC)
and CDWs. Indeed, our results of the order-parameter dynamics at half filling can be translated to those for SC and CDW since
there is a symmetry between the repulsive and attractive Hubbard models at half filling, via the Shiba transformation
(spin-dependent particle-hole exchange).\cite{Shiba1972}
Away from half filling, without this symmetry, the dynamics of SC and CDW phases remains to be explored.
The interaction quench physics in the SC phase with dilute filling may have direct relevance to cold-atom experiments.
A switching from the SC to the CDW phase (or vice versa) would be another interesting topic 
in connection with related pump-probe experiments.\cite{Fausti2011}
The character of the nonthermal critical point that may govern the transient order-parameter dynamics
has not been fully revealed. While in this paper we focused on the interaction ramps, an interesting question
is if the same universal behavior is observed when we consider other types of perturbations (e.g., a system driven by electric fields).
The characterization of the nonthermal critical point (if exists) in the intermediate and strong-coupling regime
is also desired and is left as a future problem. 

\acknowledgements

We thank M. Eckstein, A. Georges, and S. Biermann for fruitful discussions.
We also thank M. Eckstein and A. Koga for providing some of the QMC data which were used to test the perturbative solvers.
The calculations were carried out on the UniFr cluster.
N.T. was supported by a 
Grant-in-Aid for Scientific Research on Innovative Areas ``Materials Design through Computics: Complex Correlation
and Non-equilibrium Dynamics'' (Grant No.~25104709),
and a Grant-in-Aid for Young Scientists (B) (Grant No.~25800192).
P.W. acknowledges support from FP7/ERC starting grant No. 278023.

\appendix

\section{Numerical implementation of the nonequilibrium Dyson equation}
\label{appendix:Dyson}

In this appendix, we describe our numerical implementation of the solution of the nonequilibrium Dyson equation on the Kadanoff-Baym contour $\mathcal{C}$ (Fig.~\ref{contour}).
The Dyson equation that one encounters in the nonequilibrium DMFT calculation is an integral-differential equation of the form (\ref{intdiff eq})
\begin{align}
[i\partial_t-\epsilon(t)] G(t,t')-\int_{\mathcal C} d\bar{t}\, \Sigma(t,\bar{t}) G(\bar{t},t')
&=
\delta_{\mathcal C}(t,t'),
\label{appendix:intdiff}
\end{align}
or an integral equation of the form (\ref{integral eq})
\begin{align}
G(t,t')-\int_{\mathcal C} d\bar{t}\, K(t,\bar{t}) G(\bar{t},t')
&=
G_0(t,t').
\label{appendix:int}
\end{align}
Here $K$, and 
$G_0$ 
are given contour-ordered functions, and we have to solve these equations for $G$.
Let us assume that $\Sigma$ and $K$ do not contain $\delta$ functions. If there is a $\delta$ function inside $\Sigma$
such as a Hartree term, one can include it in $\epsilon(t)$.
Using the Langreth rule,\cite{Langreth1976} one can decompose Eq.~(\ref{appendix:intdiff}) into the components of the contour functions
to rewrite
\begin{align}
[i\partial_t-\epsilon(t)] G^R(t,t')-\int_{t'}^t d\bar{t}\, \Sigma^R(t,\bar{t})G^R(\bar{t},t')
&=
\delta(t,t'),
\label{appendix:Dyson retarded}
\\
[i\partial_t-\epsilon(t)] G^{\lh}(t,\tau ')-\int_0^t d\bar{t}\, \Sigma^R(t,\bar{t})G^{\lh}(\bar{t},\tau ')
&=
Q^{\lh}(t,t'),
\\
[i\partial_t-\epsilon(t)] G^<(t,t')-\int_0^t d\bar{t}\, \Sigma^R(t,\bar{t})G^<(\bar{t},t')
&=
Q^<(t,t'),
\end{align}
where $G^M$, $G^R$, $G^{\lh}$, and $G^<$ are the Matsubara, retarded, left-mixing, and lesser components of $G$
(the same notation is applied to other contour functions), respectively, and we have defined
\begin{align}
Q^{\lh}(t,t')
&=
\int_0^\beta d\bar{\tau}\, \Sigma^{\lh}(t,\bar{\tau})G^M(\bar{\tau},\tau '),
\\
Q^<(t,t')
&=
\int_0^{t'} d\bar{t}\, \Sigma^<(t,\bar{t})G^A(\bar{t},t')
-i\int_0^\beta d\bar{\tau}\, K^{\lh}(t,\bar{\tau})G^{\rh}(\bar{t},t').
\end{align}
We can also decompose Eq.~(\ref{appendix:int}) into
\begin{align}
G^R(t,t')-\int_{t'}^t d\bar{t}\, K^R(t,\bar{t})G^R(\bar{t},t')
&=
G_0^R(t,t'),
\\
G^{\lh}(t,\tau ')-\int_0^t d\bar{t}\, K^R(t,\bar{t})G^{\lh}(\bar{t},\tau ')
&=
R^{\lh}(t,\tau '),
\\
G^<(t,t')-\int_0^t d\bar{t}\, K^R(t,\bar{t})G^<(\bar{t},t')
&=
R^<(t,t'),
\end{align}
where we have defined
\begin{align}
R^{\lh}(t,\tau ')
&=
G_0^{\lh}(t,\tau ')+\int_0^\beta d\bar{\tau}\, K^{\lh}(t,\bar{\tau})G^M(\bar{\tau},\tau '),
\\
R^<(t,t')
&=
G_0^<(t,t')+\int_0^{t'}d\bar{t}\, K^<(t,\bar{t})G^A(\bar{t},t')
\nonumber
\\
&\quad
-i\int_0^\beta d\bar{\tau}\, K^{\lh}(t,\bar{\tau})G^{\rh}(\bar{\tau},t').
\end{align}
By fixing the second time argument of the contour functions ($t'$ or $\tau '$),
one finds that these equations can be expressed as 
\begin{align}
&i\frac{d}{dt} g(t)-\epsilon(t)g(t)-\int_0^t d\bar{t}\, \Sigma^R(t,\bar{t})g(\bar{t})=q(t),
\label{Volterra intdiff}
\\
&g(t)-\int_0^t d\bar{t}\, K^R(t,\bar{t})g(\bar{t})=r(t),
\label{Volterra int}
\end{align}
where $g(t)=G(t,\ast)$, $q(t)=Q(t,\ast)$, and $r(t)=R(t,\ast)$.
Both of them are categorized as Volterra equations of the second kind,
for which various numerical algorithms exist in the literature.\cite{NumericalRecipesC,LinzBook,BrunnervanderHouwenBook}
Here we present one of them, namely the implicit Runge-Kutta method (or the collocation method), which is employed throughout the paper. 
For the implementation, we discretize the time with equal spacing, $t_i=i\times\Delta t$ ($i=0,1,\dots, N$), with $\Delta t=t_{\rm max}/N$.
It is crucial to employ higher-order schemes to accurately simulate the long-time evolution. The $n$th-order scheme has 
systematic errors of $O(N(\Delta t)^{n+1})=O(t_{\rm max}(\Delta t)^n)$. Typically we require $n\ge 2$ to control the error. In the following,
we present the second-order and fourth-order schemes.

\subsection{Second-order scheme}

We first treat the Volterra integral-differential equation (\ref{Volterra intdiff}). We set the initial condition, $g(t_0)=g(0)$,
from which we can solve it step by step on the discretized time grid. That is, once $g(t_i)$ ($i=0,1,\dots, n-1$)
and $g'(t_{n-1})$ have been obtained, we use these values to determine $g(t_n)$ and $g'(t_n)$ on the next step.
To get $g(t_n)$, we express the difference between $g(t_n)$ and $g(t_{n-1})$ 
by an integral, which is evaluated by the trapezoid integral formula
\begin{align}
g(t_n)-g(t_{n-1})
&=
\int_{t_{n-1}}^{t_n} d\bar{t}\, g'(\bar{t})
\approx
\frac{\Delta t}{2}[g'(t_{n-1})+g'(t_n)].
\label{appendix:2nd1}
\end{align}
Here $g'(t_{n-1})$ is already know from the previous calculation, and $g'(t_n)$ is evaluated from Eq.~(\ref{Volterra intdiff}),
\begin{align}
ig'(t_n)
&=
q(t_n)+\epsilon(t_n)g(t_n)+\int_0^{t_n} d\bar{t}\, \Sigma^R(t_n,\bar{t})g(\bar{t})
\nonumber
\\
&\approx
q(t_n)+\epsilon(t_n)g(t_n)+\Delta t\sum_{i=0}^n w_{n,i}\Sigma^R(t_n,t_i)g(t_i),
\label{appendix:2nd2}
\end{align}
where we again approximated the integral by the trapezoid rule with weights
\begin{align}
w_{n,i}
&=
\begin{cases}
1/2 & i=0, n \\
1 & 1\le n \le n-1
\end{cases}.
\end{align}
Equations (\ref{appendix:2nd1}) and (\ref{appendix:2nd2}) provide a set of linear equations for $g(t_n)$,
which we can solve to obtain
\begin{align}
g(t_n)
&=
\left[1+i\frac{\Delta t}{2}\epsilon(t_n)+i\frac{(\Delta t)^2}{2}w_{n,n}\Sigma^R(t_n,t_n)\right]^{-1}
\nonumber
\\
&\quad\times\bigg\{
g(t_{n-1})+\frac{\Delta t}{2}\bigg[g'(t_{n-1})+q(t_n)
\nonumber
\\
&\quad
-i\Delta t\sum_{i=0}^{n-1}w_{n,i}\Sigma^R(t_n,t_i)g(t_i)\bigg]
\bigg\}.
\label{appendix:2nd3}
\end{align}
$g'(t_n)$ is derived from Eq.~(\ref{appendix:2nd2}) with $g(t_n)$ substituted by the result (\ref{appendix:2nd3}).
The numerical errors are of $O(t_{\rm max}(\Delta t)^2)$. To avoid repeated calculations of the sums in Eqs.~(\ref{appendix:2nd2})
and (\ref{appendix:2nd3}), one should store them in memory.

The same technique can be applied to the Volterra integral equation (\ref{Volterra int}). We evaluate the integral in Eq.~(\ref{Volterra int})
by the trapezoid rule and find 
\begin{align}
g(t_n)
&\approx
r(t_n)+\Delta t\sum_{i=0}^n w_{n,i}K^R(t_n,t_i)g(t_i)
\nonumber
\\
&=
\frac{\displaystyle r(t_n)+\Delta t\sum_{i=0}^{n-1} w_{n,i}K^R(t_n,t_i)g(t_i)}{1-\Delta t\, w_{n,n} K^R(t_n,t_n)}.
\label{appendix:2nd4}
\end{align}
It often holds that $K(t_n,t_n)=0$, in which case the denominator in Eq.~(\ref{appendix:2nd4}) becomes unity.

\subsection{Fourth-order scheme}

One can derive the fourth-order scheme by replacing the numerical integral formula (trapezoid rule) used in the second-order approximation
with higher-order formulas. We need at least three points to evaluate integrals with higher order algorithms. 
Let us consider the case of $n\ge 2$ first. In this case, we evaluate the difference between $g(t_n)$ and $g(t_{0})$,
\begin{align}
g(t_n)-g(t_{0})
&=
\int_{t_{0}}^{t_n} d\bar{t}\, g'(\bar{t})
\approx
\Delta t\sum_{i=0}^n w_{n,i} g'(t_i),
\label{g(t_n)-g(t_0)}
\end{align}
where one can use Simpson's rule for $n=2$,
\begin{align}
w_{2,i}
&=
\begin{cases}
1/3 & i=0, 2,\\
4/3 & i=1,
\end{cases}
\label{Simpson}
\end{align}
Simpson's $3/8$ rule for $n=3$,
\begin{align}
w_{3,i}
&=
\begin{cases}
3/8 & i=0, 3,\\
9/8 & i=1, 2,
\end{cases}
\end{align}
the composite Simpson's rule for $n=4$,
\begin{align}
w_{4,i}
&=
\begin{cases}
1/3 & i=0, 4,\\
4/3 & i=1, 3,\\
2/3 & i=2,
\end{cases}
\end{align}
and the fourth-order Gregory's rule for $n\ge 5$,
\begin{align}
w_{n,i}
&=
\begin{cases}
3/8 & i=0, n, \\
7/6 & i=1, n-1,\\
23/24 & i=2, n-2,\\
1 & 3\le i \le n-3.
\end{cases}
\label{Gregory}
\end{align}
$g'(t_n)$ in Eq.~(\ref{g(t_n)-g(t_0)}) is calculated by Eq.~(\ref{appendix:2nd2}) with an appropriate higher-order integral formula. 
In the same way as in the second-order scheme, one arrives at a set of linear equations
for $g(t_n)$, which is solved as
\begin{align}
g(t_n)
&=
\left[1+i\Delta t\, w_{n,n}\epsilon(t_n)+i(\Delta t\, w_{n,n})^2 \Sigma^R(t_n,t_n)\right]^{-1}
\nonumber
\\
&\quad\times\bigg\{
g(t_0)+\Delta t \sum_{i=0}^{n-1} w_{n,i} g'(t_i)
\nonumber
\\
&\quad
-i\Delta t\, w_{n,n}\bigg[q(t_n)+\Delta t \sum_{i=0}^{n-1} w_{n,i} \Sigma^R(t_n,t_i)g(t_i)\bigg]
\bigg\}.
\end{align}
Here one needs $g'(t_i)$ ($0\le i\le n-1$), which has to be stored. To avoid that, we employ the relation
\begin{align}
g(t_{n-1})-g(t_0)\approx \Delta t\sum_{i=0}^{n-1} w_{n-1,i}g'(t_i).
\end{align}
With this, we obtain the final expression for the fourth-order scheme,
\begin{align}
g(t_n)
&=
\left[1+i\Delta t\, w_{n,n}\epsilon(t_n)+i(\Delta t\, w_{n,n})^2 \Sigma^R(t_n,t_n)\right]^{-1}
\nonumber
\\
&\quad\times\bigg\{
g(t_{n-1})+\Delta t \sum_{i=0}^{n-1} (w_{n,i}-w_{n-1,i}) g'(t_i)
\nonumber
\\
&\quad
-i\Delta t\, w_{n,n}\bigg[q(t_n)+\Delta t \sum_{i=0}^{n-1} w_{n,i} \Sigma^R(t_n,t_i)g(t_i)\bigg]
\bigg\}.
\end{align}
Note that $w_{n,i}-w_{n-1,i}=0$ for $0\le i\le n-4$ if we use the fourth-order Gregory's rule (\ref{Gregory}) for sufficiently large $n$.
Hence what one has to store are $g'(t_{n-1})$, $g'(t_{n-2})$ and $g'(t_{n-3})$.
The numerical errors are suppressed to $O(N(\Delta t)^5)=O(t_{\rm max}(\Delta t)^4)$. For the Volterra integral equation (\ref{Volterra int}),
one can use the same expression as in the second-order scheme (\ref{appendix:2nd4}),
in combination with the higher-order integral formulas.

The remaining task is to get the starting value $g(t_1)$. Since the higher-order integral formulas need at least three points on the grid,
the above approach cannot be directly applied for $n=1$.
One possible solution is to take a very fine grid on the interval $t_0\le t\le t_1$,
and use a lower-order integral formula (e.g., trapezoid rule). This approach is simple and straightforward, but
the complication is that one has to change the time steps.
Another way, which we adopted in the paper, is to take the middle point,\cite{LinzBook}
$t_{1/2}=\Delta t/2$, and apply Simpson's rule to the integral from $t_0$ to $t_1$,
\begin{align}
g(t_1)-g(t_0)
&\approx
\frac{\Delta t}{6}[g'(t_0)+4g'(t_{1/2})+g'(t_1)].
\label{appendix:4th1}
\end{align}
The value at the middle point is obtained from a quadratic interpolation,
\begin{align}
g'(t_{1/2})
\approx
\frac{3}{8}g'(t_0)+\frac{3}{4}g'(t_1)-\frac{1}{8}g'(t_2),
\label{appendix:4th2}
\end{align}
which has an error of $O((\Delta t)^3)$ for the smooth function $g(t)$. Since $g'(t_{1/2})$ is multiplied with $\Delta t$ in Eq.~(\ref{appendix:4th1}),
the overall error is of $O((\Delta t)^4)$, which is compatible with the fourth-order scheme.
$g'(t_0)$ is known from the initial condition.
$g'(t_2)$ is evaluated from Eq.~(\ref{appendix:2nd2}) as in the second-order scheme.
$g'(t_1)$ is also evaluated from Eq.~(\ref{appendix:2nd2}) with Simpson's rule taking the middle point,
\begin{align}
&ig'(t_1)
\approx
q(t_1)+\epsilon(t_1)G(t_1)
\nonumber
\\
&
+\frac{\Delta t}{6}[\Sigma^R(t_1,t_0)g(t_0)+4\Sigma^R(t_1,t_{1/2})g(t_{1/2})
+\Sigma^R(t_1,t_1)g(t_1)].
\label{appendix:4th3}
\end{align}
One again uses the quadratic interpolation to get the middle-point values,
\begin{align}
g(t_{1/2})
&\approx
\frac{3}{8}g(t_0)+\frac{3}{4}g(t_1)-\frac{1}{8}g(t_2),
\label{appendix:4th4}
\\
\Sigma^R(t_1,t_{1/2})
&\approx
\frac{3}{8}\Sigma^R(t_1,t_0)+\frac{3}{4}\Sigma^R(t_1,t_1)+\frac{1}{8}\Sigma^A(t_1,t_2).
\label{appendix:4th5}
\end{align}
In Eq.~(\ref{appendix:4th5}), the last term on the right-hand side would be $-\frac{1}{8}\Sigma^R(t_1,t_2)$
according to the standard interpolation. However, the retarded self-energy $\Sigma^R(t,t')$ is discontinuous 
at $t=t'$ due to causality, so that the interpolation cannot be applied in this form. Instead of $\Sigma^R(t,t')$, 
we can take $\Sigma^R(t,t')-\Sigma^A(t,t')=\Sigma^>(t,t')-\Sigma^<(t,t')$, which is smooth at $t=t'$, and is identical to $\Sigma^R(t,t')$ for $t\ge t'$.
Thus, the last term in Eq.~(\ref{appendix:4th5}) should be $+\frac{1}{8}\Sigma^A(t_1,t_2)$.
From Eq.~(\ref{appendix:4th4}), it turns out that to get $g(t_1)$ we need $g(t_2)$. On the other hand, $g(t_2)$ can be determined
from $g(t_0)$ and $g(t_1)$. In total, we can couple these equations for $g(t_1)$ and $g(t_2)$ to determine both simultaneously. 

In this way, one can calculate the starting values for every component of $G(t,t')$, except for $G^R(t_2,t_1)$.
If we follow the above strategy, $G^R(t_2,t_1)$ is determined in combination with $G^R(t_3,t_1)$, which is, however,
not available at the first step ($n=2$). As we see below, there is a way to get around this, and $G^R(t_2,t_1)$
is determined from the knowledge of $G^R(t_1,t_0)$ and $G^R(t_2,t_0)$. 

We first rewrite the difference between $G^R(t_2,t_1)$ and $G^R(t_1,t_1)$
using the trapezoid rule and the middle point $t_{3/2}$, 
\begin{align}
&G^R(t_2,t_1)-G^R(t_1,t_1)
\nonumber
\\
&\approx
\frac{\Delta t}{6}[{G^R}'(t_1,t_1)+4{G^R}'(t_{3/2},t_1)+{G^R}'(t_2,t_1)].
\end{align}
Note that the prime indicates the derivative with respect to the first time argument. ${G^R}'(t_1,t_1)$ is given by the initial condition.
${G^R}'(t_{3/2},t_1)$ is given by the quadratic interpolation,
\begin{align}
{G^R}'(t_{3/2},t_1)
\approx
\frac{3}{8}{G^R}'(t_2,t_1)+\frac{3}{4}{G^R}'(t_1,t_1)+\frac{1}{8}{G^A}'(t_0,t_1),
\label{appendix:4th6}
\end{align}
where we have used the fact that ${G^R}'(t,t')-{G^A}'(t,t')={G^>}'(t,t')-{G^<}'(t,t')$ is a smooth function.
${G^R}'(t_2,t_1)$ is determined by Eq.~(\ref{appendix:Dyson retarded}); i.e.,
\begin{align}
i{G^R}'(t_2,t_1)
&\approx
\epsilon(t_2)G^R(t_2,t_1)+\frac{\Delta t}{6}[\Sigma^R(t_2,t_1)G^R(t_1,t_1)
\nonumber
\\
&\quad
+4\Sigma^R(t_2,t_{3/2})G^R(t_{3/2},t_1)+\Sigma^R(t_2,t_2)G^R(t_2,t_1)].
\end{align}
Using appropriate interpolations, we can substitute $\Sigma^R(t_2,t_{3/2})$ and $G^R(t_{3/2},t_1)$
with integer-point values. To evaluate ${G^A}'(t_0,t_1)$ in Eq.~(\ref{appendix:4th6}), we use the conjugate Dyson equation,
\begin{align}
[i\partial_{t}-\epsilon(t)]G^A(t,t')-\int_{t}^{t'} d\bar{t}\, \Sigma^A(t,\bar{t})G^A(\bar{t},t')=\delta(t,t').
\end{align}
[Note that ${G^A}'(t,t')\neq {G^R}'(t',t)^\ast$.]
With this, ${G^A}'(t_0,t_1)$ is given by
\begin{align}
i{G^A}'(t_0,t_1)
&\approx
\epsilon(t_0)G^A(t_0,t_1)
+\frac{\Delta t}{6}[\Sigma^A(t_0,t_0)G^A(t_0,t_1)
\nonumber
\\
&\quad
+4\Sigma^A(t_0,t_{1/2})G^A(t_{1/2},t_1)+\Sigma^A(t_0,t_1)G^A(t_1,t_1)].
\end{align}
Again, we employ quadratic interpolations to evaluate $\Sigma^A(t_0,t_{1/2})$ and $G^A(t_{1/2},t_1)$.
Thus, we have obtained a set of linear equations to determine $G^R(t_2,t_1)$ from known functions.

The same technique is applicable to the Volterra integral equation (\ref{Volterra int}) to get the starting values at fourth order.
In this case, one has to be particularly careful when one uses a quadratic interpolation like
\begin{align}
K^R(t_1,t_{1/2})
\approx
\frac{3}{8}K^R(t_1,t_0)+\frac{3}{4}K^R(t_1,t_1)+\frac{1}{8}K^A(t_1,t_2).
\end{align}
The contour function $K$ is usually defined as a convolution of two contour functions such as $K=G_0\ast \Sigma$,
which implies that $K^A(t,t')\neq K^R(t',t')^\ast$. Instead of taking the complex conjugate,
one should go back to the original definition of the convolution, 
\begin{align}
K^A(t,t')
&=
\int_t^{t'} d\bar{t}\, \Sigma^A(t,\bar{t})G_0^A(\bar{t},t').
\end{align}
From this, we have
\begin{align}
K^A(t_1,t_2)
&\approx
\frac{\Delta t}{6}[\Sigma^A(t_1,t_1)G_0^A(t_1,t_2)+4\Sigma^A(t_1,t_{3/2})G^A(t_{3/2},t_2)
\nonumber
\\
&\quad
+\Sigma^A(t_1,t_2)G^A(t_2,t_2)].
\end{align}

To summarize, we have overviewed the fourth-order implicit Runge-Kutta method for the Volterra integral(-differential) equations.
The higher-order scheme is quite important when one tackles the problem of calculating the long-time evolution within nonequilibrium DMFT,
keeping the errors down to $O(N(\Delta t)^5)=O(t_{\rm max}(\Delta t)^4)$.
The subtle issue of computing the starting values in the fourth-order scheme can be overcome with 
the middle-point approach and quadratic interpolations.

\bibliographystyle{apsrev}
\bibliography{perturbation}

\begin{thebibliography}{89}
\expandafter\ifx\csname natexlab\endcsname\relax\def\natexlab#1{#1}\fi
\expandafter\ifx\csname bibnamefont\endcsname\relax
  \def\bibnamefont#1{#1}\fi
\expandafter\ifx\csname bibfnamefont\endcsname\relax
  \def\bibfnamefont#1{#1}\fi
\expandafter\ifx\csname citenamefont\endcsname\relax
  \def\citenamefont#1{#1}\fi
\expandafter\ifx\csname url\endcsname\relax
  \def\url#1{\texttt{#1}}\fi
\expandafter\ifx\csname urlprefix\endcsname\relax\def\urlprefix{URL }\fi
\providecommand{\bibinfo}[2]{#2}
\providecommand{\eprint}[2][]{\url{#2}}

\bibitem[{\citenamefont{Ogasawara et~al.}(2000)\citenamefont{Ogasawara, Ashida,
  Motoyama, Eisaki, Uchida, Tokura, Ghosh, Shukla, Mazumdar, and
  Kuwata-Gonokami}}]{Ogasawara2000}
\bibinfo{author}{\bibfnamefont{T.}~\bibnamefont{Ogasawara}},
  \bibinfo{author}{\bibfnamefont{M.}~\bibnamefont{Ashida}},
  \bibinfo{author}{\bibfnamefont{N.}~\bibnamefont{Motoyama}},
  \bibinfo{author}{\bibfnamefont{H.}~\bibnamefont{Eisaki}},
  \bibinfo{author}{\bibfnamefont{S.}~\bibnamefont{Uchida}},
  \bibinfo{author}{\bibfnamefont{Y.}~\bibnamefont{Tokura}},
  \bibinfo{author}{\bibfnamefont{H.}~\bibnamefont{Ghosh}},
  \bibinfo{author}{\bibfnamefont{A.}~\bibnamefont{Shukla}},
  \bibinfo{author}{\bibfnamefont{S.}~\bibnamefont{Mazumdar}}, \bibnamefont{and}
  \bibinfo{author}{\bibfnamefont{M.}~\bibnamefont{Kuwata-Gonokami}},
  \bibinfo{journal}{Phys. Rev. Lett.} \textbf{\bibinfo{volume}{85}},
  \bibinfo{pages}{2204} (\bibinfo{year}{2000}).

\bibitem[{\citenamefont{Iwai et~al.}(2003)\citenamefont{Iwai, Ono, Maeda,
  Matsuzaki, Kishida, Okamoto, and Tokura}}]{Iwai2003}
\bibinfo{author}{\bibfnamefont{S.}~\bibnamefont{Iwai}},
  \bibinfo{author}{\bibfnamefont{M.}~\bibnamefont{Ono}},
  \bibinfo{author}{\bibfnamefont{A.}~\bibnamefont{Maeda}},
  \bibinfo{author}{\bibfnamefont{H.}~\bibnamefont{Matsuzaki}},
  \bibinfo{author}{\bibfnamefont{H.}~\bibnamefont{Kishida}},
  \bibinfo{author}{\bibfnamefont{H.}~\bibnamefont{Okamoto}}, \bibnamefont{and}
  \bibinfo{author}{\bibfnamefont{Y.}~\bibnamefont{Tokura}},
  \bibinfo{journal}{Phys. Rev. Lett.} \textbf{\bibinfo{volume}{91}},
  \bibinfo{pages}{057401} (\bibinfo{year}{2003}).

\bibitem[{\citenamefont{Perfetti et~al.}(2006)\citenamefont{Perfetti, Loukakos,
  Lisowski, Bovensiepen, Berger, Biermann, Cornaglia, Georges, and
  Wolf}}]{Perfetti2006}
\bibinfo{author}{\bibfnamefont{L.}~\bibnamefont{Perfetti}},
  \bibinfo{author}{\bibfnamefont{P.~A.} \bibnamefont{Loukakos}},
  \bibinfo{author}{\bibfnamefont{M.}~\bibnamefont{Lisowski}},
  \bibinfo{author}{\bibfnamefont{U.}~\bibnamefont{Bovensiepen}},
  \bibinfo{author}{\bibfnamefont{H.}~\bibnamefont{Berger}},
  \bibinfo{author}{\bibfnamefont{S.}~\bibnamefont{Biermann}},
  \bibinfo{author}{\bibfnamefont{P.~S.} \bibnamefont{Cornaglia}},
  \bibinfo{author}{\bibfnamefont{A.}~\bibnamefont{Georges}}, \bibnamefont{and}
  \bibinfo{author}{\bibfnamefont{M.}~\bibnamefont{Wolf}},
  \bibinfo{journal}{Phys. Rev. Lett.} \textbf{\bibinfo{volume}{97}},
  \bibinfo{pages}{067402} (\bibinfo{year}{2006}).

\bibitem[{\citenamefont{Okamoto et~al.}(2007)\citenamefont{Okamoto, Matsuzaki,
  Wakabayashi, Takahashi, and Hasegawa}}]{Okamoto2007}
\bibinfo{author}{\bibfnamefont{H.}~\bibnamefont{Okamoto}},
  \bibinfo{author}{\bibfnamefont{H.}~\bibnamefont{Matsuzaki}},
  \bibinfo{author}{\bibfnamefont{T.}~\bibnamefont{Wakabayashi}},
  \bibinfo{author}{\bibfnamefont{Y.}~\bibnamefont{Takahashi}},
  \bibnamefont{and} \bibinfo{author}{\bibfnamefont{T.}~\bibnamefont{Hasegawa}},
  \bibinfo{journal}{Phys. Rev. Lett.} \textbf{\bibinfo{volume}{98}},
  \bibinfo{pages}{037401} (\bibinfo{year}{2007}).

\bibitem[{\citenamefont{Wall et~al.}(2011)\citenamefont{Wall, Brida, Clark,
  Ehrke, Jaksch, Ardavan, Bonora, Uemura, Takahashi, Hasegawa
  et~al.}}]{Wall2011}
\bibinfo{author}{\bibfnamefont{S.}~\bibnamefont{Wall}},
  \bibinfo{author}{\bibfnamefont{D.}~\bibnamefont{Brida}},
  \bibinfo{author}{\bibfnamefont{S.~R.} \bibnamefont{Clark}},
  \bibinfo{author}{\bibfnamefont{H.~P.} \bibnamefont{Ehrke}},
  \bibinfo{author}{\bibfnamefont{D.}~\bibnamefont{Jaksch}},
  \bibinfo{author}{\bibfnamefont{A.}~\bibnamefont{Ardavan}},
  \bibinfo{author}{\bibfnamefont{S.}~\bibnamefont{Bonora}},
  \bibinfo{author}{\bibfnamefont{H.}~\bibnamefont{Uemura}},
  \bibinfo{author}{\bibfnamefont{Y.}~\bibnamefont{Takahashi}},
  \bibinfo{author}{\bibfnamefont{T.}~\bibnamefont{Hasegawa}},
  \bibnamefont{et~al.}, \bibinfo{journal}{Nat. Phys.}
  \textbf{\bibinfo{volume}{7}}, \bibinfo{pages}{114} (\bibinfo{year}{2011}).

\bibitem[{\citenamefont{Bloch et~al.}(2008)\citenamefont{Bloch, Dalibard, and
  Zwerger}}]{BlochDalibardZwerger2008}
\bibinfo{author}{\bibfnamefont{I.}~\bibnamefont{Bloch}},
  \bibinfo{author}{\bibfnamefont{J.}~\bibnamefont{Dalibard}}, \bibnamefont{and}
  \bibinfo{author}{\bibfnamefont{W.}~\bibnamefont{Zwerger}},
  \bibinfo{journal}{Rev. Mod. Phys.} \textbf{\bibinfo{volume}{80}},
  \bibinfo{pages}{885} (\bibinfo{year}{2008}).

\bibitem[{\citenamefont{J{\"o}rdens et~al.}(2008)\citenamefont{J{\"o}rdens,
  Strohmaier, G{\"u}nter, Moritz, and Esslinger}}]{Joerdens2008}
\bibinfo{author}{\bibfnamefont{R.}~\bibnamefont{J{\"o}rdens}},
  \bibinfo{author}{\bibfnamefont{N.}~\bibnamefont{Strohmaier}},
  \bibinfo{author}{\bibfnamefont{K.}~\bibnamefont{G{\"u}nter}},
  \bibinfo{author}{\bibfnamefont{H.}~\bibnamefont{Moritz}}, \bibnamefont{and}
  \bibinfo{author}{\bibfnamefont{T.}~\bibnamefont{Esslinger}},
  \bibinfo{journal}{Nature} \textbf{\bibinfo{volume}{455}},
  \bibinfo{pages}{204} (\bibinfo{year}{2008}).

\bibitem[{\citenamefont{Schneider et~al.}(2008)\citenamefont{Schneider,
  Hackerm{\"u}ller, Will, Best, Bloch, Costi, Helmes, Rasch, and
  Rosch}}]{Schneider2008}
\bibinfo{author}{\bibfnamefont{U.}~\bibnamefont{Schneider}},
  \bibinfo{author}{\bibfnamefont{L.}~\bibnamefont{Hackerm{\"u}ller}},
  \bibinfo{author}{\bibfnamefont{S.}~\bibnamefont{Will}},
  \bibinfo{author}{\bibfnamefont{T.}~\bibnamefont{Best}},
  \bibinfo{author}{\bibfnamefont{I.}~\bibnamefont{Bloch}},
  \bibinfo{author}{\bibfnamefont{T.~A.} \bibnamefont{Costi}},
  \bibinfo{author}{\bibfnamefont{R.~W.} \bibnamefont{Helmes}},
  \bibinfo{author}{\bibfnamefont{D.}~\bibnamefont{Rasch}}, \bibnamefont{and}
  \bibinfo{author}{\bibfnamefont{A.}~\bibnamefont{Rosch}},
  \bibinfo{journal}{Science} \textbf{\bibinfo{volume}{322}},
  \bibinfo{pages}{1520} (\bibinfo{year}{2008}).

\bibitem[{\citenamefont{Schmitt et~al.}(2008)\citenamefont{Schmitt, Kirchmann,
  Bovensiepen, Moore, Rettig, Krenz, Chu, Ru, Perfetti, Lu
  et~al.}}]{Schmitt2008}
\bibinfo{author}{\bibfnamefont{F.}~\bibnamefont{Schmitt}},
  \bibinfo{author}{\bibfnamefont{P.~S.} \bibnamefont{Kirchmann}},
  \bibinfo{author}{\bibfnamefont{U.}~\bibnamefont{Bovensiepen}},
  \bibinfo{author}{\bibfnamefont{R.~G.} \bibnamefont{Moore}},
  \bibinfo{author}{\bibfnamefont{L.}~\bibnamefont{Rettig}},
  \bibinfo{author}{\bibfnamefont{M.}~\bibnamefont{Krenz}},
  \bibinfo{author}{\bibfnamefont{J.-H.} \bibnamefont{Chu}},
  \bibinfo{author}{\bibfnamefont{N.}~\bibnamefont{Ru}},
  \bibinfo{author}{\bibfnamefont{L.}~\bibnamefont{Perfetti}},
  \bibinfo{author}{\bibfnamefont{D.~H.} \bibnamefont{Lu}},
  \bibnamefont{et~al.}, \bibinfo{journal}{Science}
  \textbf{\bibinfo{volume}{321}}, \bibinfo{pages}{1649} (\bibinfo{year}{2008}).

\bibitem[{\citenamefont{Hellmann et~al.}(2010)\citenamefont{Hellmann, Beye,
  Sohrt, Rohwer, Sorgenfrei, Redlin, Kall\"ane, Marczynski-B\"uhlow, Hennies,
  Bauer et~al.}}]{Hellmann2010}
\bibinfo{author}{\bibfnamefont{S.}~\bibnamefont{Hellmann}},
  \bibinfo{author}{\bibfnamefont{M.}~\bibnamefont{Beye}},
  \bibinfo{author}{\bibfnamefont{C.}~\bibnamefont{Sohrt}},
  \bibinfo{author}{\bibfnamefont{T.}~\bibnamefont{Rohwer}},
  \bibinfo{author}{\bibfnamefont{F.}~\bibnamefont{Sorgenfrei}},
  \bibinfo{author}{\bibfnamefont{H.}~\bibnamefont{Redlin}},
  \bibinfo{author}{\bibfnamefont{M.}~\bibnamefont{Kall\"ane}},
  \bibinfo{author}{\bibfnamefont{M.}~\bibnamefont{Marczynski-B\"uhlow}},
  \bibinfo{author}{\bibfnamefont{F.}~\bibnamefont{Hennies}},
  \bibinfo{author}{\bibfnamefont{M.}~\bibnamefont{Bauer}},
  \bibnamefont{et~al.}, \bibinfo{journal}{Phys. Rev. Lett.}
  \textbf{\bibinfo{volume}{105}}, \bibinfo{pages}{187401}
  (\bibinfo{year}{2010}).

\bibitem[{\citenamefont{Petersen et~al.}(2011)\citenamefont{Petersen, Kaiser,
  Dean, Simoncig, Liu, Cavalieri, Cacho, Turcu, Springate, Frassetto
  et~al.}}]{Petersen2011}
\bibinfo{author}{\bibfnamefont{J.~C.} \bibnamefont{Petersen}},
  \bibinfo{author}{\bibfnamefont{S.}~\bibnamefont{Kaiser}},
  \bibinfo{author}{\bibfnamefont{N.}~\bibnamefont{Dean}},
  \bibinfo{author}{\bibfnamefont{A.}~\bibnamefont{Simoncig}},
  \bibinfo{author}{\bibfnamefont{H.~Y.} \bibnamefont{Liu}},
  \bibinfo{author}{\bibfnamefont{A.~L.} \bibnamefont{Cavalieri}},
  \bibinfo{author}{\bibfnamefont{C.}~\bibnamefont{Cacho}},
  \bibinfo{author}{\bibfnamefont{I.~C.~E.} \bibnamefont{Turcu}},
  \bibinfo{author}{\bibfnamefont{E.}~\bibnamefont{Springate}},
  \bibinfo{author}{\bibfnamefont{F.}~\bibnamefont{Frassetto}},
  \bibnamefont{et~al.}, \bibinfo{journal}{Phys. Rev. Lett.}
  \textbf{\bibinfo{volume}{107}}, \bibinfo{pages}{177402}
  (\bibinfo{year}{2011}).

\bibitem[{\citenamefont{Matsunaga and Shimano}(2012)}]{Matsunaga2012}
\bibinfo{author}{\bibfnamefont{R.}~\bibnamefont{Matsunaga}} \bibnamefont{and}
  \bibinfo{author}{\bibfnamefont{R.}~\bibnamefont{Shimano}},
  \bibinfo{journal}{Phys. Rev. Lett.} \textbf{\bibinfo{volume}{109}},
  \bibinfo{pages}{187002} (\bibinfo{year}{2012}).

\bibitem[{\citenamefont{Fausti et~al.}(2011)\citenamefont{Fausti, Tobey, Dean,
  Kaiser, Dienst, Hoffmann, Pyon, Takayama, Takagi, and
  Cavalleri}}]{Fausti2011}
\bibinfo{author}{\bibfnamefont{D.}~\bibnamefont{Fausti}},
  \bibinfo{author}{\bibfnamefont{R.~I.} \bibnamefont{Tobey}},
  \bibinfo{author}{\bibfnamefont{N.}~\bibnamefont{Dean}},
  \bibinfo{author}{\bibfnamefont{S.}~\bibnamefont{Kaiser}},
  \bibinfo{author}{\bibfnamefont{A.}~\bibnamefont{Dienst}},
  \bibinfo{author}{\bibfnamefont{M.~C.} \bibnamefont{Hoffmann}},
  \bibinfo{author}{\bibfnamefont{S.}~\bibnamefont{Pyon}},
  \bibinfo{author}{\bibfnamefont{T.}~\bibnamefont{Takayama}},
  \bibinfo{author}{\bibfnamefont{H.}~\bibnamefont{Takagi}}, \bibnamefont{and}
  \bibinfo{author}{\bibfnamefont{A.}~\bibnamefont{Cavalleri}},
  \bibinfo{journal}{Science} \textbf{\bibinfo{volume}{331}},
  \bibinfo{pages}{189} (\bibinfo{year}{2011}).

\bibitem[{Kai()}]{Kaiser2012}
\bibinfo{note}{S. Kaiser, D. Nicoletti, C. R. Hunt, W. Hu, I. Gierz, H. Y. Liu,
  M. Le Tacon, T. Loew, D. Haug, B. Keimer, and A. Cavalleri, arXiv:1205.4661.}

\bibitem[{\citenamefont{Demsar et~al.}(1999)\citenamefont{Demsar,
  Biljakovi\ifmmode~\acute{c}\else \'{c}\fi{}, and Mihailovic}}]{Demsar1999}
\bibinfo{author}{\bibfnamefont{J.}~\bibnamefont{Demsar}},
  \bibinfo{author}{\bibfnamefont{K.}~\bibnamefont{Biljakovi\ifmmode~\acute{c}\%
else \'{c}\fi{}}}, \bibnamefont{and}
  \bibinfo{author}{\bibfnamefont{D.}~\bibnamefont{Mihailovic}},
  \bibinfo{journal}{Phys. Rev. Lett.} \textbf{\bibinfo{volume}{83}},
  \bibinfo{pages}{800} (\bibinfo{year}{1999}).

\bibitem[{\citenamefont{Yusupov et~al.}(2010)\citenamefont{Yusupov, Mertelj,
  Kabanov, Brazovskii, Kusar, Chu, Fisher, and Mihailovic}}]{Yusupov2010}
\bibinfo{author}{\bibfnamefont{R.}~\bibnamefont{Yusupov}},
  \bibinfo{author}{\bibfnamefont{T.}~\bibnamefont{Mertelj}},
  \bibinfo{author}{\bibfnamefont{V.~V.} \bibnamefont{Kabanov}},
  \bibinfo{author}{\bibfnamefont{S.}~\bibnamefont{Brazovskii}},
  \bibinfo{author}{\bibfnamefont{P.}~\bibnamefont{Kusar}},
  \bibinfo{author}{\bibfnamefont{J.-H.} \bibnamefont{Chu}},
  \bibinfo{author}{\bibfnamefont{I.~R.} \bibnamefont{Fisher}},
  \bibnamefont{and}
  \bibinfo{author}{\bibfnamefont{D.}~\bibnamefont{Mihailovic}},
  \bibinfo{journal}{Nat. Phys.} \textbf{\bibinfo{volume}{6}},
  \bibinfo{pages}{681} (\bibinfo{year}{2010}).

\bibitem[{\citenamefont{Torchinsky et~al.}(2013)\citenamefont{Torchinsky,
  Mahmood, Bollinger, Bo\v{z}ovi\'c, and Gedik}}]{Torchinsky2013}
\bibinfo{author}{\bibfnamefont{D.~H.} \bibnamefont{Torchinsky}},
  \bibinfo{author}{\bibfnamefont{F.}~\bibnamefont{Mahmood}},
  \bibinfo{author}{\bibfnamefont{A.~T.} \bibnamefont{Bollinger}},
  \bibinfo{author}{\bibfnamefont{I.}~\bibnamefont{Bo\v{z}ovi\'c}},
  \bibnamefont{and} \bibinfo{author}{\bibfnamefont{N.}~\bibnamefont{Gedik}},
  \bibinfo{journal}{Nat. Mater.} \textbf{\bibinfo{volume}{12}},
  \bibinfo{pages}{387} (\bibinfo{year}{2013}).

\bibitem[{\citenamefont{Matsunaga et~al.}(2013)\citenamefont{Matsunaga, Hamada,
  Makise, Uzawa, Terai, Wang, and Shimano}}]{Matsunaga2013}
\bibinfo{author}{\bibfnamefont{R.}~\bibnamefont{Matsunaga}},
  \bibinfo{author}{\bibfnamefont{Y.~I.} \bibnamefont{Hamada}},
  \bibinfo{author}{\bibfnamefont{K.}~\bibnamefont{Makise}},
  \bibinfo{author}{\bibfnamefont{Y.}~\bibnamefont{Uzawa}},
  \bibinfo{author}{\bibfnamefont{H.}~\bibnamefont{Terai}},
  \bibinfo{author}{\bibfnamefont{Z.}~\bibnamefont{Wang}}, \bibnamefont{and}
  \bibinfo{author}{\bibfnamefont{R.}~\bibnamefont{Shimano}},
  \bibinfo{journal}{Phys. Rev. Lett.} \textbf{\bibinfo{volume}{111}},
  \bibinfo{pages}{057002} (\bibinfo{year}{2013}).

\bibitem[{\citenamefont{Kibble}(1976)}]{Kibble1976}
\bibinfo{author}{\bibfnamefont{T.~W.~B.} \bibnamefont{Kibble}},
  \bibinfo{journal}{J. Phys. A} \textbf{\bibinfo{volume}{9}},
  \bibinfo{pages}{1387} (\bibinfo{year}{1976}).

\bibitem[{\citenamefont{Zurek}(1985)}]{Zurek1985}
\bibinfo{author}{\bibfnamefont{W.~H.} \bibnamefont{Zurek}},
  \bibinfo{journal}{Nature (London)} \textbf{\bibinfo{volume}{317}},
  \bibinfo{pages}{505} (\bibinfo{year}{1985}).

\bibitem[{\citenamefont{Berges et~al.}(2004)\citenamefont{Berges, Bors\'anyi,
  and Wetterich}}]{BergesBorsanyiWetterich2004}
\bibinfo{author}{\bibfnamefont{J.}~\bibnamefont{Berges}},
  \bibinfo{author}{\bibfnamefont{S.}~\bibnamefont{Bors\'anyi}},
  \bibnamefont{and}
  \bibinfo{author}{\bibfnamefont{C.}~\bibnamefont{Wetterich}},
  \bibinfo{journal}{Phys. Rev. Lett.} \textbf{\bibinfo{volume}{93}},
  \bibinfo{pages}{142002} (\bibinfo{year}{2004}).

\bibitem[{\citenamefont{Moeckel and Kehrein}(2008)}]{MoeckelKehrein2008}
\bibinfo{author}{\bibfnamefont{M.}~\bibnamefont{Moeckel}} \bibnamefont{and}
  \bibinfo{author}{\bibfnamefont{S.}~\bibnamefont{Kehrein}},
  \bibinfo{journal}{Phys. Rev. Lett.} \textbf{\bibinfo{volume}{100}},
  \bibinfo{pages}{175702} (\bibinfo{year}{2008}).

\bibitem[{\citenamefont{Eckstein et~al.}(2009)\citenamefont{Eckstein, Kollar,
  and Werner}}]{EcksteinKollarWerner2009}
\bibinfo{author}{\bibfnamefont{M.}~\bibnamefont{Eckstein}},
  \bibinfo{author}{\bibfnamefont{M.}~\bibnamefont{Kollar}}, \bibnamefont{and}
  \bibinfo{author}{\bibfnamefont{P.}~\bibnamefont{Werner}},
  \bibinfo{journal}{Phys. Rev. Lett.} \textbf{\bibinfo{volume}{103}},
  \bibinfo{pages}{056403} (\bibinfo{year}{2009}).

\bibitem[{\citenamefont{Berges et~al.}(2008)\citenamefont{Berges, Rothkopf, and
  Schmidt}}]{BergesRothkoptSchmidt2008}
\bibinfo{author}{\bibfnamefont{J.}~\bibnamefont{Berges}},
  \bibinfo{author}{\bibfnamefont{A.}~\bibnamefont{Rothkopf}}, \bibnamefont{and}
  \bibinfo{author}{\bibfnamefont{J.}~\bibnamefont{Schmidt}},
  \bibinfo{journal}{Phys. Rev. Lett.} \textbf{\bibinfo{volume}{101}},
  \bibinfo{pages}{041603} (\bibinfo{year}{2008}).

\bibitem[{\citenamefont{Werner et~al.}(2012)\citenamefont{Werner, Tsuji, and
  Eckstein}}]{WernerTsujiEckstein2012}
\bibinfo{author}{\bibfnamefont{P.}~\bibnamefont{Werner}},
  \bibinfo{author}{\bibfnamefont{N.}~\bibnamefont{Tsuji}}, \bibnamefont{and}
  \bibinfo{author}{\bibfnamefont{M.}~\bibnamefont{Eckstein}},
  \bibinfo{journal}{Phys. Rev. B} \textbf{\bibinfo{volume}{86}},
  \bibinfo{pages}{205101} (\bibinfo{year}{2012}).

\bibitem[{\citenamefont{Tsuji et~al.}(2013)\citenamefont{Tsuji, Eckstein, and
  Werner}}]{TsujiEcksteinWerner2012}
\bibinfo{author}{\bibfnamefont{N.}~\bibnamefont{Tsuji}},
  \bibinfo{author}{\bibfnamefont{M.}~\bibnamefont{Eckstein}}, \bibnamefont{and}
  \bibinfo{author}{\bibfnamefont{P.}~\bibnamefont{Werner}},
  \bibinfo{journal}{Phys. Rev. Lett.} \textbf{\bibinfo{volume}{110}},
  \bibinfo{pages}{136404} (\bibinfo{year}{2013}).

\bibitem[{\citenamefont{Hohenberg and Halperin}(1977)}]{HohenbergHalperin1977}
\bibinfo{author}{\bibfnamefont{P.~C.} \bibnamefont{Hohenberg}}
  \bibnamefont{and} \bibinfo{author}{\bibfnamefont{B.~I.}
  \bibnamefont{Halperin}}, \bibinfo{journal}{Rev. Mod. Phys.}
  \textbf{\bibinfo{volume}{49}}, \bibinfo{pages}{435} (\bibinfo{year}{1977}).

\bibitem[{\citenamefont{Polkovnikov et~al.}(2011)\citenamefont{Polkovnikov,
  Sengupta, Silva, and Vengalattore}}]{Polkovnikov2011}
\bibinfo{author}{\bibfnamefont{A.}~\bibnamefont{Polkovnikov}},
  \bibinfo{author}{\bibfnamefont{K.}~\bibnamefont{Sengupta}},
  \bibinfo{author}{\bibfnamefont{A.}~\bibnamefont{Silva}}, \bibnamefont{and}
  \bibinfo{author}{\bibfnamefont{M.}~\bibnamefont{Vengalattore}},
  \bibinfo{journal}{Rev. Mod. Phys.} \textbf{\bibinfo{volume}{83}},
  \bibinfo{pages}{863} (\bibinfo{year}{2011}).

\bibitem[{\citenamefont{Schmidt and Monien}()}]{SchmidtMonien2002}
\bibinfo{author}{\bibfnamefont{P.}~\bibnamefont{Schmidt}} \bibnamefont{and}
  \bibinfo{author}{\bibfnamefont{H.}~\bibnamefont{Monien}},
  \bibinfo{note}{arXiv:cond-mat/0202046 (unpublished)}.

\bibitem[{\citenamefont{Freericks et~al.}(2006)\citenamefont{Freericks,
  Turkowski, and Zlati\ifmmode~\acute{c}\else
  \'{c}\fi{}}}]{FreericksTurkowskiZlatic2006}
\bibinfo{author}{\bibfnamefont{J.~K.} \bibnamefont{Freericks}},
  \bibinfo{author}{\bibfnamefont{V.~M.} \bibnamefont{Turkowski}},
  \bibnamefont{and}
  \bibinfo{author}{\bibfnamefont{V.}~\bibnamefont{Zlati\ifmmode~\acute{c}\else
  \'{c}\fi{}}}, \bibinfo{journal}{Phys. Rev. Lett.}
  \textbf{\bibinfo{volume}{97}}, \bibinfo{pages}{266408}
  (\bibinfo{year}{2006}).

\bibitem[{\citenamefont{Aoki et~al.}()\citenamefont{Aoki, Tsuji, Eckstein,
  Kollar, Oka, and Werner}}]{AokiTsujiEcksteinKollarOkaWerner2013}
\bibinfo{author}{\bibfnamefont{H.}~\bibnamefont{Aoki}},
  \bibinfo{author}{\bibfnamefont{N.}~\bibnamefont{Tsuji}},
  \bibinfo{author}{\bibfnamefont{M.}~\bibnamefont{Eckstein}},
  \bibinfo{author}{\bibfnamefont{M.}~\bibnamefont{Kollar}},
  \bibinfo{author}{\bibfnamefont{T.}~\bibnamefont{Oka}}, \bibnamefont{and}
  \bibinfo{author}{\bibfnamefont{P.}~\bibnamefont{Werner}}, \bibinfo{note}{to
  be published in Rev. Mod. Phys. (arXiv:1310.5329)}.

\bibitem[{\citenamefont{Georges et~al.}(1996)\citenamefont{Georges, Kotliar,
  Krauth, and Rozenberg}}]{GeorgesKotliarKrauthRozenberg1996}
\bibinfo{author}{\bibfnamefont{A.}~\bibnamefont{Georges}},
  \bibinfo{author}{\bibfnamefont{G.}~\bibnamefont{Kotliar}},
  \bibinfo{author}{\bibfnamefont{W.}~\bibnamefont{Krauth}}, \bibnamefont{and}
  \bibinfo{author}{\bibfnamefont{M.~J.} \bibnamefont{Rozenberg}},
  \bibinfo{journal}{Rev. Mod. Phys.} \textbf{\bibinfo{volume}{68}},
  \bibinfo{pages}{13} (\bibinfo{year}{1996}).

\bibitem[{\citenamefont{Metzner and Vollhardt}(1989)}]{MetznerVollhardt1989}
\bibinfo{author}{\bibfnamefont{W.}~\bibnamefont{Metzner}} \bibnamefont{and}
  \bibinfo{author}{\bibfnamefont{D.}~\bibnamefont{Vollhardt}},
  \bibinfo{journal}{Phys. Rev. Lett.} \textbf{\bibinfo{volume}{62}},
  \bibinfo{pages}{324} (\bibinfo{year}{1989}).

\bibitem[{\citenamefont{M\"uhlbacher and Rabani}(2008)}]{MuehlbacherRabani2008}
\bibinfo{author}{\bibfnamefont{L.}~\bibnamefont{M\"uhlbacher}}
  \bibnamefont{and} \bibinfo{author}{\bibfnamefont{E.}~\bibnamefont{Rabani}},
  \bibinfo{journal}{Phys. Rev. Lett.} \textbf{\bibinfo{volume}{100}},
  \bibinfo{pages}{176403} (\bibinfo{year}{2008}).

\bibitem[{\citenamefont{Werner et~al.}(2009)\citenamefont{Werner, Oka, and
  Millis}}]{WernerOkaMillis2009}
\bibinfo{author}{\bibfnamefont{P.}~\bibnamefont{Werner}},
  \bibinfo{author}{\bibfnamefont{T.}~\bibnamefont{Oka}}, \bibnamefont{and}
  \bibinfo{author}{\bibfnamefont{A.~J.} \bibnamefont{Millis}},
  \bibinfo{journal}{Phys. Rev. B} \textbf{\bibinfo{volume}{79}},
  \bibinfo{pages}{035320} (\bibinfo{year}{2009}).

\bibitem[{\citenamefont{Schir\'o and Fabrizio}(2009)}]{SchiroFabrizio2009}
\bibinfo{author}{\bibfnamefont{M.}~\bibnamefont{Schir\'o}} \bibnamefont{and}
  \bibinfo{author}{\bibfnamefont{M.}~\bibnamefont{Fabrizio}},
  \bibinfo{journal}{Phys. Rev. B} \textbf{\bibinfo{volume}{79}},
  \bibinfo{pages}{153302} (\bibinfo{year}{2009}).

\bibitem[{\citenamefont{Gull et~al.}(2011)\citenamefont{Gull, Millis,
  Lichtenstein, Rubtsov, Troyer, and
  Werner}}]{GullMillisLichtensteinRubtsovTroyerPhilipp2011}
\bibinfo{author}{\bibfnamefont{E.}~\bibnamefont{Gull}},
  \bibinfo{author}{\bibfnamefont{A.~J.} \bibnamefont{Millis}},
  \bibinfo{author}{\bibfnamefont{A.~I.} \bibnamefont{Lichtenstein}},
  \bibinfo{author}{\bibfnamefont{A.~N.} \bibnamefont{Rubtsov}},
  \bibinfo{author}{\bibfnamefont{M.}~\bibnamefont{Troyer}}, \bibnamefont{and}
  \bibinfo{author}{\bibfnamefont{P.}~\bibnamefont{Werner}},
  \bibinfo{journal}{Rev. Mod. Phys.} \textbf{\bibinfo{volume}{83}},
  \bibinfo{pages}{349} (\bibinfo{year}{2011}).

\bibitem[{\citenamefont{Eckstein and Werner}(2010)}]{EcksteinWerner2010}
\bibinfo{author}{\bibfnamefont{M.}~\bibnamefont{Eckstein}} \bibnamefont{and}
  \bibinfo{author}{\bibfnamefont{P.}~\bibnamefont{Werner}},
  \bibinfo{journal}{Phys. Rev. B} \textbf{\bibinfo{volume}{82}},
  \bibinfo{pages}{115115} (\bibinfo{year}{2010}).

\bibitem[{\citenamefont{Arrigoni et~al.}(2013)\citenamefont{Arrigoni, Knap, and
  von~der Linden}}]{Arrigoni2013}
\bibinfo{author}{\bibfnamefont{E.}~\bibnamefont{Arrigoni}},
  \bibinfo{author}{\bibfnamefont{M.}~\bibnamefont{Knap}}, \bibnamefont{and}
  \bibinfo{author}{\bibfnamefont{W.}~\bibnamefont{von~der Linden}},
  \bibinfo{journal}{Phys. Rev. Lett.} \textbf{\bibinfo{volume}{110}},
  \bibinfo{pages}{086403} (\bibinfo{year}{2013}).

\bibitem[{\citenamefont{Gramsch et~al.}()\citenamefont{Gramsch, Balzer,
  Eckstein, and Kollar}}]{Gramsch2013}
\bibinfo{author}{\bibfnamefont{C.}~\bibnamefont{Gramsch}},
  \bibinfo{author}{\bibfnamefont{K.}~\bibnamefont{Balzer}},
  \bibinfo{author}{\bibfnamefont{M.}~\bibnamefont{Eckstein}}, \bibnamefont{and}
  \bibinfo{author}{\bibfnamefont{M.}~\bibnamefont{Kollar}},
  \bibinfo{note}{arXiv:1306.6315}.

\bibitem[{Abr()}]{AbrikosovGorkovDzyaloshinskiBook}
\bibinfo{note}{A. A. Abrikosov, L. P. Gorkov, and I. E. Dzyaloshinski, {\it
  Methods of Quantum Field Theory in Statistical Physics} (Dover, New York,
  1975).}

\bibitem[{\citenamefont{Rammer}(2007)}]{RammerBook}
\bibinfo{author}{\bibfnamefont{J.}~\bibnamefont{Rammer}},
  \emph{\bibinfo{title}{Quantum Field Theory of Non-equilibrium States}}
  (\bibinfo{publisher}{Cambridge University Press},
  \bibinfo{address}{Cambridge}, \bibinfo{year}{2007}).

\bibitem[{\citenamefont{Kamenev}(2011)}]{KamenevBook}
\bibinfo{author}{\bibfnamefont{A.}~\bibnamefont{Kamenev}},
  \emph{\bibinfo{title}{{Field Theory of Non-Equilibrium Systems}}}
  (\bibinfo{publisher}{Cambridge University Press},
  \bibinfo{address}{Cambridge, UK}, \bibinfo{year}{2011}).

\bibitem[{Yos()}]{YosidaYamada}
\bibinfo{note}{K. Yosida and K. Yamada, Prog. Theor. Phys. Suppl. {\bf 46}, 244
  (1970); K. Yamada, Prog. Theor. Phys. {\bf 53}, 970 (1975); K. Yosida and K.
  Yamada, {\it ibid}. {\bf 53}, 1286 (1975).}

\bibitem[{Hor()}]{HorvaticZlatic}
\bibinfo{note}{B. Horvati\'{c} and V. Zlati\'{c}, Phys. Status Solidi B {\bf
  99}, 251 (1980); V. Zlati\'{c} and B. Horvati\'{c}, Phys. Rev. B {\bf 28},
  6904 (1983); V. Zlati\'{c}, B. Horvati\'{c}, and D. \v{S}ok\v{c}evi\'{c}, Z.
  Phys. B {\bf 59}, 151 (1985).}

\bibitem[{\citenamefont{Georges and Kotliar}(1992)}]{GeorgesKotliar1992}
\bibinfo{author}{\bibfnamefont{A.}~\bibnamefont{Georges}} \bibnamefont{and}
  \bibinfo{author}{\bibfnamefont{G.}~\bibnamefont{Kotliar}},
  \bibinfo{journal}{Phys. Rev. B} \textbf{\bibinfo{volume}{45}},
  \bibinfo{pages}{6479} (\bibinfo{year}{1992}).

\bibitem[{\citenamefont{Zhang et~al.}(1993)\citenamefont{Zhang, Rozenberg, and
  Kotliar}}]{ZhangRozenbergKotliar1993}
\bibinfo{author}{\bibfnamefont{X.~Y.} \bibnamefont{Zhang}},
  \bibinfo{author}{\bibfnamefont{M.~J.} \bibnamefont{Rozenberg}},
  \bibnamefont{and} \bibinfo{author}{\bibfnamefont{G.}~\bibnamefont{Kotliar}},
  \bibinfo{journal}{Phys. Rev. Lett.} \textbf{\bibinfo{volume}{70}},
  \bibinfo{pages}{1666} (\bibinfo{year}{1993}).

\bibitem[{\citenamefont{Freericks}(1994)}]{Freericks1994}
\bibinfo{author}{\bibfnamefont{J.~K.} \bibnamefont{Freericks}},
  \bibinfo{journal}{Phys. Rev. B} \textbf{\bibinfo{volume}{50}},
  \bibinfo{pages}{403} (\bibinfo{year}{1994}).

\bibitem[{\citenamefont{Freericks and Jarrell}(1994)}]{FreericksJarrell1994}
\bibinfo{author}{\bibfnamefont{J.~K.} \bibnamefont{Freericks}}
  \bibnamefont{and} \bibinfo{author}{\bibfnamefont{M.}~\bibnamefont{Jarrell}},
  \bibinfo{journal}{Phys. Rev. B} \textbf{\bibinfo{volume}{50}},
  \bibinfo{pages}{6939} (\bibinfo{year}{1994}).

\bibitem[{\citenamefont{Gebhard et~al.}(2003)\citenamefont{Gebhard, Jeckelmann,
  Mahlert, Nishimoto, and Noack}}]{GebhardJeckelmannMahlertNishimotoNoack2003}
\bibinfo{author}{\bibfnamefont{F.}~\bibnamefont{Gebhard}},
  \bibinfo{author}{\bibfnamefont{E.}~\bibnamefont{Jeckelmann}},
  \bibinfo{author}{\bibfnamefont{S.}~\bibnamefont{Mahlert}},
  \bibinfo{author}{\bibfnamefont{S.}~\bibnamefont{Nishimoto}},
  \bibnamefont{and} \bibinfo{author}{\bibfnamefont{R.}~\bibnamefont{Noack}},
  \bibinfo{journal}{Eur. Phys. J. B} \textbf{\bibinfo{volume}{36}},
  \bibinfo{pages}{491} (\bibinfo{year}{2003}).

\bibitem[{Her()}]{HershfieldDaviesWilkins}
\bibinfo{note}{S. Hershfield, J. H. Davies, and J. W. Wilkins, Phys. Rev. Lett.
  {\bf 67}, 3720 (1991); Phys. Rev. B {\bf 46}, 7046 (1992).}

\bibitem[{Fuj()}]{FujiiUeda}
\bibinfo{note}{T. Fujii and K. Ueda, Phys. Rev. B {\bf 68}, 155310 (2003); J.
  Phys. Soc. Jpn. {\bf 74}, 127 (2005).}

\bibitem[{\citenamefont{Eckstein et~al.}(2010)\citenamefont{Eckstein, Kollar,
  and Werner}}]{EcksteinKollarWerner2010}
\bibinfo{author}{\bibfnamefont{M.}~\bibnamefont{Eckstein}},
  \bibinfo{author}{\bibfnamefont{M.}~\bibnamefont{Kollar}}, \bibnamefont{and}
  \bibinfo{author}{\bibfnamefont{P.}~\bibnamefont{Werner}},
  \bibinfo{journal}{Phys. Rev. B} \textbf{\bibinfo{volume}{81}},
  \bibinfo{pages}{115131} (\bibinfo{year}{2010}).

\bibitem[{\citenamefont{Aron et~al.}(2012)\citenamefont{Aron, Kotliar, and
  Weber}}]{AronKotliarWeber2012}
\bibinfo{author}{\bibfnamefont{C.}~\bibnamefont{Aron}},
  \bibinfo{author}{\bibfnamefont{G.}~\bibnamefont{Kotliar}}, \bibnamefont{and}
  \bibinfo{author}{\bibfnamefont{C.}~\bibnamefont{Weber}},
  \bibinfo{journal}{Phys. Rev. Lett.} \textbf{\bibinfo{volume}{108}},
  \bibinfo{pages}{086401} (\bibinfo{year}{2012}).

\bibitem[{\citenamefont{Amaricci et~al.}(2012)\citenamefont{Amaricci, Weber,
  Capone, and Kotliar}}]{Amaricci2012}
\bibinfo{author}{\bibfnamefont{A.}~\bibnamefont{Amaricci}},
  \bibinfo{author}{\bibfnamefont{C.}~\bibnamefont{Weber}},
  \bibinfo{author}{\bibfnamefont{M.}~\bibnamefont{Capone}}, \bibnamefont{and}
  \bibinfo{author}{\bibfnamefont{G.}~\bibnamefont{Kotliar}},
  \bibinfo{journal}{Phys. Rev. B} \textbf{\bibinfo{volume}{86}},
  \bibinfo{pages}{085110} (\bibinfo{year}{2012}).

\bibitem[{\citenamefont{Tsuji et~al.}(2012)\citenamefont{Tsuji, Oka, Aoki, and
  Werner}}]{TsujiOkaAokiWerner2012}
\bibinfo{author}{\bibfnamefont{N.}~\bibnamefont{Tsuji}},
  \bibinfo{author}{\bibfnamefont{T.}~\bibnamefont{Oka}},
  \bibinfo{author}{\bibfnamefont{H.}~\bibnamefont{Aoki}}, \bibnamefont{and}
  \bibinfo{author}{\bibfnamefont{P.}~\bibnamefont{Werner}},
  \bibinfo{journal}{Phys. Rev. B} \textbf{\bibinfo{volume}{85}},
  \bibinfo{pages}{155124} (\bibinfo{year}{2012}).

\bibitem[{Kad()}]{KadanoffBaymBook}
\bibinfo{note}{L. P. Kadanoff and G. Baym, {\it Quantum Statistical Mechanics}
  (W. A. Benjamin, New York, 1962).}

\bibitem[{\citenamefont{Keldysh}(1965)}]{Keldysh1964}
\bibinfo{author}{\bibfnamefont{L.~V.} \bibnamefont{Keldysh}},
  \bibinfo{journal}{Zh. Eksp. Teor. Fiz.} \textbf{\bibinfo{volume}{47}},
  \bibinfo{pages}{1515} (\bibinfo{year}{1965}), \bibinfo{note}{[Sov. Phys. JETP
  {\bf 20}, 1018 (1965)]}.

\bibitem[{Num()}]{NumericalRecipesC}
\bibinfo{note}{W. H. Press, S. A. Teukolsky, W. T. Vetterling, and B. P.
  Flannery, {\it Numerical Recipes in C}, 2nd ed. (Cambridge University Press,
  Cambridge, 1992).}

\bibitem[{Lin()}]{LinzBook}
\bibinfo{note}{P. Linz, {\it Analytical and Numerical Methods for Volterra
  Equations}, (SIAM, Philadelphia, 1985).}

\bibitem[{Bru()}]{BrunnervanderHouwenBook}
\bibinfo{note}{H. Brunner and P. J. van der Houwen, {\it The Numerical Solution
  of Volterra Equations} (North-Holland, Amsterdam, 1986).}

\bibitem[{\citenamefont{Rubtsov et~al.}(2005)\citenamefont{Rubtsov, Savkin, and
  Lichtenstein}}]{Rubtsov2005}
\bibinfo{author}{\bibfnamefont{A.~N.} \bibnamefont{Rubtsov}},
  \bibinfo{author}{\bibfnamefont{V.~V.} \bibnamefont{Savkin}},
  \bibnamefont{and} \bibinfo{author}{\bibfnamefont{A.~I.}
  \bibnamefont{Lichtenstein}}, \bibinfo{journal}{Phys. Rev. B}
  \textbf{\bibinfo{volume}{72}}, \bibinfo{pages}{035122}
  (\bibinfo{year}{2005}).

\bibitem[{\citenamefont{Werner et~al.}(2010)\citenamefont{Werner, Oka,
  Eckstein, and Millis}}]{WernerOkaEcksteinMillis2010}
\bibinfo{author}{\bibfnamefont{P.}~\bibnamefont{Werner}},
  \bibinfo{author}{\bibfnamefont{T.}~\bibnamefont{Oka}},
  \bibinfo{author}{\bibfnamefont{M.}~\bibnamefont{Eckstein}}, \bibnamefont{and}
  \bibinfo{author}{\bibfnamefont{A.~J.} \bibnamefont{Millis}},
  \bibinfo{journal}{Phys. Rev. B} \textbf{\bibinfo{volume}{81}},
  \bibinfo{pages}{035108} (\bibinfo{year}{2010}).

\bibitem[{Mah()}]{MahanBook}
\bibinfo{note}{G. D. Mahan, {\it Many-Particle Physics}, 3rd ed. (Plenum, New
  York, 2000).}

\bibitem[{\citenamefont{Baym and Kadanoff}(1961)}]{BaymKadanoff1961}
\bibinfo{author}{\bibfnamefont{G.}~\bibnamefont{Baym}} \bibnamefont{and}
  \bibinfo{author}{\bibfnamefont{L.~P.} \bibnamefont{Kadanoff}},
  \bibinfo{journal}{Phys. Rev.} \textbf{\bibinfo{volume}{124}},
  \bibinfo{pages}{287} (\bibinfo{year}{1961}).

\bibitem[{\citenamefont{Kajueter and Kotliar}(1996)}]{KajueterKotliar1996}
\bibinfo{author}{\bibfnamefont{H.}~\bibnamefont{Kajueter}} \bibnamefont{and}
  \bibinfo{author}{\bibfnamefont{G.}~\bibnamefont{Kotliar}},
  \bibinfo{journal}{Phys. Rev. Lett.} \textbf{\bibinfo{volume}{77}},
  \bibinfo{pages}{131} (\bibinfo{year}{1996}).

\bibitem[{\citenamefont{Koga and Werner}(2011)}]{Koga2011}
\bibinfo{author}{\bibfnamefont{A.}~\bibnamefont{Koga}} \bibnamefont{and}
  \bibinfo{author}{\bibfnamefont{P.}~\bibnamefont{Werner}},
  \bibinfo{journal}{Phys. Rev. A} \textbf{\bibinfo{volume}{84}},
  \bibinfo{pages}{023638} (\bibinfo{year}{2011}).

\bibitem[{\citenamefont{Keller et~al.}(1999)\citenamefont{Keller, Metzner, and
  Schollw\"ock}}]{KellerMetznerSchollwoeck1999}
\bibinfo{author}{\bibfnamefont{M.}~\bibnamefont{Keller}},
  \bibinfo{author}{\bibfnamefont{W.}~\bibnamefont{Metzner}}, \bibnamefont{and}
  \bibinfo{author}{\bibfnamefont{U.}~\bibnamefont{Schollw\"ock}},
  \bibinfo{journal}{Phys. Rev. B} \textbf{\bibinfo{volume}{60}},
  \bibinfo{pages}{3499} (\bibinfo{year}{1999}).

\bibitem[{\citenamefont{Keller et~al.}(2001)\citenamefont{Keller, Metzner, and
  Schollw\"ock}}]{KellerMetznerSchollwoeck2001}
\bibinfo{author}{\bibfnamefont{M.}~\bibnamefont{Keller}},
  \bibinfo{author}{\bibfnamefont{W.}~\bibnamefont{Metzner}}, \bibnamefont{and}
  \bibinfo{author}{\bibfnamefont{U.}~\bibnamefont{Schollw\"ock}},
  \bibinfo{journal}{Phys. Rev. Lett.} \textbf{\bibinfo{volume}{86}},
  \bibinfo{pages}{4612} (\bibinfo{year}{2001}).

\bibitem[{\citenamefont{Moeckel and Kehrein}(2010)}]{MoeckelKehrein2010}
\bibinfo{author}{\bibfnamefont{M.}~\bibnamefont{Moeckel}} \bibnamefont{and}
  \bibinfo{author}{\bibfnamefont{S.}~\bibnamefont{Kehrein}},
  \bibinfo{journal}{New J. Phys.} \textbf{\bibinfo{volume}{12}},
  \bibinfo{pages}{055016} (\bibinfo{year}{2010}).

\bibitem[{\citenamefont{Schir\'o and Fabrizio}(2010)}]{SchiroFabrizio2010}
\bibinfo{author}{\bibfnamefont{M.}~\bibnamefont{Schir\'o}} \bibnamefont{and}
  \bibinfo{author}{\bibfnamefont{M.}~\bibnamefont{Fabrizio}},
  \bibinfo{journal}{Phys. Rev. Lett.} \textbf{\bibinfo{volume}{105}},
  \bibinfo{pages}{076401} (\bibinfo{year}{2010}).

\bibitem[{\citenamefont{Schir\'o and Fabrizio}(2011)}]{SchiroFabrizio2011}
\bibinfo{author}{\bibfnamefont{M.}~\bibnamefont{Schir\'o}} \bibnamefont{and}
  \bibinfo{author}{\bibfnamefont{M.}~\bibnamefont{Fabrizio}},
  \bibinfo{journal}{Phys. Rev. B} \textbf{\bibinfo{volume}{83}},
  \bibinfo{pages}{165105} (\bibinfo{year}{2011}).

\bibitem[{\citenamefont{Kollar et~al.}(2011)\citenamefont{Kollar, Wolf, and
  Eckstein}}]{KollarWolfEckstein2011}
\bibinfo{author}{\bibfnamefont{M.}~\bibnamefont{Kollar}},
  \bibinfo{author}{\bibfnamefont{F.~A.} \bibnamefont{Wolf}}, \bibnamefont{and}
  \bibinfo{author}{\bibfnamefont{M.}~\bibnamefont{Eckstein}},
  \bibinfo{journal}{Phys. Rev. B} \textbf{\bibinfo{volume}{84}},
  \bibinfo{pages}{054304} (\bibinfo{year}{2011}).

\bibitem[{\citenamefont{Hamerla and Uhrig}(2013)}]{HamerlaUhrig2013}
\bibinfo{author}{\bibfnamefont{S.~A.} \bibnamefont{Hamerla}} \bibnamefont{and}
  \bibinfo{author}{\bibfnamefont{G.~S.} \bibnamefont{Uhrig}},
  \bibinfo{journal}{Phys. Rev. B} \textbf{\bibinfo{volume}{87}},
  \bibinfo{pages}{064304} (\bibinfo{year}{2013}).

\bibitem[{\citenamefont{Hamerla and Uhrig}()}]{HamerlaUhrig2013b}
\bibinfo{author}{\bibfnamefont{S.}~\bibnamefont{Hamerla}} \bibnamefont{and}
  \bibinfo{author}{\bibfnamefont{G.}~\bibnamefont{Uhrig}},
  \bibinfo{note}{arXiv:1307.3438}.

\bibitem[{\citenamefont{Stark and Kollar}()}]{StarkKollar2013}
\bibinfo{author}{\bibfnamefont{M.}~\bibnamefont{Stark}} \bibnamefont{and}
  \bibinfo{author}{\bibfnamefont{M.}~\bibnamefont{Kollar}},
  \bibinfo{note}{arXiv:1308.1610}.

\bibitem[{\citenamefont{M{\"u}ller-Hartmann}(1989)}]{Muller-Hartmann1989b}
\bibinfo{author}{\bibfnamefont{E.}~\bibnamefont{M{\"u}ller-Hartmann}},
  \bibinfo{journal}{Z. Phys. B: Condens. Matter} \textbf{\bibinfo{volume}{76}},
  \bibinfo{pages}{211} (\bibinfo{year}{1989}).

\bibitem[{\citenamefont{Werner et~al.}(2005)\citenamefont{Werner, Parcollet,
  Georges, and Hassan}}]{WernerParcolletGeorgesHassan2005}
\bibinfo{author}{\bibfnamefont{F.}~\bibnamefont{Werner}},
  \bibinfo{author}{\bibfnamefont{O.}~\bibnamefont{Parcollet}},
  \bibinfo{author}{\bibfnamefont{A.}~\bibnamefont{Georges}}, \bibnamefont{and}
  \bibinfo{author}{\bibfnamefont{S.~R.} \bibnamefont{Hassan}},
  \bibinfo{journal}{Phys. Rev. Lett.} \textbf{\bibinfo{volume}{95}},
  \bibinfo{pages}{056401} (\bibinfo{year}{2005}).

\bibitem[{\citenamefont{Schmid}(1966)}]{Schmid1966}
\bibinfo{author}{\bibfnamefont{A.}~\bibnamefont{Schmid}},
  \bibinfo{journal}{Phys. Kondens. Mater.} \textbf{\bibinfo{volume}{5}},
  \bibinfo{pages}{302} (\bibinfo{year}{1966}).

\bibitem[{\citenamefont{Abrahams and Tsuneto}(1966)}]{AbrahamsTsuneto1966}
\bibinfo{author}{\bibfnamefont{E.}~\bibnamefont{Abrahams}} \bibnamefont{and}
  \bibinfo{author}{\bibfnamefont{T.}~\bibnamefont{Tsuneto}},
  \bibinfo{journal}{Phys. Rev.} \textbf{\bibinfo{volume}{152}},
  \bibinfo{pages}{416} (\bibinfo{year}{1966}).

\bibitem[{\citenamefont{S\'a~de Melo et~al.}(1993)\citenamefont{S\'a~de Melo,
  Randeria, and Engelbrecht}}]{SadeMelo1993}
\bibinfo{author}{\bibfnamefont{C.~A.~R.} \bibnamefont{S\'a~de Melo}},
  \bibinfo{author}{\bibfnamefont{M.}~\bibnamefont{Randeria}}, \bibnamefont{and}
  \bibinfo{author}{\bibfnamefont{J.~R.} \bibnamefont{Engelbrecht}},
  \bibinfo{journal}{Phys. Rev. Lett.} \textbf{\bibinfo{volume}{71}},
  \bibinfo{pages}{3202} (\bibinfo{year}{1993}).

\bibitem[{\citenamefont{Barankov et~al.}(2004)\citenamefont{Barankov, Levitov,
  and Spivak}}]{Barankov2004}
\bibinfo{author}{\bibfnamefont{R.~A.} \bibnamefont{Barankov}},
  \bibinfo{author}{\bibfnamefont{L.~S.} \bibnamefont{Levitov}},
  \bibnamefont{and} \bibinfo{author}{\bibfnamefont{B.~Z.}
  \bibnamefont{Spivak}}, \bibinfo{journal}{Phys. Rev. Lett.}
  \textbf{\bibinfo{volume}{93}}, \bibinfo{pages}{160401}
  (\bibinfo{year}{2004}).

\bibitem[{\citenamefont{Anderson}(1958)}]{Anderson1958}
\bibinfo{author}{\bibfnamefont{P.~W.} \bibnamefont{Anderson}},
  \bibinfo{journal}{Phys. Rev.} \textbf{\bibinfo{volume}{112}},
  \bibinfo{pages}{1900} (\bibinfo{year}{1958}).

\bibitem[{\citenamefont{Yuzbashyan et~al.}(2005)\citenamefont{Yuzbashyan,
  Altshuler, Kuznetsov, and Enolskii}}]{Yuzbashyan2005}
\bibinfo{author}{\bibfnamefont{E.~A.} \bibnamefont{Yuzbashyan}},
  \bibinfo{author}{\bibfnamefont{B.~L.} \bibnamefont{Altshuler}},
  \bibinfo{author}{\bibfnamefont{V.~B.} \bibnamefont{Kuznetsov}},
  \bibnamefont{and} \bibinfo{author}{\bibfnamefont{V.~Z.}
  \bibnamefont{Enolskii}}, \bibinfo{journal}{Phys. Rev. B}
  \textbf{\bibinfo{volume}{72}}, \bibinfo{pages}{220503}
  (\bibinfo{year}{2005}).

\bibitem[{\citenamefont{Warner and Leggett}(2005)}]{WarnerLeggett2005}
\bibinfo{author}{\bibfnamefont{G.~L.} \bibnamefont{Warner}} \bibnamefont{and}
  \bibinfo{author}{\bibfnamefont{A.~J.} \bibnamefont{Leggett}},
  \bibinfo{journal}{Phys. Rev. B} \textbf{\bibinfo{volume}{71}},
  \bibinfo{pages}{134514} (\bibinfo{year}{2005}).

\bibitem[{\citenamefont{Barankov and Levitov}(2006)}]{Barankov2006}
\bibinfo{author}{\bibfnamefont{R.~A.} \bibnamefont{Barankov}} \bibnamefont{and}
  \bibinfo{author}{\bibfnamefont{L.~S.} \bibnamefont{Levitov}},
  \bibinfo{journal}{Phys. Rev. Lett.} \textbf{\bibinfo{volume}{96}},
  \bibinfo{pages}{230403} (\bibinfo{year}{2006}).

\bibitem[{\citenamefont{Yuzbashyan and Dzero}(2006)}]{Yuzbashyan2006}
\bibinfo{author}{\bibfnamefont{E.~A.} \bibnamefont{Yuzbashyan}}
  \bibnamefont{and} \bibinfo{author}{\bibfnamefont{M.}~\bibnamefont{Dzero}},
  \bibinfo{journal}{Phys. Rev. Lett.} \textbf{\bibinfo{volume}{96}},
  \bibinfo{pages}{230404} (\bibinfo{year}{2006}).

\bibitem[{\citenamefont{Shiba}(1972)}]{Shiba1972}
\bibinfo{author}{\bibfnamefont{H.}~\bibnamefont{Shiba}},
  \bibinfo{journal}{Prog. Theor. Phys.} \textbf{\bibinfo{volume}{48}},
  \bibinfo{pages}{2171} (\bibinfo{year}{1972}).

\bibitem[{\citenamefont{Langreth}(1976)}]{Langreth1976}
\bibinfo{author}{\bibfnamefont{D.~C.} \bibnamefont{Langreth}},
  \emph{\bibinfo{title}{Linear and Nonlinear Electron Transport in Solids}}
  (\bibinfo{publisher}{Plenum Press}, \bibinfo{address}{New York and London},
  \bibinfo{year}{1976}), \bibinfo{note}{edited by J. T. Devreese and V. E. van
  Doren}.

\end{thebibliography}

\end{document}